\documentclass[12pt,oneside,reqno]{amsart}
\usepackage[T1]{fontenc}
\usepackage[latin9]{inputenc}
\usepackage[letterpaper]{geometry}
\usepackage{bm}
\usepackage{amsthm}
\usepackage{amstext}
\usepackage{hyperref}
\usepackage{breakurl}
\usepackage{amssymb}
\usepackage{graphicx}
\usepackage{setspace}
\usepackage{esint}
\usepackage{rotating}
\usepackage[authoryear]{natbib}

\makeatletter

\providecommand{\tabularnewline}{\\}

\theoremstyle{plain}

\newtheorem{lemma}{Lemma}[section]

\newtheorem{algthm}{Algorithm}[section]

\makeatother

\begin{document}

\title{Efficient Bayesian Multivariate Surface Regression}

\author{Feng Li}
\author{Mattias Villani}

\thanks{Li (corresponding author): Department of Statistics, Stockholm University,
SE-106 91 Stockholm, Sweden. E-mail: \href{mailto:feng.li@stat.su.se}{\texttt{feng.li@stat.su.se}}.
Villani: Division of Statistics, Department of Computer and Information
Science, Linköping University, SE-581 83 Linköping, Sweden. E-mail:
\href{mailto:mattias.villani@liu.se}{\texttt{mattias.villani@liu.se}}.}
\begin{abstract}
Methods for choosing a fixed set of knot locations in additive spline
models are fairly well established in the statistical literature.
While most of these methods are in principle directly extendable to
non-additive surface models, they are less likely to be successful
in that setting because of the curse of dimensionality, especially
when there are more than a couple of covariates. We propose a regression
model for a multivariate Gaussian response that combines both additive
splines and interactive splines, and a highly efficient MCMC algorithm
that updates all the knot locations jointly. We use shrinkage priors
to avoid overfitting with different estimated shrinkage factors for
the additive and surface part of the model, and also different shrinkage
parameters for the different response variables. This makes it possible
for the model to adapt to varying degrees of nonlinearity in different
parts of the data in a parsimonious way. Simulated data and an application
to firm leverage data show that the approach is computationally efficient,
and that allowing for freely estimated knot locations can offer a
substantial improvement in out-of-sample predictive performance.

\end{abstract}
\maketitle

\section{Introduction}

\label{sec:Introduction}

Flexible models of the regression function $\mathrm{E}(y|x)$ has
been an active research field for decades, see e.g. \citet{ruppert2003semiparametric}
for a recent textbook introduction and further references. Intensive
research was initially devoted to kernel regression methods \citep{Nada1964,watson1964smooth,gasser1979kernel},
and later followed by a large literature on spline regression modeling.
A spline is a linear regression on a set of nonlinear basis functions
of the original regressors. Each basis function is defined from a
knot in regressor space and the knots determine the points of flexibility
of the fitted regression function. This gives rise to a locally adaptable
model with continuity at the knots.

The most widely used models assume additivity in the regressors, i.e.
$\mathrm{E}(y|x_{1},...,x_{q})=\sum_{j=1}^{q}f_{j}(x_{j})$, where
$f_{j}(x_{j})$ is a spline function for the $j$th regressor \citep{hastie1990generalized}.
Assuming additivity is clearly a very convenient simplification, but
it is also somewhat unnatural to make such a strong assumption in
an otherwise very flexible model. This has motivated research on surface
models with interactions between regressors. One line of research
extends the additive models by including higher-order interactions
of the spline basis functions, see e.g. the structured ANOVA approach
or the tensor product basis in \citet{hastie2009elements}. The multivariate
adaptive regression splines (MARS) introduced in \citet{friedman1991multivariate}
is a version of the tensor product spline with interactions sequentially
entering the model using a greedy algorithm. Regression trees \citep{breiman1984classification}
is another popular class of models, with the BART model in \citet{chipman2010bart}
as its most prominent Bayesian member. Our paper follows a recent
strand of literature that models surfaces using radial basis functions
splines, see e.g. \citet{buhmann2003radial}. A radial basis function
is defined in $\mathbb{R}^{q}$ and has a value that depends only
on the distance from a covariate vector ($\bm{x}$) to its $q$-dimensional
knot ($\bm{\xi}$), e.g. the cubic radial basis $\left\Vert \bm{x}-\bm{\xi}\right\Vert ^{3}$,
where $\bm{x}=(x_{1},...,x_{q})'$, $\bm{\xi}=(\xi_{1},...,\xi_{q})'$
and $\left\Vert \cdot\right\Vert $ is the Euclidean norm. The model
is again linear in the basis expanded space.

The basic challenge in spline regression is the choice of knot locations.
This problem is clearly much harder for a general surface than it
is for additive models since any manageable set of $q$-dimensional
knots are necessarily sparse in $\mathbb{R}^{q}$ when $q$ is moderate
or large, a manifestation of the curse of dimensionality. The state-of-the-art
inferential procedures place the knots at the centroids from a clustering
of the regressor observations. The selected knot locations are kept
fixed throughout the analysis. To prevent overfitting, Bayesian variable
selection methods are used to automatically remove or downweight the
influence of the knots using Markov chain Monte Carlo (MCMC) methods
\citep{SmithKohn1996}. The reversible jump MCMC (RJMCMC) in for example
\citet{denisonbayesian} treats the number of knots as unknown subject
to an upper bound, but the location of the knots are still fixed throughout
the analysis.

Using a fixed set of knot locations is impractical when estimating
a surface with more than a few regressors. An algorithm that can move
the knots rapidly over the regressor space is expected to be a clear
improvement. All previous attempts have focused on efficient selection
of fixed knots, and have paid little attention to moving the knots.
The otherwise very elaborate RJMCMC approaches in \citet{dimatteo2001bayesian},
\citet{denisonmallickSmith1998}, \citet{RazulFitzgeraldAndrieu2003}
and \citet{holmesmallick2003JASA} all include a very simple MCMC
update where a single knot is re-located using a Metropolis random
walk step with a proposal variance that is the same for all knots.
There are typically strong dependencies between the knots, and local
one-knot-at-a-time moves will lead to slow convergence of the algorithm
and inability to escape from local modes, see Section \ref{sub:Algorithms-and-prior}
for some evidence. This is especially true in the surface case with
more than a couple of regressors.

The main contribution in this paper is a highly efficient MCMC algorithm
for the Gaussian multivariate surface regression where the locations
of all knots are updated jointly. Rapid mixing of the knot locations
is obtained from the following two features of our algorithm. First,
the knots are simulated from a marginal posterior where the high-dimensional
regression coefficients have been integrated out analytically. Second,
the knots' proposal distribution is tailored to the posterior distribution
using the posterior gradient, which we derive in compact analytical
form and evaluate efficiently by a careful use of sparsity. We use
a shrinkage prior on the regression coefficients to prevent overfitting,
where the shrinkage hyperparameters are treated as unknowns and are
estimated in a separate updating step. Also this step is tailored
to the posterior using the gradient in analytical form.

Even a highly efficient MCMC algorithm is likely to have problems
exploring the joint posterior of many surface knots in a high-dimensional
covariate space. To deal with this, our model is decomposed into three
parts: i) the original covariates entering in linear form, ii) additive
spline basis functions and iii) radial basis functions for capturing
the remaining part of the surface and interactions. The idea is to
let the additive part of the model capture the bulk of the nonlinearities
so that the radial basis functions can focus exclusively on modeling
the interactions. This way we can keep the number of knots in the
interaction part of the model to a minimum, which is beneficial for
MCMC convergence. We use separate shrinkage priors for the three parts
of the model. Moreover, we also allow for separate shrinkage parameters
in each response equation. This gives us an extremely flexible yet
potentially parsimonious model where we can shrink out e.g. the surface
part of the model in a subset of the response equations.

Our MCMC scheme is designed for a fixed number of knots, and we select
the number of knots by Bayesian cross-validation of the log predictive
score using parallel computing, see Section \ref{sec:model-comp}.
This has the disadvantage of not accounting for the uncertainty regarding
the number of knots as is done in RJMCMC schemes, but the benefits
are substantially more robustness to variations in the prior and improved
MCMC efficiency.

We illustrate our algorithm on simulated and real data, and compare
the predictive performance of the models using Bayesian cross-validation
techniques. We find that the free knots model constantly outperforms
the model with fixed knots. Additionally, we find it is easier to
obtain better fitting result by combining additive knots and surface
knots in the model.

\section{Bayesian multivariate surface regression}

\label{sec:MultiSurfaceRegression}

\subsection{The model}

\label{sub:Model} Our proposed model is a Gaussian multivariate regression
with three sets of covariates:
\begin{equation}
\begin{gathered}\bm{Y}=\bm{X}_{o}\bm{B}_{o}+\bm{X}_{a}(\bm{\xi}_{a})\bm{B}_{a}+\bm{X}_{s}(\bm{\xi}_{s})\bm{B}_{s}+\bm{E},\end{gathered}
\label{eq:model}
\end{equation}
where $\bm{Y}(n\times p)$ contains $n$ observations on $p$ response
variables, and the rows of $\bm{E}$ are error vectors assumed to
be iid $\mathrm{N}_{p}(\bm{0},\bm{\Sigma})$. The matrix $\bm{X}_{o}(n\times q_{o})$
contains the original regressors (first column is a vector of ones
for the intercept) and $\bm{B}_{o}$ holds the corresponding regression
coefficients. The $q_{a}$ columns of the matrix $\bm{X}_{a}(\bm{\xi}_{a})$
are additive splines functions of the covariates in $\bm{X}_{o}$.
Our notation makes it clear that $\bm{X}_{a}$ depends on the knots
$\bm{\xi}_{a}$. Note that the knots in the additive part of the model
are scalars, and that our model allows for unequal number of knots
in the different covariates. Finally, $\bm{X}_{s}(\bm{\xi}_{s})$
contains the surface, or interaction, part of the model. The knots
in $\bm{\xi}_{s}$ are $q_{o}$-dimensional vectors. Note how this
decomposition makes it possible for the additive part of the model
to capture the main part of the nonlinearities so that the number
of knots in $\bm{X}_{s}$ is kept to a minimum. We will refer to the
three different parts of the model as the\emph{ linear component},
the \emph{additive component} and the \emph{surface component}, respectively.
We will refer to $\bm{\xi}_{a}$ and $\bm{\xi}_{s}$ as the additive
and surface knots, respectively. Likewise, $\bm{B}_{a}$ and $\bm{B}_{s}$
are the additive and surface coefficients.

There are a large number of different spline bases that one can use
for the additive part of the model. The menu of choices for the surface
basis is more limited, see \citet{denisonbayesian} for a survey of
the most commonly used bases. We will use thin-plate splines for illustration,
but our approach can be used with any basis with trivial changes,
see Section \ref{sec:MCMC} and Appendix \ref{AppendixGradients}
for computational details. The thin-plate spline basis in the surface
case is of the form
\begin{equation}
{\bm{x}_{sj}(\bm{\xi}_{sj})}=\Vert\bm{x}_{o}-\bm{\xi}_{sj}\Vert^{2}\ln\Vert\bm{x}_{o}-\bm{\xi}_{sj}\Vert,~j=1,...,q_{s},\label{eq:multiRadial}
\end{equation}
where $\bm{x}_{o}$ is one of the original data points and $\bm{\xi}_{sj}$
is the $j$th $q_{o}$-dimensional surface knot. The univariate thin-plate
basis used in the additive part is a special case of the multivariate
thin-plate in (\ref{eq:multiRadial}) where both the data point and
the knot are one-dimensional.

For notational convenience, we sometimes write model \eqref{eq:model}
in compact form
\[
\bm{Y}=\bm{X}\bm{B}+\bm{E},
\]
where $\bm{X}=[\bm{X}_{o},\bm{X}_{a},\bm{X}_{s}]$ is the $n\times q$
design matrix ($q=q_{o}+q_{a}+q_{s}$) and $\bm{B}=[\bm{B}_{o}^{'},\bm{B}_{a}^{'},\bm{B}_{s}^{'}]^{'}$.
Define also $\bm{b}_{i}=\mathrm{vec}\bm{B}_{i}$ as the vectorization
of the coefficients matrix $\bm{B}_{i}$, and $\bm{b}=[\bm{b}_{o}^{'},\bm{b}_{a}^{'},\bm{b}_{s}^{'}]^{'}$.

For a given set of fixed knot locations, the model in \eqref{eq:model}
is linear in the regression coefficients $\bm{B}$. As explained in
the Introduction, the great challenge with spline models is the choice
of knot locations. This is especially true in the surface case where
the curse of dimensionality makes it really hard to distribute the
multi-dimensional knots in $\mathbb{R}^{q_{o}}$ in an effective way.
To get a fair coverage of knots in the covariate space, a recommended
approach is to place the knots at the cluster centers from some clustering
algorithm, e.g. $k$-means clustering or using a mixture of multivariate
normals, see \citet{SmithKohn1996} and \citet{denisonmallickSmith1998}.
This typically leads to many redundant knots (since the response variables
are not used to aid the clustering) which is a source of overfitting.
One solution is to remove (downweight) the knots by Bayesian variable
selection \citep{SmithKohn1996}, possibly in a RJMCMC approach, see
e.g. \citet{dimatteo2001bayesian} and \citet{denisonbayesian}. Nevertheless,
using a set of pre-determined knots is unlikely to work well in the
surface case with more than a handful of regressors.

We will treat the knot locations in $\bm{\xi}_{a}$ and $\bm{\xi}_{s}$
as unknown parameters to be estimated. This is in principle straightforward
from a Bayesian point of view, but great care is needed in the actual
implementation of the posterior computations. We propose an efficient
MCMC scheme for sampling from the joint posterior of the all knot
locations and the regression coefficients, see Section \ref{sec:MCMC}
for details. The model is clearly highly (over)parametrized and in
need of some regularization of the parameters. The two main regularization
techniques in Bayesian analysis are shrinkage priors and variable
(knot) selection priors. Variable selection can in principle be incorporated
in the analysis, but would be computationally demanding since the
number of gradient evaluations needed in our MCMC algorithm would
increase dramatically. This is important since evaluating the gradient
with respect to the knots is time-consuming as the knot locations
enter the likelihood in a very complicated nonlinear way; see Section
\ref{sec:mcmc-details} for details. Moreover, part of the attraction
of variable selection is that they also provide interpretable measures
of variable importance; this is much less interesting here since the
covariates correspond to knot locations, which are not interesting
in themselves. We have therefore chosen to achieving parsimony with
shrinkage priors that pull the regression coefficients towards zero
(or any other reference point if so desired), see Section \ref{sub:Prior}
for details. We allow for separate shrinkage parameters for the linear,
additive and surface parts of the model, and separate shrinkage parameters
for the $p$ responses within each of the three model parts. The shrinkage
parameters are treated as unknowns and estimated, so that, for example,
the surface part can be shrunk towards zero if this agrees with the
data. Allowing the knots to move freely in covariate space introduces
a knot switching problem similar to the well-known label switching
problem in mixture models. The likelihood is invariant to a switch
of two knot locations and their regression coefficients. This lack
of identification is not important if our aim is to model the regression
surface $\mathrm{E}(\bm{y}|\bm{x})$, without regard to the posterior
of the individual knot locations \citep{geweke2007computational}.
Also, the MCMC draws of the knot locations can also be used to construct
heat maps in covariate space to represent the density of knots in
a certain regions, see Section \ref{sec:Application}. Such heat maps
are clearly also immune to the knot switching problem.

\subsection{The prior}

\label{sub:Prior} We now introduce an easily specified shrinkage
prior for the three sets of regression coefficients $\bm{B}_{o}$,
$\bm{B}_{a}$ and $\bm{B}_{s}$ and the covariance matrix $\bm{\Sigma}$,
conditional on the knots. The prior for $\bm{b}$ and $\bm{\Sigma}$
are set as
\[
\begin{split}\mathrm{vec}\bm{B}_{i}|\bm{\Sigma},~\bm{\lambda}_{i} & \sim\mathrm{N}\left(\bm{\mu}_{i},~\bm{\Lambda}_{i}^{1/2}\bm{\Sigma}\bm{\Lambda}_{i}^{1/2}\otimes\bm{P}_{i}^{-1}\right),~i\in\{o,a,s\},\\
\bm{\Sigma} & \sim\mathrm{IW}\left(n_{0}\bm{S}_{0},~n_{0}\right),
\end{split}
\]
with prior independence between the $\bm{B}_{i}$. The prior mean
of $\mathrm{vec}\bm{B}_{i}$ is $\bm{\mu}_{i}$, which we set to zero
in our shrinkage prior. $\bm{\Lambda}_{i}=\mathrm{diag}(\bm{\lambda}_{i})=\mathrm{diag}(\lambda_{i,1},...,\lambda_{i,p})$,
$\bm{P}_{i}$ is a positive definite symmetric matrix. $\mathrm{IW}(~\cdot~)$
denotes the inverse Wishart distribution, with location matrix $\bm{S}_{0}$
and degrees of freedom $n_{0}$. $\bm{P}_{i}$ is typically either
the identity matrix or $\bm{P}_{i}=\bm{X}'_{i}\bm{X}_{i}$. The latter
choice has been termed a \emph{g}-prior by \citet{zellner1986assessing}
and has the advantage of automatically adjusting for the different
scales of the covariates. Setting $\lambda_{i}=n$ makes the information
content of the prior equivalent to a single data point and is usually
called the unit information prior. The choice of $\bm{P}_{i}=\bm{I}_{q_{i}}$
can prevent the design matrix from falling into singularity problem
when some of the basis functions are highly correlated, which can
easily happen with many spline knots. See also the discussion in \citet{denisonbayesian}.
Our default choice is therefore $\bm{P}_{o}=\bm{X}_{o}'\bm{X}_{o}$,
$\bm{P}_{a}=\bm{I}_{q_{a}}$ and $\bm{P}_{s}=\bm{I}_{q_{s}}$. Other
shrinkage priors on the regression coefficients can be used in our
approach, for example the Laplace distribution leading to the popular
Lasso \citep{tibshirani1996regression}, but they will typically not
allow us to integrate out the regression coefficents analytically,
see Section \ref{sec:posterior}. The optimal choice of shrinkage
prior depends on the unknown data generating model (a normal prior
is better when all coefficients have roughly the same magnitude; Lasso
is better when many coefficients are close to zero, but some are really
large etc).

We also estimate the shrinkage parameters, $\bm{\lambda}_{o}$, $\bm{\lambda}_{a}$
and $\bm{\lambda}_{s}$ via a Bayesian approach. Note that our prior
constructions for $\bm{B}$ allow for separate shrinkage of the linear,
additive and surface components. This gives us automatic regularization/shrinkage
of the regression coefficients and helps to avoid problems with overfitting.
Our MCMC scheme in Section ~\ref{sec:MCMC} allows for a user-specified
prior on $\lambda_{ij}$, for $i\in\{o,a,s\}$ and $j=1,2,...,p$
of essentially any functional form. However the default prior of $\lambda_{ij}$
in this paper follows a log normal distribution with mean of $n/2$
and standard deviation of $n/2$ in order to ensure that both tight
and flat shrinkages are attainable within one standard deviation in
the prior. For computational convenience, we use a log link for $\lambda_{ij}$
and make inference on $\log(\lambda_{ij})$. As a result the preceding
prior on $\lambda_{ij}$ yields a normal prior for $\log(\lambda_{ij})$
with mean $[\log(n)-3/2\cdot\log(2)]$ and variance $\log(2)$.

We use the same number of additive knots for each covariate in the
simulations and the application in Section \ref{sec:simulation} and
\ref{sec:Application}, but it should be clear that our approach also
permits unequal number of knots in the different covariates. There
is no particular requirements for the prior on the knots, but a vague
prior should permit the knots to move freely in covariate space. Our
default prior assumes independent knot locations following a normal
distribution. The mean of the knots comes from the centers of a \emph{k}-means
clustering of the covariates. In the additive case, the prior variance
of all the knots in the $k$th covariate is $c^{2}(\bm{a}'\bm{a})^{-1}$,
where $\bm{a}$ is the $k$th column of $\bm{X}_{o}$. Similarly,
the prior covariance matrix of a surface knot is $c^{2}(\bm{X}_{o}'\bm{X}_{o})^{-1}$.
We use $c^{2}=n$ as the default setting.

The hyperparameter $\bm{S}_{0}$ in the $\mathrm{IW}$ prior for $\bm{\Sigma}$
is set equal to the estimated error covariance matrix from the fitted
linear model $\hat{\bm{Y}}=\bm{X}_{o}\hat{\bm{B}_{o}}$. A small degrees
of freedom ($n_{0}$) gives diffuse prior on $\bm{\Sigma}$ and $n_{0}=10$
is set as the default.

For notational convenience and further computational implementation,
we write the prior for the regression coefficients in condensed form
as $\bm{b}|\bm{\Sigma},\bm{\lambda}\sim\mathrm{N}\left(\bm{\mu}^{*},\bm{\Sigma}_{\bm{b}}\right)$
where $\bm{\lambda}=(\bm{\lambda}'_{o},\bm{\lambda}'_{a},\bm{\lambda}'_{s})'$,
$\bm{\mu}^{*}=(\bm{\mu}'_{o},\bm{\mu}'_{a},\bm{\mu}'_{s})'$, $\bm{\Sigma}_{\bm{b}}=(\bm{\Lambda}^{1/2}\bm{\Sigma}_{K}\bm{\Lambda}^{1/2})\divideontimes\bm{P}^{-1}$,
$\bm{\Lambda}=\mathrm{diag}(\bm{\lambda})$, $\bm{\Sigma}_{K}$ is
a three-block diagonal matrix with $\bm{\Sigma}$ on each block, $\bm{P}=\mathrm{diag}(\bm{P}_{o},\bm{P}_{a},\bm{P}_{s})$
is a block diagonal matrix and $\bm{A}\divideontimes\bm{C}$ denotes
the Khatri-Rao product \citep{khatri1968solutions} which is Kronecker
product of the corresponding blocks of matrices $\bm{A}$ and $\bm{C}$.
It will also be convenient to define $\bm{\beta}=\mathrm{vec}\bm{B}$.
Note that $\bm{b}$ and $\bm{\beta}$ contain the same elements with
two different stacking orders. As a result, $\bm{\beta}|\bm{\Sigma},\bm{\lambda}\sim\mathrm{N}\left(\bm{\mu},\bm{\Sigma}_{\bm{\beta}}\right)$
where $\bm{\mu}$ and $\bm{\Sigma}_{\bm{\beta}}$ essentially have
the same entries as $\bm{\mu^{*}}$ and $\bm{\Sigma}_{\bm{b}}$ have,
respectively (Section \ref{sec:compremarks}).

\section{The posterior inference}

\label{sec:MCMC}

\subsection{The posterior}

\label{sec:posterior}

The posterior distribution can be decomposed as
\[
p(\bm{B},\bm{\Sigma},\bm{\xi},\bm{\lambda}|\bm{Y},\bm{X})=p(\bm{B}|\bm{\xi},\bm{\lambda},\bm{\Sigma},\bm{Y},\bm{X})p(\bm{\xi},\bm{\lambda},\bm{\Sigma}|\bm{Y},\bm{X}),
\]
where
\[
\mathrm{vec}\bm{B}|\bm{\xi},\bm{\lambda},\bm{\Sigma},\bm{Y},\bm{X}\sim\mathrm{N}(\bm{\tilde{\beta}},~\bm{\Sigma}_{\tilde{\bm{\beta}}}),
\]
${\bm{\Sigma}_{\bm{\tilde{\beta}}}}={[{\bm{\Sigma}^{-1}}\otimes\bm{X}'\bm{X}+\bm{\Sigma}_{\bm{\beta}}^{-1}]^{-1}}$
, $\bm{\tilde{\beta}}=\mathrm{vec}\bm{\tilde{B}}=\bm{\Sigma}_{\bm{\tilde{\beta}}}[{\mathrm{vec}({\bm{X}'\bm{Y}{\bm{\Sigma}^{-1}}})+\bm{\Sigma}_{\bm{\beta}}^{-1}\bm{\mu}}]$
\citep{zellner1971introduction}, and
\begin{equation}
\begin{split}p\left({\bm{\xi},\bm{\lambda},\bm{\Sigma}|\bm{Y},\bm{X}}\right)=~ & c\times p(\bm{\xi},\bm{\lambda})\times|\bm{\Sigma}_{\bm{\beta}}|^{-1/2}|\bm{\Sigma}|^{-(n+{n_{0}}+p+1)/2}|\bm{\Sigma}_{\bm{\tilde{\beta}}}|^{-1/2}\\
 & \times\exp\left\{ {-\frac{1}{2}\left[{\mathrm{tr}{\bm{\Sigma}^{-1}}\left({{n_{0}}{\bm{S}_{0}}+n\bm{\tilde{S}}}\right)+\left({\bm{\tilde{\beta}}-\bm{\mu}}\right)'{\bm{\Sigma}_{\bm{\beta}}^{-1}}\left({\bm{\tilde{\beta}}-\bm{\mu}}\right)}\right]}\right\}
\end{split}
\end{equation}
where $\bm{\tilde{S}}=(\bm{Y}-\bm{X}\bm{\tilde{B}})'(\bm{Y}-\bm{X}\bm{\tilde{B}})/n$,
$c={2^{-({n_{0}}+n+q)p/2}}{\pi^{-p(n+q)/2}}\Gamma_{p}^{-1}({n_{0}}/2)|{n_{0}}{\bm{S_{0}}}{|^{{n_{0}}/2}}$,
$\Gamma_{p}(a)=\pi^{p(p-1)/4}\prod_{j=1}^{p}\Gamma\left[a+(1-j)/2\right]$
is the multivariate gamma function. It is important to note that it
is in general not possible to integrate out $\bm{\Sigma}$ analytically
in our model. This is a consequence of using different shrinkage factors
for the different responses and on the original, additive and surface
parts of the model (the prior covariance matrix of $\bm{B}$ does
not have a Kronecker structure). Only in the special case with a univariate
response ($p=1$) can we integrate out $\bm{\Sigma}$ analytically,
since $\bm{\Sigma}$ is then a scalar. To obtain a uniform treatment
of the models and their gradients, we have chosen to not integrate
out $\bm{\Sigma}$ even for the case $p=1$. The next subsection proposes
an MCMC algorithm for sampling from the joint posterior distribution
of all parameters.

\subsection{The MCMC algorithm}

\label{sec:mcmc-details}

Our approach is to sample from $p\left({\bm{\xi},\bm{\lambda},\bm{\Sigma}|\bm{Y},\bm{X}}\right)$
using a three-block Gibbs sampling algorithm with Metropolis-Hastings
(MH) updating steps. Draws from $p(\bm{B}|\bm{\xi},\bm{\lambda},\bm{\Sigma},\bm{Y},\bm{X})$
can subsequently be obtained by direct simulation. The updating steps
of the Gibbs sampling algorithm are:
\begin{enumerate}
\item Simulate $\bm{\Sigma}$ from $p(\bm{\Sigma}|\bm{\xi},\bm{\lambda},\bm{Y},\bm{X})$.
\item Simulate $\bm{\xi}$ from $p(\bm{\xi}|\bm{\lambda},\bm{\Sigma},\bm{Y},\bm{X})$.
\item Simulate $\bm{\lambda}$ from $p(\bm{\lambda}|\bm{\xi},\bm{\Sigma},\bm{Y},\bm{X})$.
\end{enumerate}
In the special case when $p=1$
\begin{equation}
\noindent\bm{\Sigma}|\bm{\xi},\bm{\lambda},\bm{Y},\bm{X}\sim\mathrm{IW}\left({n_{0}}{\bm{S}_{0}}+n\bm{\tilde{S}}+\sum\nolimits _{i\in\{o,a,s\}}{\bm{\Lambda}_{i}^{-1/2}({{\bm{\tilde{B}}}_{i}}-{\bm{M}_{i}})'{\bm{P}_{i}}({{\bm{\tilde{B}}}_{i}}-{\bm{M}_{i}})\bm{\Lambda}_{i}^{-1/2}},~n_{0}+n\right)\label{eq:IWdensity}
\end{equation}
where $\bm{M}_{i}$ and $\bm{\tilde{B}}_{i}$ are the prior and posterior
mean of $\bm{B}_{i}$, respectively. Actually, when $p=1$, $\bm{\Sigma}$
is a scalar and the $\mathrm{IW}$ density reduces to a scaled $\chi^{2}$
distribution. When $p>1$, $p(\bm{\Sigma}|\bm{\xi},\bm{\lambda},\bm{Y},\bm{X})$
is no longer $\mathrm{IW}$, but the distribution in \eqref{eq:IWdensity}
is an excellent approximation and can be used as a very efficient
MH proposal density.

The conditional posterior distributions for $\bm{\xi}$ and $\bm{\lambda}$
in Steps (2) and (3) above are highly non-standard and we update these
parameters using Metropolis-Hastings steps with a tailored proposal,
which we now describe for a general parameter vector $\bm{\theta}$
with posterior $p(\bm{\theta}|\bm{Y})$, which could be a conditional
posterior in a Metropolis-within-Gibbs algorithm (e.g. $p(\bm{\xi}|\bm{\lambda},\bm{\Sigma},\bm{Y},\bm{X})$).
This method was originally proposed by \citet{gamerman1997sampling}
and later extended by \citet{nott2004sampling} and \citet{villani2010}.
All of these three articles are confined to a generalized linear model
(GLM) or GLM-like context where the parameters enter the likelihood
function through a scalar-valued link function. A contribution of
our paper is to show that the algorithm can be extended to models
without such a nice structure and that it retains its efficiency even
when the parameters are high-dimensional and enter the model in a
highly nonlinear way. The way the knot locations and the shrinkage
parameters are buried deep in the marginal posterior (see Equation
3.1 above) makes the necessary gradients (see below) much more involved
and numerically challenging (see Appendix \ref{AppendixGradients}).

At any given MCMC iteration we use Newton's method to iterate $R$
steps from the current point $\bm{\theta}_{c}$ in the MCMC sampling
towards the mode of $p(\bm{\theta}|\bm{Y})$, to obtain $\bm{\hat{\theta}}$
and the Hessian at $\bm{\hat{\theta}}$. Note that $\bm{\hat{\theta}}$
may not be the mode but is typically close to it already after a few
Newton iterations since the previously accepted $\bm{\theta}$ is
used as the initial value; setting $R=1,2$ or $3$ is therefore usually
sufficient. This makes the algorithm very fast. Having obtained good
approximations of the posterior mode and covariance matrix from the
Newton iterations, the proposal $\bm{\theta}_{p}$ is now drawn from
the multivariate $\bm{t}$-distribution with $\nu>2$ degrees of freedom:
\[
\bm{\theta}_{p}|\bm{\theta}_{c}\sim\bm{t}\left[\bm{\hat{\theta}},~\left.-\left(\frac{\partial^{2}\ln p(\bm{\theta}|\bm{Y})}{\partial\bm{\theta}\partial\bm{\theta}^{\prime}}\right)^{-1}\right\vert _{\bm{\theta}=\bm{\hat{\theta}}},~\nu\right],
\]
where the second argument of the density is the covariance matrix
and $\hat{\bm{\theta}}$ is the terminal point of the $R$ Newton
steps. The Metropolis-Hastings acceptance probability is
\[
a\left(\bm{\theta}_{c}\rightarrow\bm{\theta}_{p}\right)=\min\left[1,~\frac{p(\bm{Y}|\bm{\theta}_{p})p(\bm{\theta}_{p})g(\bm{\theta}_{c}|\bm{\theta}_{p})}{p(\bm{Y}|\bm{\theta}_{c})p(\bm{\theta}_{c})g(\bm{\theta}_{p}|\bm{\theta}_{c})}\right].
\]
The proposal density at the current point $g(\bm{\theta}_{c}|\bm{\theta}_{p})$
is a multivariate $\bm{t}$-density with mode $\bm{\tilde{\theta}}$
and covariance matrix equal to the negative inverse Hessian evaluated
at $\bm{\tilde{\theta}}$, where $\bm{\tilde{\theta}}$ is the point
obtained by iterating $R$ steps with the Newton algorithm, \emph{this
time starting from} $\bm{\theta}_{p}$. The need to iterate backwards
from $\bm{\theta}_{p}$ is clearly important to fulfill the reversibility
of the Metropolis-Hastings algorithm. When the number of parameters
in $\bm{\theta}$ is large one can successively apply the algorithm
to smaller blocks of parameters in $\bm{\theta}$.

The tailored proposal distribution turns out to be hugely beneficial
for MCMC efficiency, see Section \ref{sub:Algorithms-and-prior} for
some evidence, but a naive implementation can easily make the gradient
and Hessian evaluations an insurmountable bottleneck in the computations,
and a source of numerical instability. We have found the outer product
of gradients approximation of the Hessian to work very well, so all
we need to implement efficiently are the gradient vector of $p(\bm{\xi}|\bm{\lambda},\bm{\Sigma},\bm{Y},\bm{X})$
and $p(\bm{\lambda}|\bm{\xi},\bm{\Sigma},\bm{Y},\bm{X})$. Appendix
\ref{AppendixGradients} gives compact analytical expression for these
two gradient vectors, and shows how to exploit sparsity to obtain
fast and stable gradient evaluations. Our gradient evaluations can
easily be orders of magnitudes faster than state-of-the-art numerical
derivatives, and substantially more stable numerically. For example,
already in a relatively small-dimensional model in Section \ref{sec:Application}
with only four covariates, $20$ surface knots and $4$ additive knots,
the analytical gradient for the knot parameters are more than $40$
times faster compared to a numerical gradient with tolerance of $10^{-3}$.
Since the gradient evaluations accounts for 70-90\% of total computing
time, this is clearly an important advantage.

\subsection{Model comparison}

\label{sec:model-comp}

The number of knots is determined via the $D$-fold out-of-sample
log predictive density score (LPDS), defined as
\[
\frac{1}{D}\sum\nolimits _{d=1}^{D}\ln p(\tilde{\bm{Y}}_{d}|\tilde{\bm{Y}}_{-d},\bm{X}),
\]
where $\tilde{\bm{Y}}_{d}$ is an $(n_{d}\times p)$-dimensional matrix
containing the $n_{d}$ observations in the $d$th testing sample
and $\tilde{\bm{Y}}_{-d}$ denotes the training observations used
for estimation. If we assume that the observations are independent
conditional on $\bm{\theta}$, then
\[
p(\tilde{\bm{Y}}_{d}|\tilde{\bm{Y}}_{-d},\bm{X})=\int\!\prod\nolimits _{i\in\tau_{d}}p(\bm{y}_{i}|\bm{\theta},\bm{x}_{i})p(\bm{\theta}|\tilde{\bm{Y}}_{-d})\mathrm{d}\bm{\theta},
\]
where $\tau_{d}$ is the index set for the observations in $\tilde{\bm{Y}}_{d}$,
and the LPDS is easily computed by averaging $\prod_{i\in\tau_{d}}p(\bm{y}_{i}|\bm{\theta},\bm{x}_{i})$
over the posterior draws from $p(\bm{\theta}|\tilde{\bm{Y}}_{-d})$.
This requires sampling from each of the $D$ posteriors $p(\bm{\theta}|\tilde{\bm{Y}}_{-d})$
for $d=1,...,D$, but these MCMC runs can all be run in isolation
from each other and are therefore ideal for straightforward parallel
computing on widely available multi-core processors. The main advantage
for choosing LPDS instead of the marginal likelihood is that the LPDS
is not nearly as sensitive to the choice of prior as the marginal
likelihood, see e.g. \citet*{kass1993bayes} and \citet*{richardson1997bayesian}
for a general discussion. The marginal likelihood can also lead to
poor predictive inference when the true data generating process is
not included in the class of compared models, see e.g. \citet{geweke2011optimal}
for an illuminating perspective. The main disadvantage of using the
LPDS for selecting the number of knots is that, unlike the marginal
likelihood and RJMCMC, there is no rigorous way of including the uncertainty
regarding the number of knots in the final inferences. The dataset
is systematically partitioned into five folds in our firm leverage
application in Section \ref{sec:Application}.

\section{Simulations}

\label{sec:simulation}

As discussed in the Introduction, the most commonly used approach
for spline regression modeling is to use a large number of fixed knots
and to use shrinkage priors or Bayesian variable selection to avoid
overfitting \citep{denisonbayesian}. We compare the performance of
the traditional fixed knots approach to our approach with freely estimated
knot locations using simulated data with different number of covariates
and for varying degrees of nonlinearity in the true surface. We use
shrinkage priors with estimated shrinkage both for the fixed and free
knot models, but no variable selection. Models with univariate and
multivariate response variables are both investigated.

\subsection{Simulation setup}

We consider data generating processes (DGP) with both univariate ($p=1$)
and bivariate ($p=2$) responses, and datasets with $q_{o}=10$ regressors
and two sample sizes, $n=200$ and $n=1000$. We first generate the
covariate matrix $\bm{X}_{o}$ from a mixture of multivariate normals
with five components. The weight for the $r$th mixture component
is $u_{r}/\sum_{l=1}^{5}u_{l}$, where $u_{1},...,u_{5}$ are independent
$\mathrm{U}(0,1)$ variables. The mean of each component is a draw
from $\mathrm{U}(-1,1)$ and the components' variances are all $0.1$.
We randomly select five observations without replacement from $\bm{X}_{o}$
as the true surface knots $\bm{\xi}_{s}$, and then create the basis
expanded design matrix $\bm{X}$ using the thin-plate radial basis
surface spline, see Section \ref{sub:Model}. The coefficients matrix
$\bm{B}$ is generated by repeating the sequence $\{-1,1\}$. The
error term $\bm{E}$ is from multivariate normal distribution with
mean zero, variance $0.1$ and covariance $0.1$. These settings guarantee
a reasonable signal-to-noise ratio.

Following \citet{wood2002bayesian}, we measure the degrees of nonlinearity
($\mathrm{DNL}$) in the DGP by the distance between the true surface
$f(\cdot)$ and the plane $\hat{g}(\cdot)$ fitted by ordinary least
squares without any knots in the model, i.e.
\begin{equation}
\mathrm{DNL}=\sqrt{n^{-1}\sum\nolimits _{i=1}^{n}[f(\bm{x}_{i})-\hat{g}(\bm{x}_{i})]^{2}}.\label{eq:estimatedDNL}
\end{equation}
A larger $\mathrm{DNL}$ indicates a DGP with stronger nonlinearity.

We generate $100$ datasets and for each dataset we fit the fixed
knots model with $5$, $10$, $15$, $20$, $25$ and $50$ surface
knots, and also the free knots model with $5$, $10$, and $15$ surface
knots. All fitted models have only linear and surface components.
The knot locations are determined by \emph{k}-means clustering. We
compare the models with respect to the mean squared loss
\begin{equation}
\mathrm{Loss}(q_{s})=\frac{1}{n^{*}}\sum\nolimits _{i=1}^{n^{*}}[f(\bm{x}_{i})-\tilde{f}(\bm{x}_{i})]^{2}\label{eq:SurfaceLoss}
\end{equation}
where $f(\cdot)$ is the true surface and $\tilde{f}(\cdot)$ is the
posterior mean surface of a given model with $q_{s}$ surface knots.
The $\mathrm{Loss}$ in \eqref{eq:SurfaceLoss} is evaluated over
a new sample of $n^{*}$ covariate vectors, and it therefore measures
out-of-sample performance of the posterior mean surface. We will here
set $n^{*}=n$. Note that the shrinkages and the covariance matrix
of the error terms are also estimated in both the fixed and free knots
models.

\subsection{Results}

\begin{figure}
\centering \includegraphics[width=0.8\textwidth]{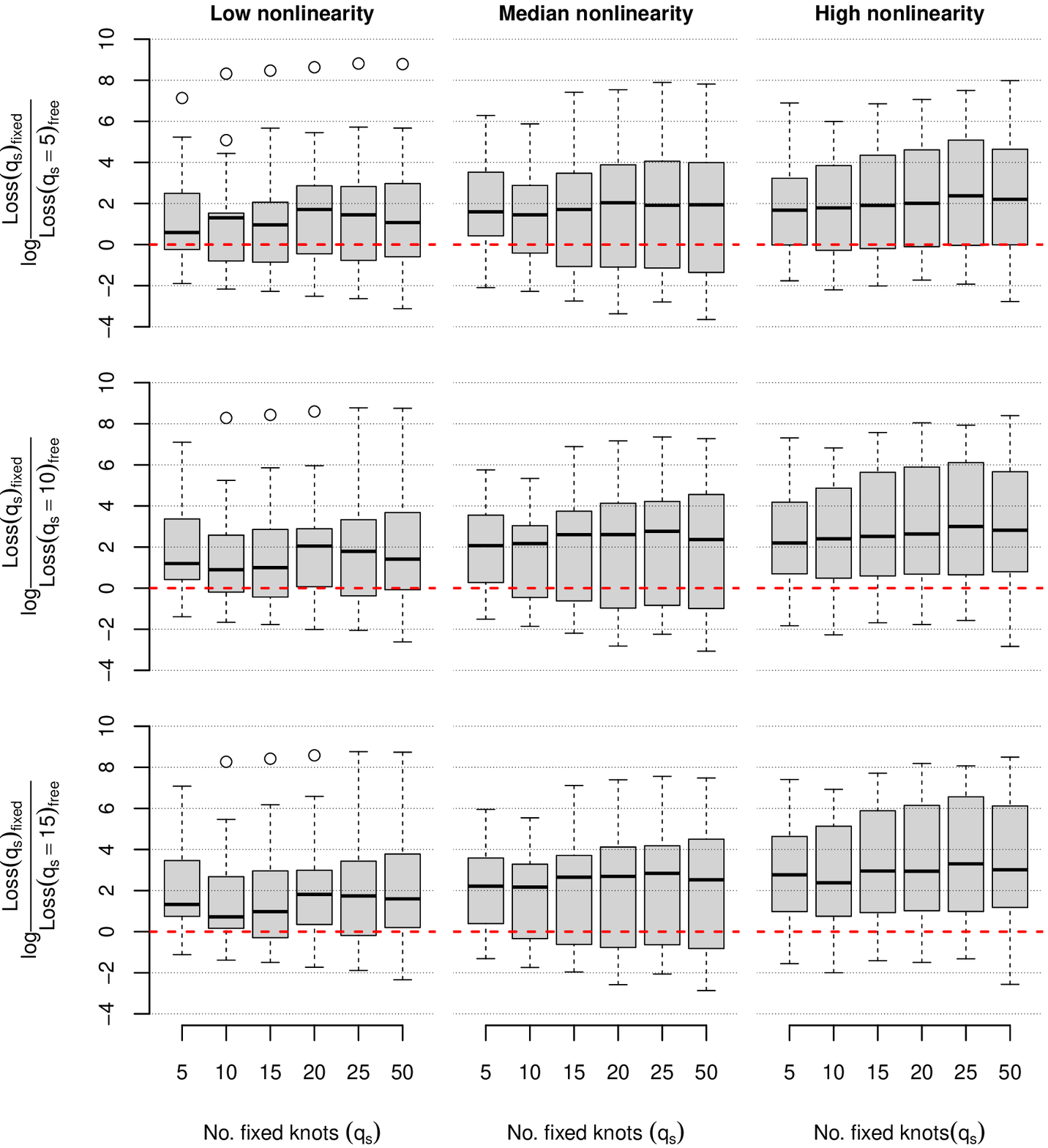} \caption{Boxplot of the log loss ratio comparing the performance of the fixed
knots model with the free knots model for the DGP with $p=2$ and
$n=200$. The three columns of the figure correspond to different
degrees of nonlinearity of the realized datasets, as measured by estimated
$\mathrm{DNL}$ in \eqref{eq:estimatedDNL}. }

\label{fig:LOSS}
\end{figure}

We present the results for $p=2$ and $n=200$. The results for $p=1$
and $n\in\{200,1000\}$, and $p=2$ and $n=1000$ are qualitatively
similar and are available upon request. The Supporting Information
documents the results for $p=2$ and $n=1000$ for a few different
model configurations. Figure \ref{fig:LOSS} displays boxplots for
the log ratio of the mean squared loss in \eqref{eq:SurfaceLoss}.
The columns of the figure represents varying degrees of nonlinearity
in the generated datasets according to the estimated $\mathrm{DNL}$
measure in equation \eqref{eq:estimatedDNL}. Each boxplot shows the
relative performance of a fixed knots model with a certain number
of knots compared to the free knots model with $5$ (top row), $10$
(middle row) and $15$ (bottom row) surface knots, respectively. The
short summary of Figure \ref{fig:LOSS} is that the free knots model
outperforms the fixed knots model in the large majority of the datasets.
This is particularly true when the data are strongly nonlinear. The
performance of the fixed knots model improves somewhat when we add
more knots, but the improvement is not dramatic. Having more fixed
knots clearly improves the chances of having knots close to the true
ones, but more knots also increase the risk of overfitting.

\begin{figure}
\centering \includegraphics[width=0.8\textwidth]{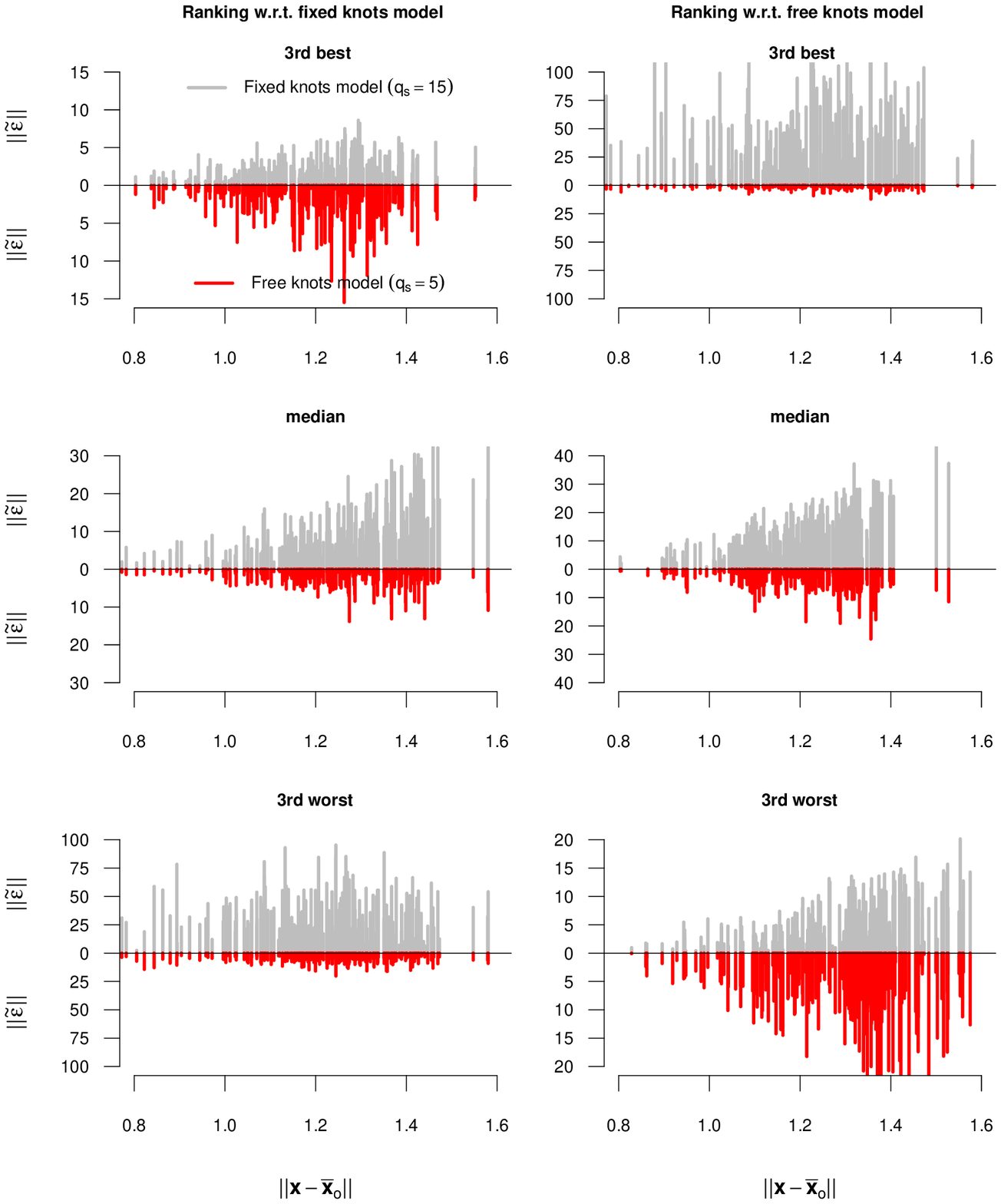}
\caption{Plotting the norm of the predictive multivariate residuals as a function
of the distance between the covariate vector and its sample mean.
The results are for the DGP with $p=2$ and $n=200$. The lines in
each subplot are the normed residuals from the model with $15$ fixed
surface knots (vertical bars above the zero line), and the model with
$5$ free knots (vertical bars below the zero line). The column to
the left shows the results for three datasets chosen when performance
is ranked according to the fixed knots model, and the right column
displays the results for three datasets chosen when performance is
ranked according to the free knots model.}

\label{fig:Resid}
\end{figure}

The aggregate results in Figure \ref{fig:LOSS} do not clearly show
how strikingly different the fixed and free knots models can perform
on a given dataset. We will now show that models with free rather
than fixed knots are much more robust across different datasets. Figure
\ref{fig:Resid} displays the Euclidean distance of the multivariate
\emph{out-of-sample} predictive residuals $\sqrt{\tilde{\bm{\varepsilon}}'\tilde{\bm{\varepsilon}}}$
for a few selected datasets as a function of the distance between
the covariate vector and the sample mean of the covariates. The normed
residuals depicted in the leftmost column are from datasets chosen
with respect to the ranking of the out-of-sample performance of the
fixed knots model. For example, the upper left subplot shows the predictive
residuals of both the model with $15$ fixed knots (vertical bars
above the zero line) and the model with $5$ free knots (vertical
bars below the zero line) on one of the datasets where the fixed knot
models outperform the free knots model by largest margin ($3$rd best
Loss in favor of fixed knots model). It is seen from this subplot
that even in this very favorably situation for the fixed knots model,
the free knots model is not generating much larger predictive residuals.
Moving down to the last row in the left hand column of Figure \ref{fig:Resid},
we see the performance of the two models when the fixed knots model
performs very poorly ($3$rd worse Loss with respect to the fixed
knots model). On this particular dataset, the free knots model does
well while the fixed knots model is a complete disaster (note the
different scales on the vertical axes of the subplots). The column
to the right in Figure \ref{fig:Resid} shows the same analysis, but
this time the datasets are chosen with respect to the ranking of the
Loss of the free knots model. Overall, Figure \ref{fig:Resid} clearly
illustrates the superior robustness of models with free knots: the
free knots model never does much worse than the fixed knots model,
but using fixed rather than free knots can lead to a dramatically
inferior predictive performance on individual datasets.

\subsection{Computing time}

The program is written in native R code and all the simulations were
performed on a Linux desktop with $2.8$ GHz CPU and $4$ GB RAM on
single instance (without parallel computing). Table \ref{tab:comptime}
shows the computing time in minutes for a single dataset. In general
the computing time increases as the size of the design matrix increases,
but it increases only marginally as we go from $p=1$ to $p=2$.

\begin{table}
\caption{Elapsed computing time (in minutes) for 5,000 iterations with a single
dataset of $10$ covariates.}

\label{tab:comptime} \centering %
\begin{tabular}{ccccccc}
\hline
 &  & \multicolumn{2}{c}{$n=200$ } &  & \multicolumn{2}{c}{$n=1000$ }\tabularnewline
\hline
No. of free surface knots  &  & $p=1$  & $p=2$  &  & $p=1$  & $p=2$ \tabularnewline
\hline
$~2$  &  & $~9$  & $~9$  &  & $16$  & $17$ \tabularnewline
$~5$  &  & $13$  & $14$  &  & $23$  & $26$ \tabularnewline
$10$  &  & $17$  & $18$  &  & $42$  & $45$ \tabularnewline
$15$  &  & $24$  & $27$  &  & $61$  & $75$ \tabularnewline
\hline
\end{tabular}
\end{table}

\section{Application to firm capital structure data}

\label{sec:Application}

\subsection{The data}

The classic paper by \citet{rajan1995we} analyze firm leverage (\textsf{leverage}\emph{
= total debt/(total debt + book value of equity)}) as a function of
its fixed assets (\textsf{tang}\emph{ = tangible assets/book value
of total assets}), its market-to-book ratio (\textsf{market2book}\emph{
= (book value of total assets - book value of equity + market value
of equity)/book value of total assets}), logarithm of sales (\textsf{LogSale})
and profit (\textsf{Profit}\emph{ = earnings before interest, taxes,
depreciation, and amortization/book value of total assets}). Strong
nonlinearities seem to be a quite general feature of balance sheet
data, but only a handful articles have suggested using nonlinear/nonparametric
models, see e.g. \citet{bastos2010nonparametric}, and \citet{villani2010}.
We use a similar data to the one in \citet{rajan1995we} which covers
$4,405$ American non-financial firms with positive sales in $1992$
and complete data records and analyze the leverage in terms of total
debt. \citet{villani2010} analyze the same data with a smooth mixture
of Beta regressions.

\begin{figure}
\centering \includegraphics[width=1\textwidth]{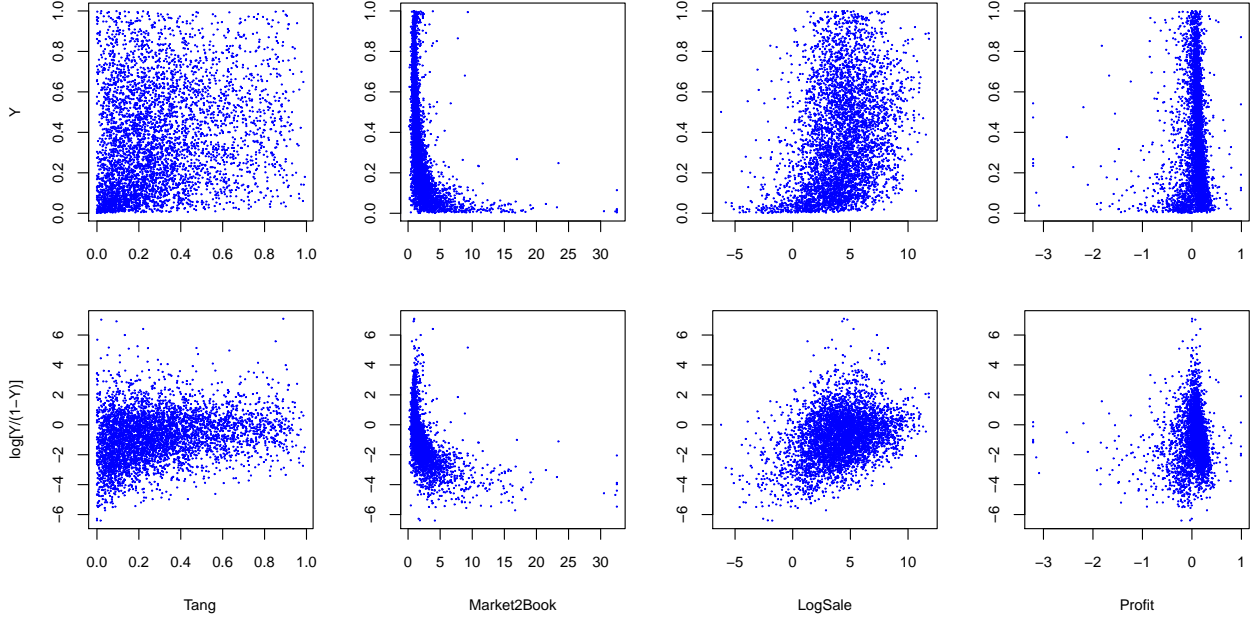} \caption{Scatter plots of the firm leverage data with leverage ($\bm{Y}$)
on both original scale (top subplots) and logit transformed scale
(bottom subplots) against each of the four covariates.}

\label{fig:rajandata}
\end{figure}

Figure~\ref{fig:rajandata} plots the response variable \textsf{leverage}
in both original scale and logit scale ($\ln[y/(1-y)]$) against each
of the four covariates. The relationships between the leverage and
the covariates are clearly highly nonlinear even when the logit transformation
is used. There are also outliers which can be seen from the subplots
with respect to covariates \textsf{Market2Book} and \textsf{Profit}.

\subsection{Models with only surface or additive components}

We first fit models that either have only a surface component or only
an additive component (both types of models also have a linear component).
Note that the shrinkage parameters are also estimated in all cases.
All four covariates are used in the estimation procedure and we use
the logit transformation of the leverage, and standardize each covariate
to have zero mean and unit variance.

\begin{figure}
\centering \includegraphics[width=0.9\textwidth]{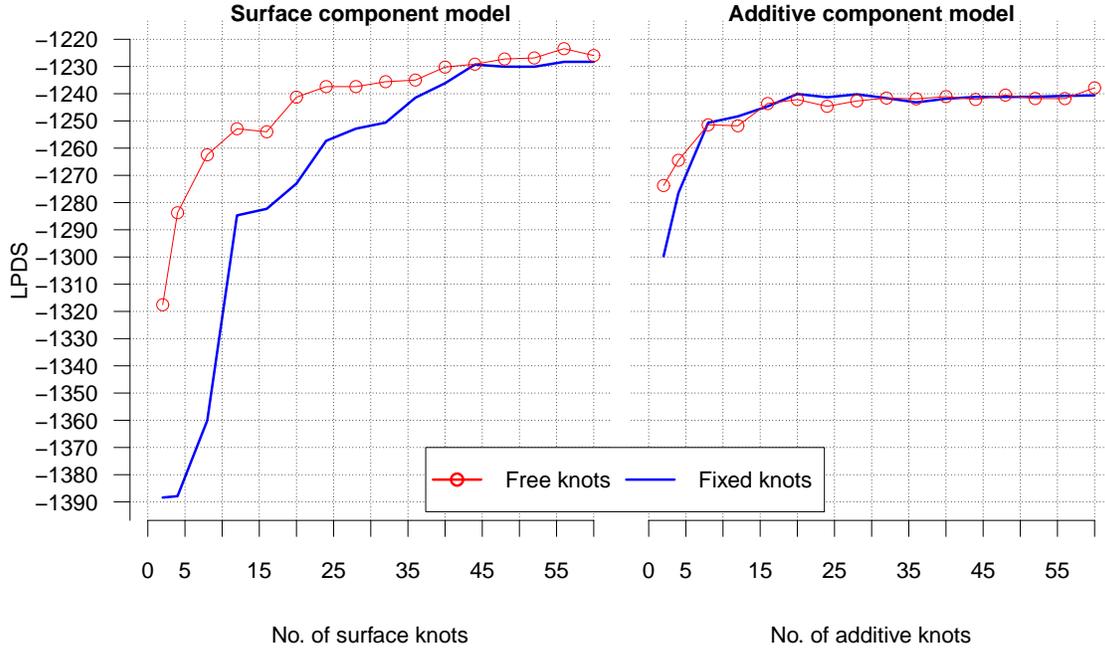}
\caption{LPDS for the firm leverage data with surface component model (left)
and additive component model (right). Note that the number of knots
in additive model is the number of spline basis functions on each
covariate.}

\label{fig:rajan_lpds1}
\end{figure}

Figure~\ref{fig:rajan_lpds1} depicts the LPDS for the surface component
model and the additive component model for both the case of fixed
and free knots. The LPDS generally improves as the number of knots
increases for both the fixed and free knots models, but seems to eventually
level off at large number of knots. The free knots model always outperforms
the fixed knots model when only a surface component is used (left
subplot). For example, the model with $12$ free surface knots is
roughly $32$ LPDS units better than the fixed knots model with the
same number of knots. This is a quite impressive improvement in out-of-sample
performance considering that the fixed knot locations are chosen with
state-of-the-art clustering methods for knot selection. The ability
to move the knots clearly also helps to keep the number of knots to
a minimum; it takes for example more than $30$ fixed surface knots
to obtain the same LPDS as a model with $12$ free surface knots.

Turning to the strictly additive models in right subplot of Figure
\ref{fig:rajan_lpds1} we see that the additive models are in general
inferior to the models with only surface knots, and that the differences
in LPDS between the fixed and free knots approaches are much smaller
here, at least for eight knots or more. The improvement in LPDS levels
off at roughly $16$ knots. It is important to note that the horizontal
axis in Figure \ref{fig:rajan_lpds1} displays the number of additive
knots \emph{in each covariate}, and the fact that we do not overfit
bear testimony to the effectiveness of the shrinkage priors.

\subsection{Models with both additive and surface components}

We now consider models with both additive and surface components.
It is worth mentioning that we draw from the joint posterior distribution
of the surface and additive knots, see Section~\ref{AppendixGradients}
for MCMC details.

\begin{figure}
\centering \includegraphics[height=0.95\textheight]{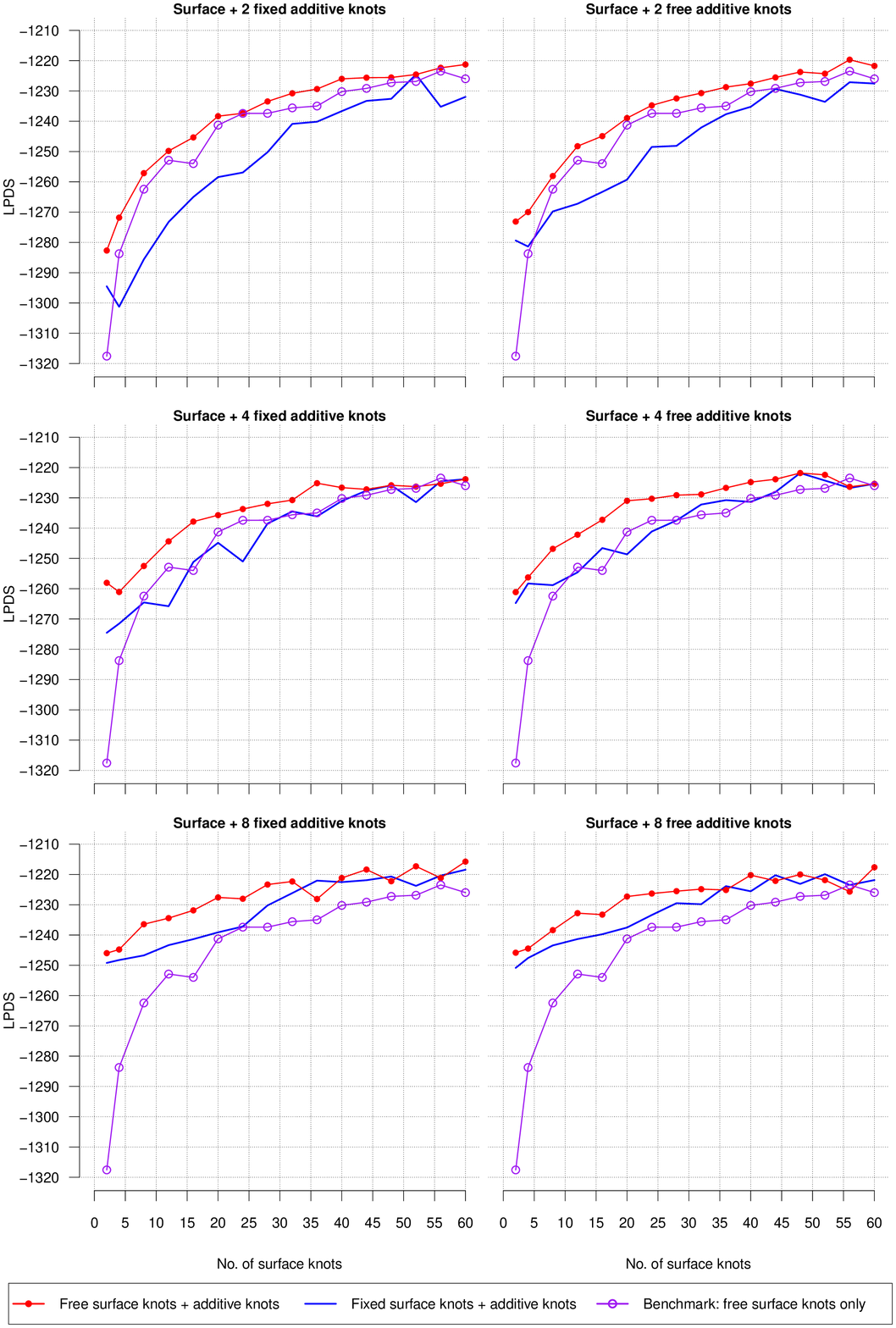}
\caption{LPDS for the firm leverage data for the free and fixed knots models
with varying number of surface and additive knots.}

\label{fig:rajan_lpds2}
\end{figure}

Figure~\ref{fig:rajan_lpds2} shows that there are generally improvements
from using both surface knots and additive knots in the same model.
For example, the model with $4$ free surface knots has an LPDS of
$-1,284$. Adding two free additive knots increases the LPDS to $-1,270$
and adding another two additive knots gives a further increase of
$14$ LPDS units. Figure \ref{fig:rajan_lpds2} also shows strong
gains from estimating the knots' locations, but the improvement in
LPDS from free knots tends to be less dramatic when more additive
knots are used to complement the surface knots. There is little or
no improvement in LPDS as the number of surface knots approaches $60$.
The results in Figure \ref{fig:rajan_lpds2} reinforces the evidence
in Figure \ref{fig:rajan_lpds1} that the shrinkage prior is very
effective in mitigating potential problems with overfitting.

\begin{figure}
\centering \includegraphics[width=0.7\textwidth]{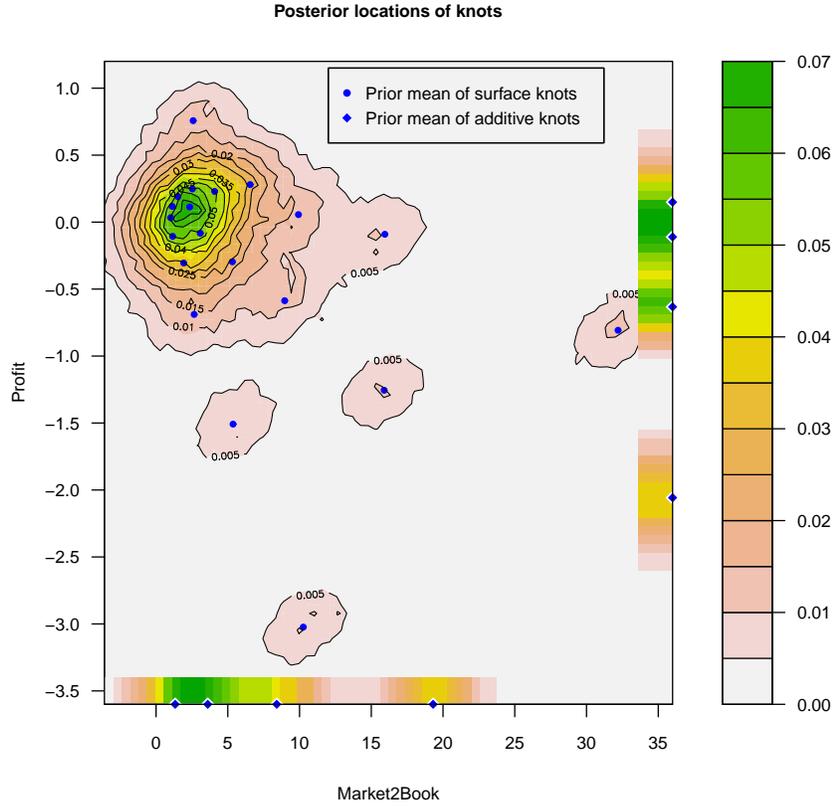}
\caption{Heat map to visualize the posterior density of the knot locations
in covariate space for model with $4$ free additive knots and $20$
free surface knots for the firm leverage dataset. The plot is constructed
by partitioning the covariate space into $70\times70$ rectangular
regions and counting the number of surface knots in each rectangle
over the MCMC draws. The posterior density of the locations of the
additive knots is constructed in a similar fashion and separate heat
maps for the additive knots in each covariate are shown just above
the horizontal axis and vertical axis, respectively.}

\label{fig:rajan_knots}
\end{figure}

To simplify the graphical presentation of the results, we choose to
illustrate the posterior inference of the knot locations in a model
with only the two covariates \textsf{Market2Book} and \textsf{Profit.}
We use $20$ surface knots and $4$ additive knots in each covariate.
The mean acceptance probabilities for the knot locations and the
shrinkage parameters in Metropolis-Hastings algorithm are $0.73$
and $0.64$, respectively, which are exceptionally large considering
that all $2\times20+2\times4=48$ knot location parameters are proposed
jointly, as are all the shrinkage parameters. The acceptance probability
in the updating step for $\bm{\Sigma}$ is $1$ since we are proposing
directly from the exact conditional posterior when $p=1$ . Because
of the knot switching problem (see Section~\ref{sec:MCMC}), it does
not make much sense to display the posterior distribution of the knot
locations directly. We instead choose to partition the covariate space
into small rectangular regions, count the frequency of knots in each
region over the MCMC iterations, and use heat maps to visualize the
density of knots in different regions of covariate space. Figure~\ref{fig:rajan_knots}
displays this knot density heat map. As expected, the estimated knot
locations are mostly concentrated in the data dense regions, particularly
in regions where the relation between the covariates and response
in the data is most nonlinear, which is seen by comparing Figure~\ref{fig:rajan_knots}
and Figure~\ref{fig:rajandata}.

\begin{figure}
\centering \mbox{ \includegraphics[width=0.5\textwidth]{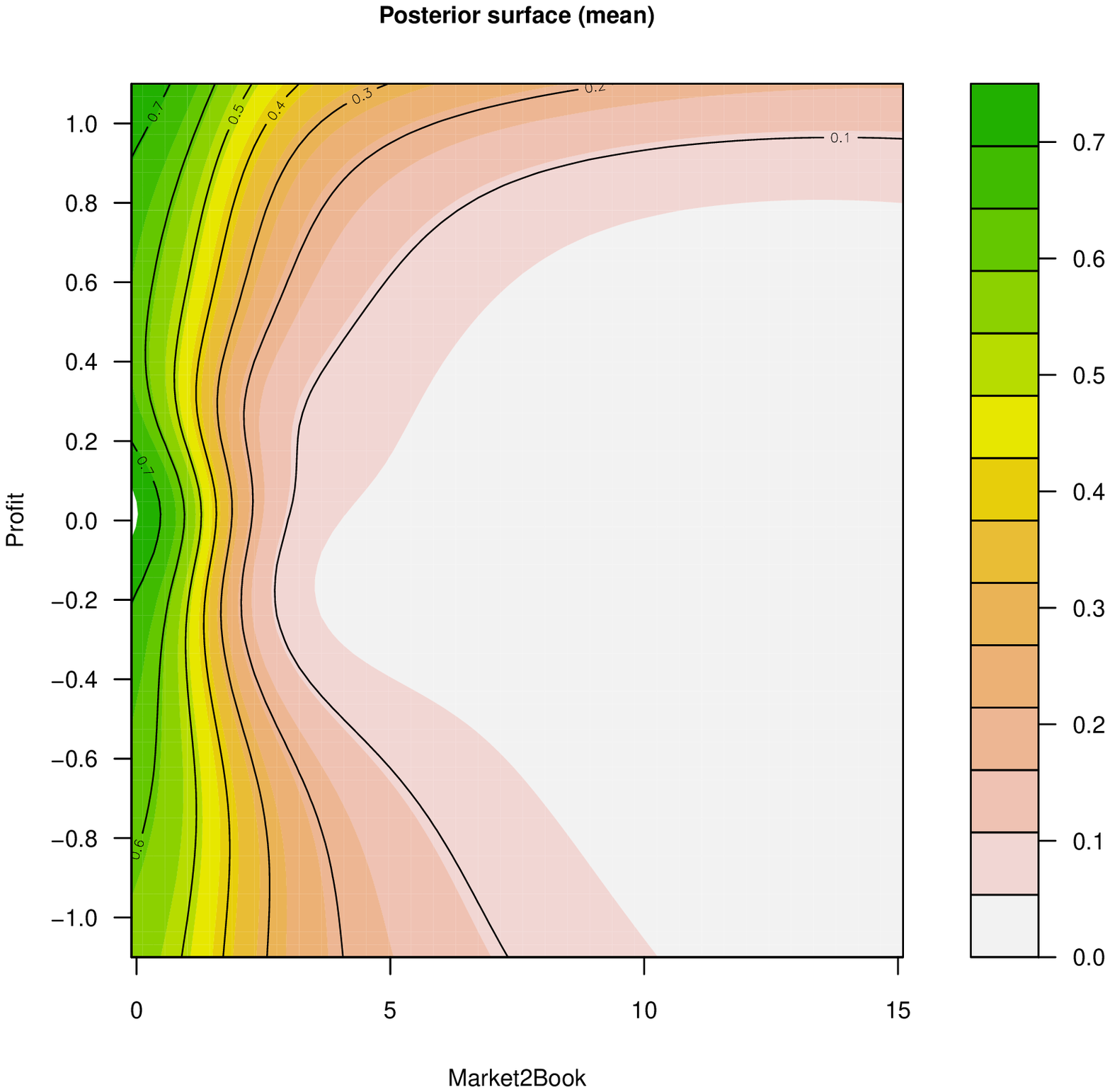}
\includegraphics[width=0.5\textwidth]{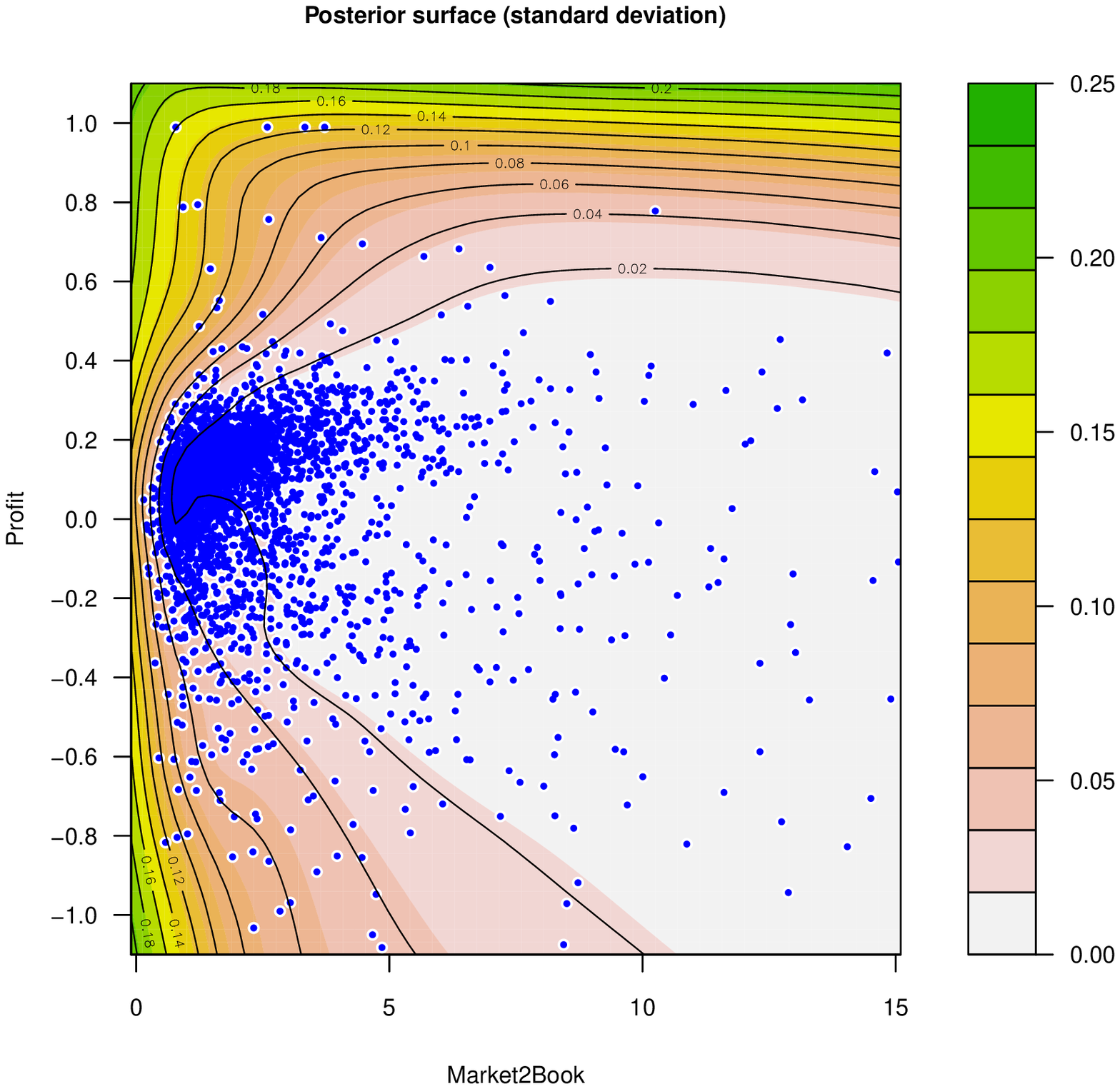}} \caption{The posterior mean (left) and standard deviation (right) of the posterior
surface for the model with $4$ free additive knots and $20$ free
surface knots for firm leverage data. The subplot to the right also
shows an overlay of the covariate observations.}

\label{fig:rajan_surface}
\end{figure}

Finally, we present the posterior surface for the firm leverage data
in Figure~\ref{fig:rajan_surface}. To enhance the visual representation,
the graphs zoom in on the region with the majority of the data observations.
Figure~\ref{fig:rajan_surface} plots the mean (left) and the standard
deviation (right) of the posterior surface. The latter object is for
brevity sometimes referred to as the \emph{posterior standard deviation
surface}. Figure~\ref{fig:rajan_surface} (right) also displays the
covariate observations to give a sense of where the data observations
are located. The Supporting Information to this article investigates
the robustness of the posterior results to variations in both the
prior mean and variance of the knot locations. The posterior heat
map of the knot locations are affected by the fairly dramatic variations
in the prior mean of the knots, and to a lesser extent by changes
in the prior variance of the knot locations, but the posterior mean
and standard deviation surfaces are robust to variations in the prior
on the knots, especially in data dense regions. The Supporting Information
also shows that the posterior is robust to changes in the prior on
the shrinkage factors.

\subsection{MCMC efficiency in the updating of the knot locations\label{sub:Algorithms-and-prior}}

In order to study the efficiency of our algorithm for sampling the
knot locations, we compare three types of MCMC updates of the knots:
i) one-knot-at-a-time updates using a random walk Metropolis proposal
with tuned variance (SRWM), ii) one-knot-at-a-time updates with
the tailored Metropolis-Hastings step (SMH) in Section \ref{sec:mcmc-details},
and iii) full block updating of all knots using the tailored Metropolis-Hastings
step (BMH) in Section \ref{sec:mcmc-details}. SRWM moves are used
in state-of-the-art RJMCMC approaches such as \citet{dimatteo2001bayesian}
and \citet{RazulFitzgeraldAndrieu2003}. Note that we are not studying
the performance of a complete RJMCMC scheme; we are here interested
in isolating this particular updating step and comparing it to our
tailored proposal. We use the inefficiency factor (IF) \citep{geweke1992evaluating}
to measure the efficiency of MCMC. The IF is a measure of the number
of draws needed to obtain the equivalent of a single independent draw.
It is defined as $\mathrm{IF}=1+2\sum_{i=1}^{\infty}\rho_{i}$ where
$\rho_{i}$ is the autocorrelation of the MCMC trajectory at lag $i$.
We also document the effective sample size per minute, i.e. $(\text{number of MCMC draws})/(\text{IF}\times\text{computing time})$
to measure the overall efficiency of the MCMC.

Table \ref{tab:kstep-vs-rv} shows the efficiency of the three knot
sampling algorithms in a model with $20$ free surface knots and $4$
additive knots in each covariate on the firm leverage data. The inefficiency
factor in Table \ref{tab:kstep-vs-rv} is the average inefficiency
of the posterior mean surface in $1000$ random chosen points in covariate
space. There is some gain from tailoring the proposal for each knot
separately, but the really striking observation from Table \ref{tab:kstep-vs-rv}
is the massive efficiency and speed gains from updating all the blocks
jointly using a tailored proposal; the effective sample size per minute
is roughly $70$ times larger when our BMH algorithm is used instead
of simple SRWM updates.

\begin{table}
\caption{Comparison of algorithms for updating the knot locations in a model
with $20$ free surface knots and $4$ additive knots in each covariate.
Firm leverage data.}

\label{tab:kstep-vs-rv} \centering %
\begin{tabular}{lrrrrrr}
\hline
 &  & \multicolumn{1}{c}{SRWM} &  & \multicolumn{1}{c}{SMH} &  & \multicolumn{1}{c}{BMH}\tabularnewline
\hline
Mean IF for the posterior mean surface  &  & $29.63$ &  & $2.70$ &  & $1.16$\tabularnewline
Mean acceptance probability  &  & $0.26$  &  & $0.62$  &  & $0.88$ \tabularnewline
Computing time (\emph{min})  &  & $388.21$ &  & $1716.07$  &  & $141.72$ \tabularnewline
Effective sample size per minute  &  & $0.87$ &  & $2.16$  &  & $60.83$ \tabularnewline
\hline
\end{tabular}
\end{table}

\section{Concluding remarks}

\label{sec:Conclusions} We have presented a general Bayesian approach
for fitting a flexible surface model for a continuous multivariate
response using a radial basis spline with freely estimated knot locations.
Our approach uses shrinkage priors to avoid overfitting. The locations
of the knots and the shrinkage parameters are treated as unknown parameters
and we propose a highly efficient MCMC algorithm for these parameters
with the coefficients of the multivariate spline integrated out analytically.
An important feature of our algorithm is that all knot locations are
sampled jointly using a Metropolis-Hastings proposal density tailored
to the conditional posterior, rather than the one-knot-at-a-time random
walk proposals used in previous literature. The same applies to the
block of shrinkage parameters. Both a simulation study and a real
application on firm leverage data show that models with free knots
have a better out-of-sample predictive performance than models with
fixed knots. Moreover, the free knots model is also more robust in
the sense that it performs consistently well across different datasets.
We also found that models that mix surface and additive spline basis
functions in the same model perform better than models with only one
of the two basis types.

Our approach can be directly used with other splines basis functions,
other priors, and it is at least in principle straightforward to augment
the model with Bayesian variable selection. Also, the assumption of
Gaussian error distribution could be easily removed by using a Dirichlet
process mixture (DPM) prior. We would still be able to integrate out
the regression coefficients if we assume a Gaussian base measure in
the DPM, see \citet{leslie2007general} for details in the univariate
case.

\section{Acknowledgements}

The authors are grateful to Paolo Giordani and Robert Kohn for stimulating
discussions and constructive suggestions. The authors thank two anonymous
referees for the helpful comments that improved the contents and presentation
of the paper. The computations were performed on resources provided
by SNIC through Uppsala Multidisciplinary Center for Advanced Computational
Science (UPPMAX) under Project p2011229.

\appendix

\section{Details of the MCMC algorithm}

\label{AppendixGradients} In this section we briefly address the
MCMC details and related computational issues. For details on matrix
manipulations and derivatives, see e.g. \citet{lutkepohl1996handbook}.
Our MCMC algorithm in Section \ref{sec:mcmc-details} only requires
the gradient of the conditional posteriors w.r.t. each parameter.
Since users can always use their own prior on the knots and shrinkages,
we will not document the gradient of any particular prior. In particular
for the normal prior, one can directly find the results in e.g. \citet{mardia1979multivariate}.
We now present the full gradients for the knot locations and the shrinkage
parameters.

\subsection{Gradient w.r.t. the knot locations}

\label{sec: grad4knots}
\[
\begin{split}\frac{{\partial\ln p\left({\bm{\xi}|\bm{\lambda},\bm{\Sigma},\bm{Y},\bm{X}}\right)}}{{\partial\bm{\xi}'}}= & ~\frac{{\partial\log p(\bm{\xi})}}{{\partial\bm{\xi}'}}-\frac{p}{2}\sum\limits _{i\in\{o,a,s\}}{\left(\mathrm{vec}{\bm{P}_{i}}\right)}'\frac{{\partial\mathrm{vec}{\bm{P}_{i}}}}{{\partial\bm{\xi}'}}\\
 & -\left({\bm{\tilde{\beta}}-\bm{\mu}}\right)'\bm{\Sigma}_{\bm{\beta}}^{-1}\frac{{\partial\bm{\tilde{\beta}}}}{{\partial\bm{\xi}'}}-\frac{1}{2}\left(\mathrm{vec}{\bm{\Sigma}_{\bm{\tilde{\beta}}}}\right)'\frac{{\partial\mathrm{vec}{\bm{[}{\Sigma}^{-1}}\otimes\bm{X}'\bm{X}]}}{{\partial\bm{\xi}'}}\hfill\\
 & -\frac{1}{2}\left(\mathrm{vec}{\bm{\Sigma}^{-1}}\right)'\left({{\bm{I}_{p}}+{\bm{K}_{p,p}}}\right)\left\{ {({\bm{I}_{p}}\otimes\bm{\tilde{E}}'\bm{X})\frac{{\partial\bm{\tilde{\beta}}}}{{\partial\bm{\xi}'}}+(\bm{\tilde{B}}'\otimes\bm{\tilde{E}}')\frac{{\partial\mathrm{vec}\bm{X}}}{{\partial\bm{\xi}'}}}\right\} \hfill\\
 & -\frac{1}{2}\left\{ \mathrm{vec}\left[{\left({\bm{\tilde{\beta}}-\bm{\mu}}\right)\left({\bm{\tilde{\beta}}-\bm{\mu}}\right)'+\bm{\Sigma}_{\bm{\tilde{\beta}}}}\right]\right\} '\frac{{\partial\mathrm{vec}\bm{\Sigma}_{\bm{\beta}}^{-1}}}{{\partial\bm{\xi}'}},\hfill
\end{split}
\]
where $\bm{\tilde{E}}=\bm{Y}-\bm{X\tilde{B}}$, $\bm{I}_{p}$ is the
identity matrix, $\bm{K}_{p,p}$ is the commutation matrix and
\[
\frac{{\partial\mathrm{vec}[{\bm{\Sigma}^{-1}}\otimes\bm{X}'\bm{X}]}}{{\partial\bm{\xi}'}}=\left({{\bm{I}_{p}}\otimes{\bm{K}_{q,p}}\otimes{\bm{I}_{q}}}\right)\left({\mathrm{vec}{\bm{\Sigma}^{-1}}\otimes{\bm{I}_{{q^{2}}}}}\right)\left({{\bm{I}_{{q^{2}}}}+{\bm{K}_{q,q}}}\right)\left({{\bm{I}_{q}}\otimes\bm{X}'}\right)\frac{{\partial\mathrm{vec}\bm{X}}}{{\partial\bm{\xi}'}},
\]
\[
\begin{split}\frac{{\partial\bm{\tilde{\beta}}}}{{\partial\bm{\xi}'}}=~ & {\bm{\Sigma}_{\bm{\tilde{\beta}}}}\left[{\left[{{\bm{\Sigma}^{-1}}\bm{Y}'\otimes{\bm{I}_{q}}}\right]{\bm{K}_{n,q}}\frac{{\partial\mathrm{vec}\bm{X}}}{{\partial\bm{\xi}'}}+\left({\bm{\mu}'\otimes{\bm{I}_{pq}}}\right)\frac{{\partial\mathrm{vec}\bm{\Sigma}_{\bm{\beta}}^{-1}}}{{\partial\bm{\xi}'}}}\right]\hfill\\
 & -\left[\left\{ {\left[{\mathrm{vec}\left({\bm{X}'\bm{Y}{\bm{\Sigma}^{-1}}}\right)+\bm{\Sigma}_{\bm{\beta}}^{-1}\bm{\mu}}\right]'{\bm{\Sigma}_{\bm{\tilde{\beta}}}}}\right\} \otimes{\bm{\Sigma}_{\bm{\tilde{\beta}}}}\right]\left[{\frac{{\partial\mathrm{vec}[{\bm{\Sigma}^{-1}}\otimes\bm{X}'\bm{X}]}}{{\partial\bm{\xi}'}}+\frac{{\partial\mathrm{vec}\bm{\Sigma}_{\bm{\beta}}^{-1}}}{{\partial\bm{\xi}'}}}\right].\hfill
\end{split}
\]

We can decompose the gradient for the design matrix w.r.t the knots
as
\[
\frac{{\partial\mathrm{vec}\bm{X}}}{{\partial\bm{\xi}'}}=\left[{\begin{array}{ll}
\bm{0}_{(nq_{o}\times l_{s})} & {\bm{0}_{(nq_{o}\times l_{a})}}\\
\partial\mathrm{vec}{\bm{X}_{s}}/\partial{\mathrm{vec}(\bm{\xi}_{s}')}' & {\bm{0}_{(nq_{s}\times l_{a})}}\\
{\bm{0}_{(nq_{a}\times l_{s})}} & \partial\mathrm{vec}{\bm{X}_{a}}/\partial{\bm{\xi}_{a}}'
\end{array}}\right]
\]
where $l_{s}$ and $l_{a}$ are numbers of parameters in the knots
locations for surface and additive component, respectively. This decomposition
makes user-specified basis functions for different components possible
and one may update the locations in a parallel mode (efficient for
small models) or batched mode (for models with many parameters). In
particular for the thin-plate spline, we have
\[
\frac{{\partial{\text{vec}}{{\bm{X}}_{i}}}}{{\partial{\bm{\xi}_{i}}^{\prime}}}=-{\left[{\begin{array}{ccc}
{({1+2\ln\Vert{{{\bm{x}}_{i}}-{\bm{\xi}_{ij}}}\Vert})({{{\bm{x}}_{i}}-{\bm{\xi}_{ij}}})} & {} & {}\\
{} & \ddots & {}\\
{} & {} & {({1+2\ln\Vert{{{\bm{x}}_{i}}-{\bm{\xi}_{ij}}}\Vert})({{{\bm{x}}_{i}}-{\bm{\xi}_{ij}}})}
\end{array}}\right]{\begin{subarray}{l}
\vspace{1.5cm}\\
i\in\{a,s\},\\
j\in\{1,...,{q_{i}}\}.
\end{subarray}}}
\]

Note that the gradient can be obtained efficiently by applying Lemma
\ref{def:CxDSigma} and Algorithm \ref{def:KX} in Section \ref{sec:compremarks}
below whenever $\partial\mathrm{vec}\bm{\Sigma}_{\bm{\beta}}^{-1}/\partial\bm{\xi}'$
and the commutation matrix appear.

\subsection{Gradient w.r.t. the shrinkage parameters}

\label{sec: grad4shrinkages}
\[
\begin{split}\frac{{\partial\ln p\left({\bm{\lambda}|\bm{\xi},\bm{\Sigma},\bm{Y},\bm{X}}\right)}}{{\partial\bm{\lambda}'}}= & ~\frac{{\partial\log p(\bm{\lambda})}}{{\partial\bm{\lambda}'}}-\frac{1}{2}\left[{{q_{o}}{\bm{\lambda}_{o}}',{q_{s}}{\bm{\lambda}_{s}}',{q_{a}}{\bm{\lambda}_{a}}'}\right]-\left(\bm{\tilde{\beta}}-\bm{\mu}\right)'\bm{\Sigma}_{\bm{\beta}}^{-1}\frac{\partial\bm{\tilde{\beta}}}{\partial\bm{\lambda}'}\\
 & -\frac{1}{2}\left(\mathrm{vec}{\bm{\Sigma}^{-1}}\right)'\left({{\bm{I}_{p}}+{\bm{K}_{p,p}}}\right)\left({{\bm{I}_{p}}\otimes\bm{\tilde{E}}'\bm{X}}\right)\frac{{\partial\bm{\tilde{\beta}}}}{{\partial\bm{\lambda}'}}\hfill\\
 & -\frac{1}{2}\mathrm{vec}\left[{\left({\bm{\tilde{\beta}}-\bm{\mu}}\right)\left({\bm{\tilde{\beta}}-\bm{\mu}}\right)'+{\mathrm{\Sigma}_{\bm{\tilde{\beta}}}}}\right]'\frac{{\partial\mathrm{vec}\bm{\Sigma}_{\bm{\beta}}^{-1}}}{{\partial\bm{\lambda}'}},\hfill
\end{split}
\]
where
\[
\frac{{\partial\bm{\tilde{\beta}}}}{{\partial\bm{\lambda}'}}=\left\{ {\left[{\left({\mathrm{vec}\left({\bm{X}'\bm{Y}{\bm{\Sigma}^{-1}}}\right)+\bm{\Sigma}_{\bm{\beta}}^{-1}\bm{\mu}}\right)'{\bm{\Sigma}_{\bm{\tilde{\beta}}}}}\right]\otimes{\bm{\Sigma}_{\bm{\beta}}}-\bm{\mu}'\otimes{\bm{\Sigma}_{\bm{\tilde{\beta}}}}}\right\} \frac{{\partial\mathrm{vec}\bm{\Sigma}_{\bm{\beta}}^{-1}}}{{\partial\bm{\lambda}'}},
\]
and $\partial\mathrm{vec}\bm{\Sigma}_{\bm{\beta}}^{-1}/\partial\bm{\lambda}'$
can be obtained efficiently by applying Lemma \ref{def:CxDSigma}
in Section \ref{sec:compremarks} and by
\[
\begin{split}\frac{{\partial\mathrm{vec}[({\bm{\bm{\Lambda}}_{i}^{-1/2}}{\bm{\Sigma}^{-1}}{\bm{\bm{\Lambda}}_{i}^{-1/2}})\otimes{\bm{P}_{i}}]}}{{\partial{\bm{\lambda}_{i}}'}}=~ & \left({{\bm{I}_{p}}\otimes{\bm{K}_{{q_{i}},p}}\otimes{\bm{I}_{{q_{i}}}}}\right)\left({{\bm{I}_{{p^{2}}}}\otimes\mathrm{vec}{\bm{P}_{i}}}\right)\left({{\bm{I}_{{p^{2}}}}+{\bm{K}_{p,p}}}\right)\hfill\\
 & \times\left({{\bm{I}_{p}}\otimes[{\bm{\bm{\Lambda}}_{i}^{-1/2}}{\bm{\Sigma}^{-1}}]}\right)\frac{\partial\mathrm{vec}\bm{\Lambda}_{i}^{-1/2}}{\partial\bm{\lambda}_{i}'},~i\in\{a,s\}.\hfill
\end{split}
\]
where $\partial\mathrm{vec}\bm{\Lambda}_{i}/\partial\bm{\lambda}_{i}'$
is $p^{2}\times p$ matrix with elements $\nabla_{j(p+1)-p,~j}=-1/2\lambda_{i,j}^{-3/2}$
for $j=1,...,p$ and zero elsewhere.

\subsection{Computational remarks}

\label{sec:compremarks} The computational implementation of gradients
in Section \ref{sec: grad4knots} and Section \ref{sec: grad4shrinkages}
is straightforward but the sparsity of some of the matrices can be
exploited in moderate to large datasets. We now present a lemma and
an algorithm that can dramatically speed up the computations. It is
convenient to define $\bm{A}(\bm{i},\textnormal{:})$ and $\bm{A}(:,\bm{j})$
as matrix operations that reorders the rows and columns of matrix
$\bm{A}$ with indices $\bm{i}$ and $\bm{j}$. Therefore, $\bm{\beta}=\bm{b}(\bm{c},:)$,
$\bm{\mu}=\bm{\mu}^{*}(\bm{c},:)$ and $\bm{\Sigma}_{\bm{\beta}}=\bm{\Sigma}_{\bm{b}}(\bm{c},\bm{c})$
for proper indices $\bm{c}$, and $|\bm{\Sigma}_{\bm{b}}|=|\bm{\Sigma}_{\bm{\beta}}|$
since permuting two rows or columns changes the sign but not the magnitude
of the determinant.

\begin{lemma} \label{def:CxDSigma} Given matrix $\bm{C}$ and the
indexing vector $\bm{z}$ such that $(\mathrm{vec}\bm{\Sigma}_{\bm{b}})(\bm{z},:)=\mathrm{vec}\bm{\Sigma}_{\bm{\beta}}$
holds, we can decompose the following gradient as
\[
\bm{C}\frac{{\partial\mathrm{vec}[\bm{\Sigma}_{\bm{\beta}}^{-1}\left(\bm{\theta}\right)]}}{{\partial\bm{\theta}}'}=\left[{{\bm{C}_{s}}\frac{{\partial\mathrm{vec}[(\bm{\Lambda}_{s}^{-1/2}\bm{\Sigma}^{-1}\bm{\Lambda}_{s}^{-1/2})\otimes{\bm{P}_{s}}]}}{{\partial{\bm{\theta}_{s}}'}},~{\bm{C}_{a}}\frac{{\partial\mathrm{vec}[({\bm{\Lambda}_{a}^{-1/2}\bm{\Sigma}^{-1}\bm{\Lambda}_{a}^{-1/2}})\otimes{\bm{P}_{a}}]}}{{\partial{\bm{\theta}_{a}}'}}}\right]
\]
where $\bm{\theta}$ is any parameter vector of the covariance matrix
$\bm{\Sigma}_{\bm{\beta}}$, $\bm{C}_{s}=\{[\bm{C}(:,\bm{z})](:,\bm{h}_{s})\}(:,\bm{z}_{s}\neq0)$,
$\bm{h}_{s}=[(p^{2}qq_{o}+1),(p^{2}qq_{o}+2),...,p^{2}q(q_{o}+q_{s})]'$,
$\bm{z}_{s}=\mathrm{vec}([\bm{0}_{pq_{s}\times pq_{o}},~\bm{1}_{pq_{s}\times pq_{s}},~\bm{0}_{pq_{s}\times pq_{a}}]')$,
$\bm{C}_{a}=\{[\bm{C}(:,\bm{z})](:,\bm{h}_{a})\}(:,\bm{z}_{a}\neq0)$,
$\bm{h}_{a}=[(p^{2}q(q_{o}+q_{s})+1),(p^{2}q(q_{o}+q_{s})+2),...,p^{2}q^{2}]'$
and $\bm{z}_{a}=\mathrm{vec}([\bm{0}_{pq_{a}\times p(q_{o}+q_{s})},~\bm{1}_{pq_{a}\times pq_{a}}]')$.
\end{lemma}

\begin{algthm} \label{def:KX} An efficient algorithm to calculate
$\bm{K}_{m,n}\bm{Q}$ (or $\bm{Q}\bm{K}_{m,n}$) where $\bm{K}_{m,n}$
is the commutation matrix and $\bm{Q}$ is any dense matrix that is
conformable to $\bm{K}_{m,n}$.
\begin{enumerate}
\item Create an $m\times n$ (or $n\times m$) matrix $\bm{T}$ and fill
it by columns with the sequence $\{1,2,...,nm\}$.
\item Obtain the indexing vector $\bm{t}=\mathrm{vec}(\bm{T}')$.
\item Return $\bm{Q}(\bm{t},:)$ (or $\bm{Q}(:,\bm{t})$).
\end{enumerate}
\end{algthm}

\bibliographystyle{agsm}
\bibliography{myBibTeX}

\end{document}


\title[Supporting information for Efficient Bayesian Multivariate Surface
Regression]{Supporting Information for\\
  Efficient Bayesian Multivariate Surface Regression}

\author{Feng Li}

\author{Mattias Villani}

\thanks{Li (corresponding author): Department of Statistics, Stockholm University,
SE-106 91 Stockholm, Sweden. E-mail: \href{mailto:feng.li@stat.su.se}{\texttt{feng.li@stat.su.se}}.
Villani: Division of Statistics, Department of Computer and Information
Science, Linköping University, SE-581 83 Linköping, Sweden. E-mail:
\href{mailto:mattias.villani@liu.se}{\texttt{mattias.villani@liu.se}}.}

\maketitle

\section{Prior Robustness\label{sec:PriorRobustness}}

This section explores the sensitivity of the posterior inferences
with respect to variations in the prior. There are clearly many aspects
of the prior to explore, but we will here focus on the sensitivity
with respect to the prior on the shrinkage factors and the prior on
the knot locations, which are the most influential priors for the
model. Since our model is very flexible and richly parametrized it
is natural to expect, or even desireable, that the posterior responds
to variations in the prior hyperparameters. But since the prior in
complex models is always hard to specify, it is hoped that moderate
changes in the prior should at least not overturn the posterior inferences.

\subsection{The prior on the shrinkage parameters}

\begin{figure}
\centering

  \begin{tabular}{rccc}
    & \small{Default variance}
    & \small{Half the default variance}
    & \small{Twice the default variance} \\

    \begin{sideways}\hspace{1.2cm} \small{Posterior knot locations}\end{sideways}
    &\includegraphics*[trim = 0 0 110 50, height=0.25\textheight]{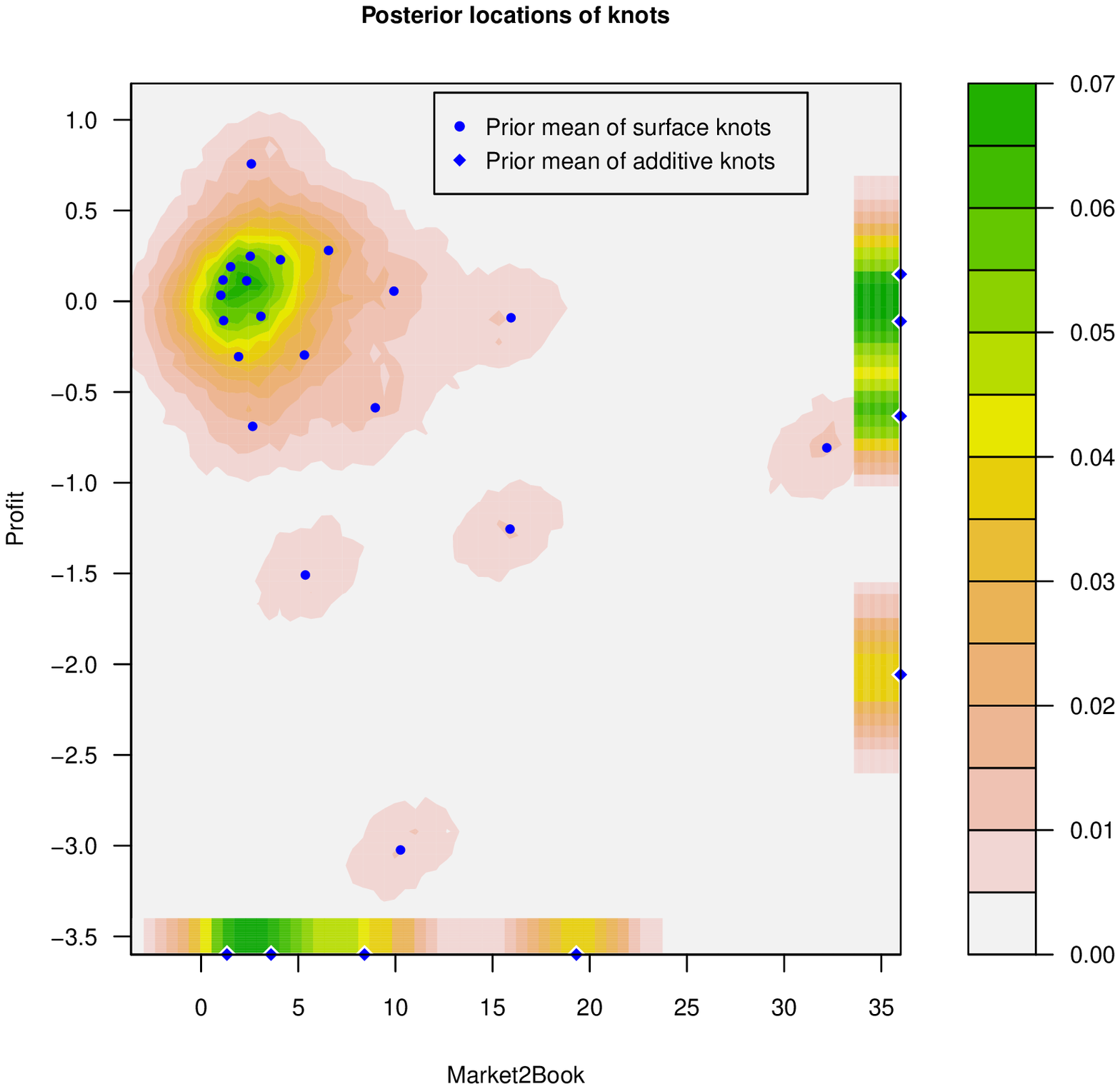}
    &\includegraphics*[trim = 30 0 110 50, height=0.25\textheight]{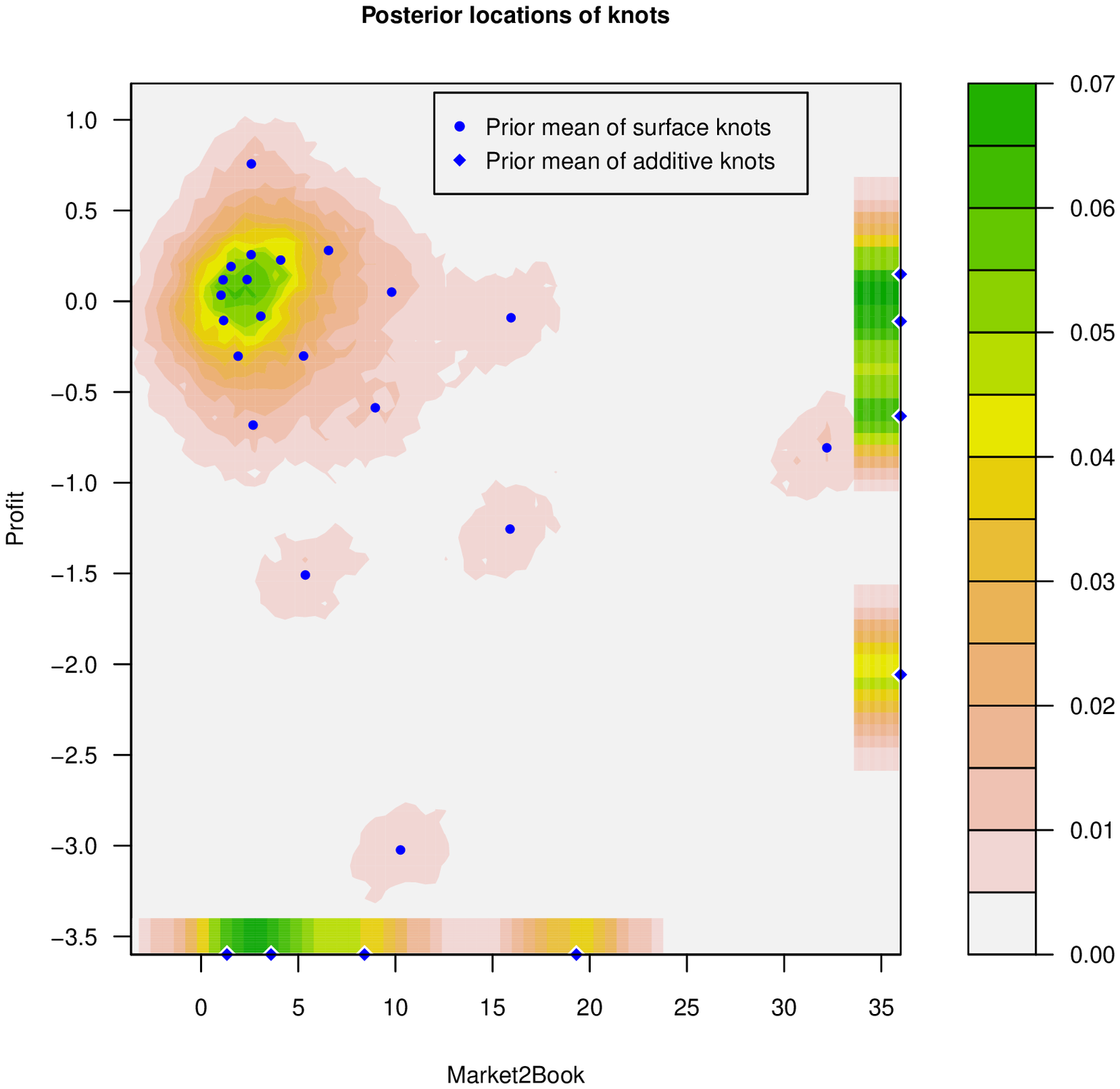}
    &\includegraphics*[trim = 30 0 0 50, height=0.25\textheight]{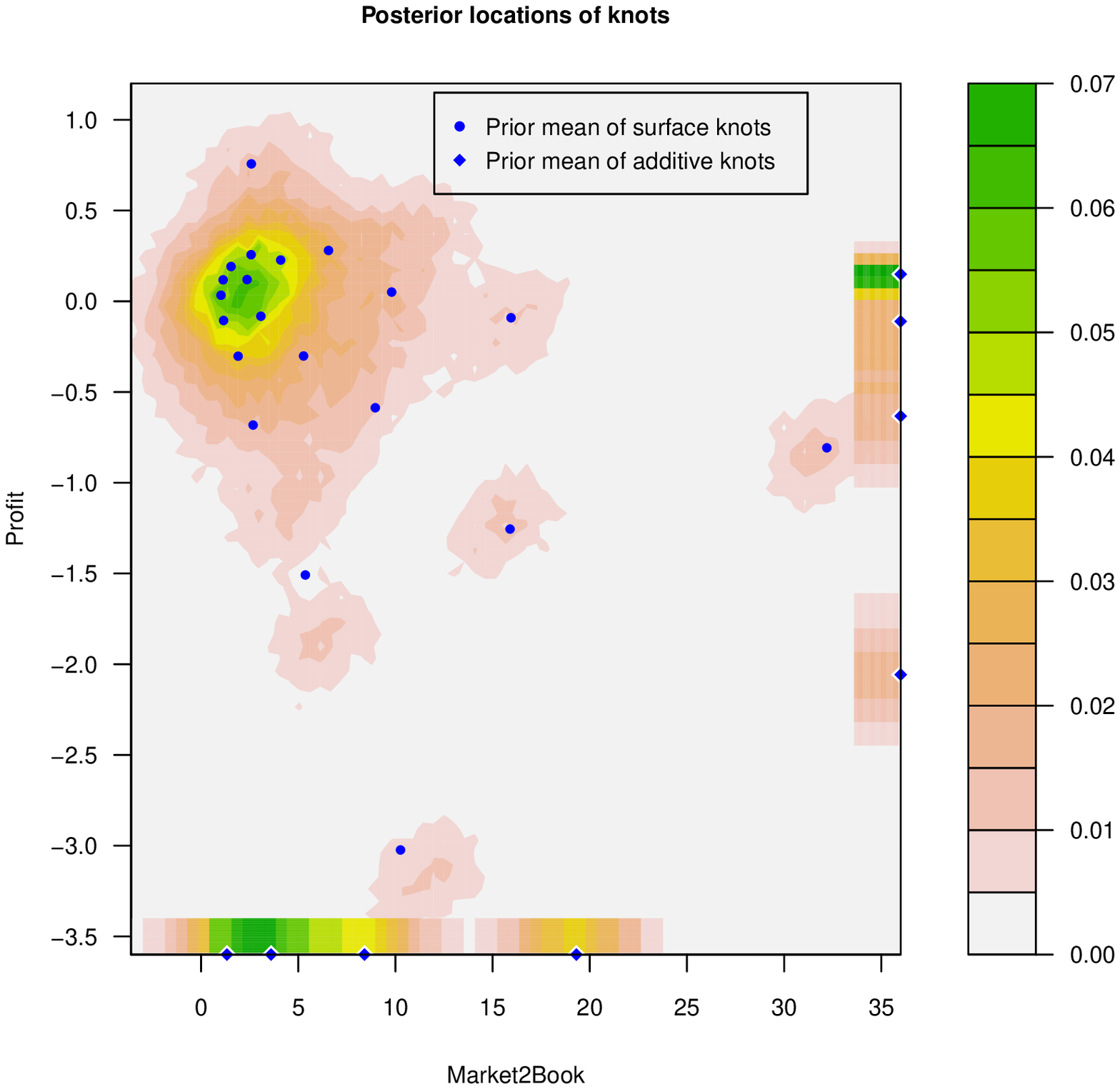}
    \\

    \begin{sideways}\hspace{0.8cm} \small{Posterior surface (mean)}\end{sideways}
    &\includegraphics*[trim = 0 0 110 50, height=0.25\textheight]{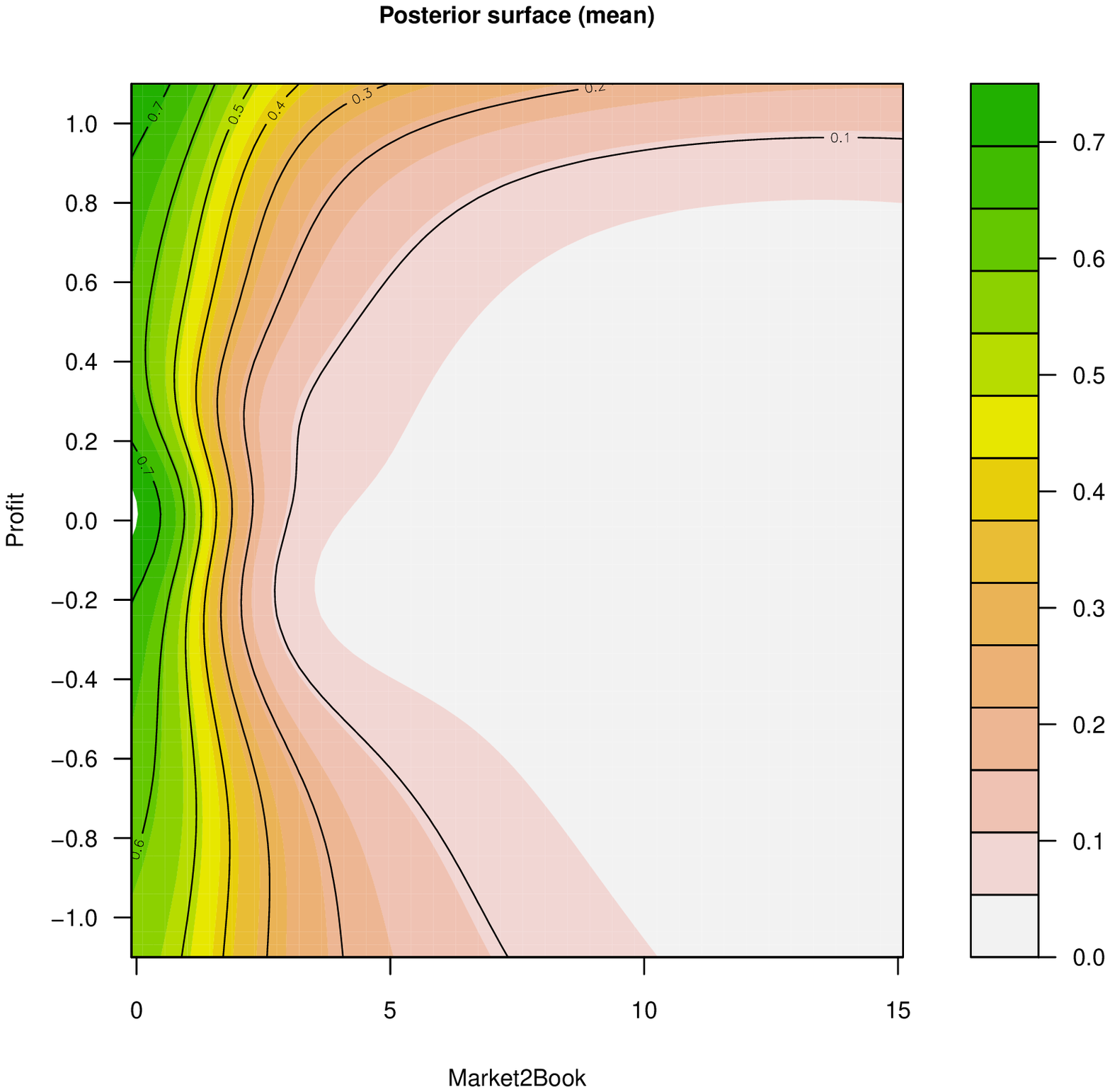}
    &\includegraphics*[trim = 30 0 110 50, height=0.25\textheight]{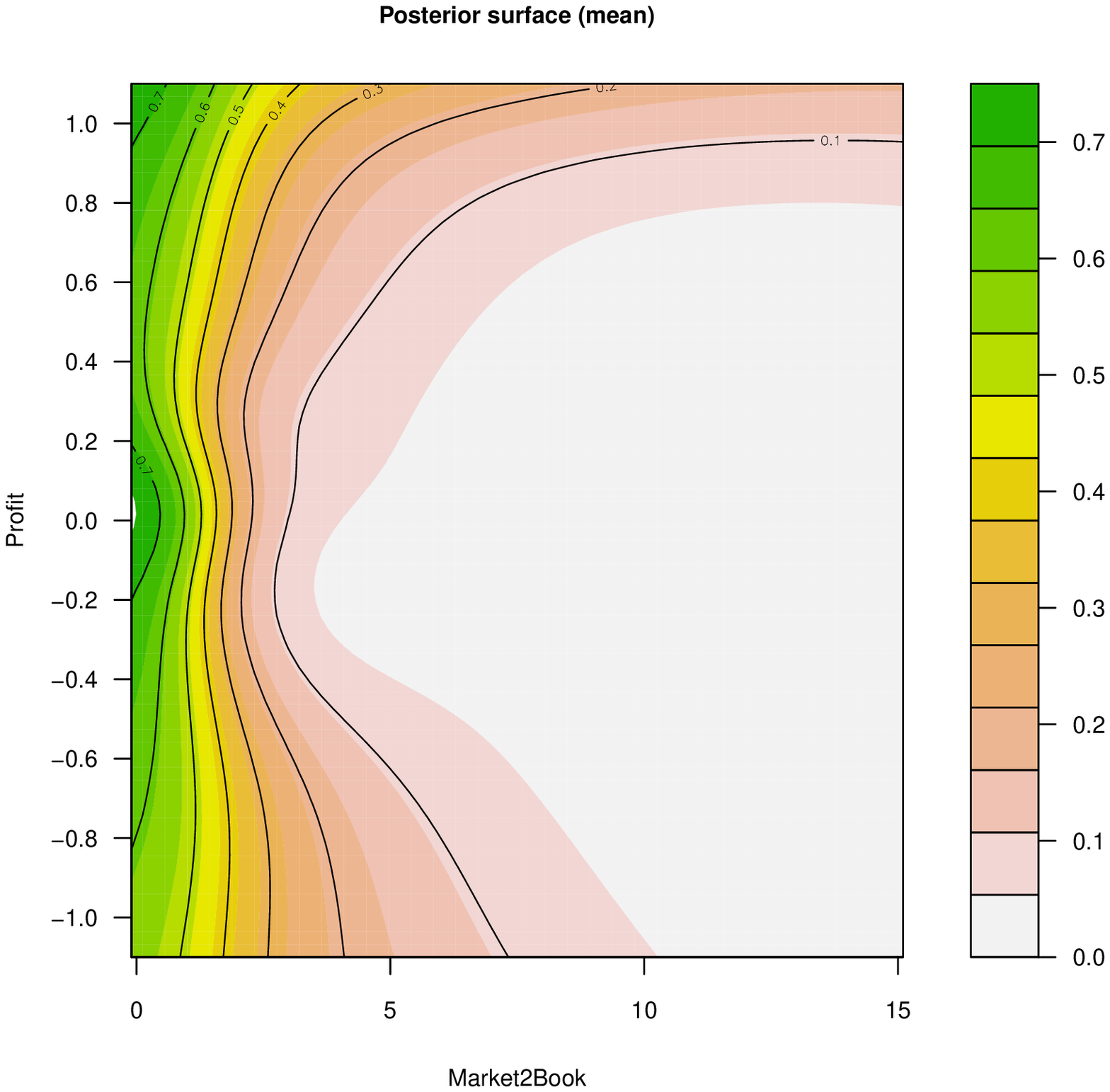}
    &\includegraphics*[trim = 30 0 0 50, height=0.25\textheight]{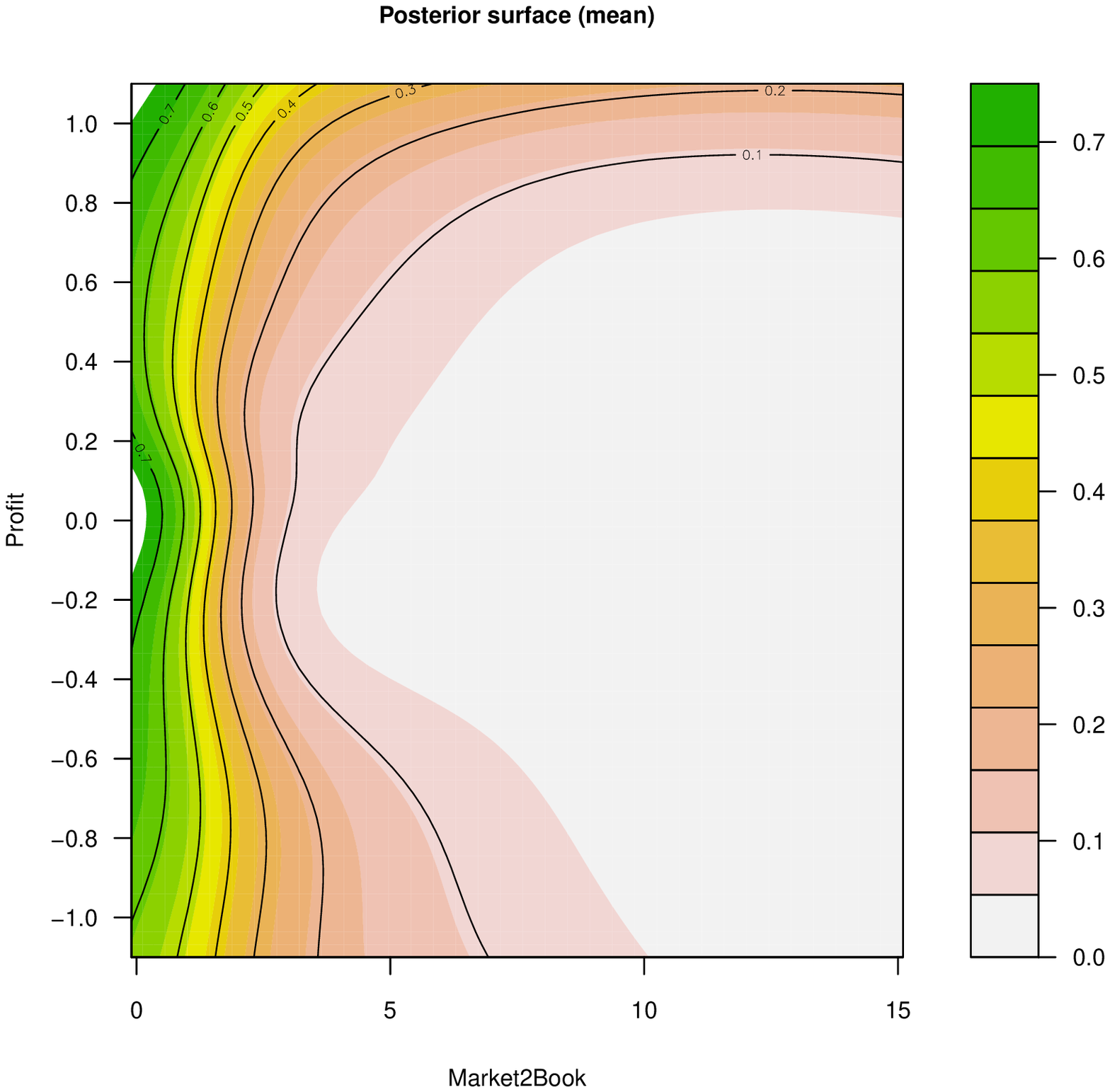}
    \\

    \begin{sideways}\hspace{0.8cm} \small{Posterior surface (sd)}\end{sideways}
    &\includegraphics*[trim = 0 0 110 50,height=0.25\textheight]{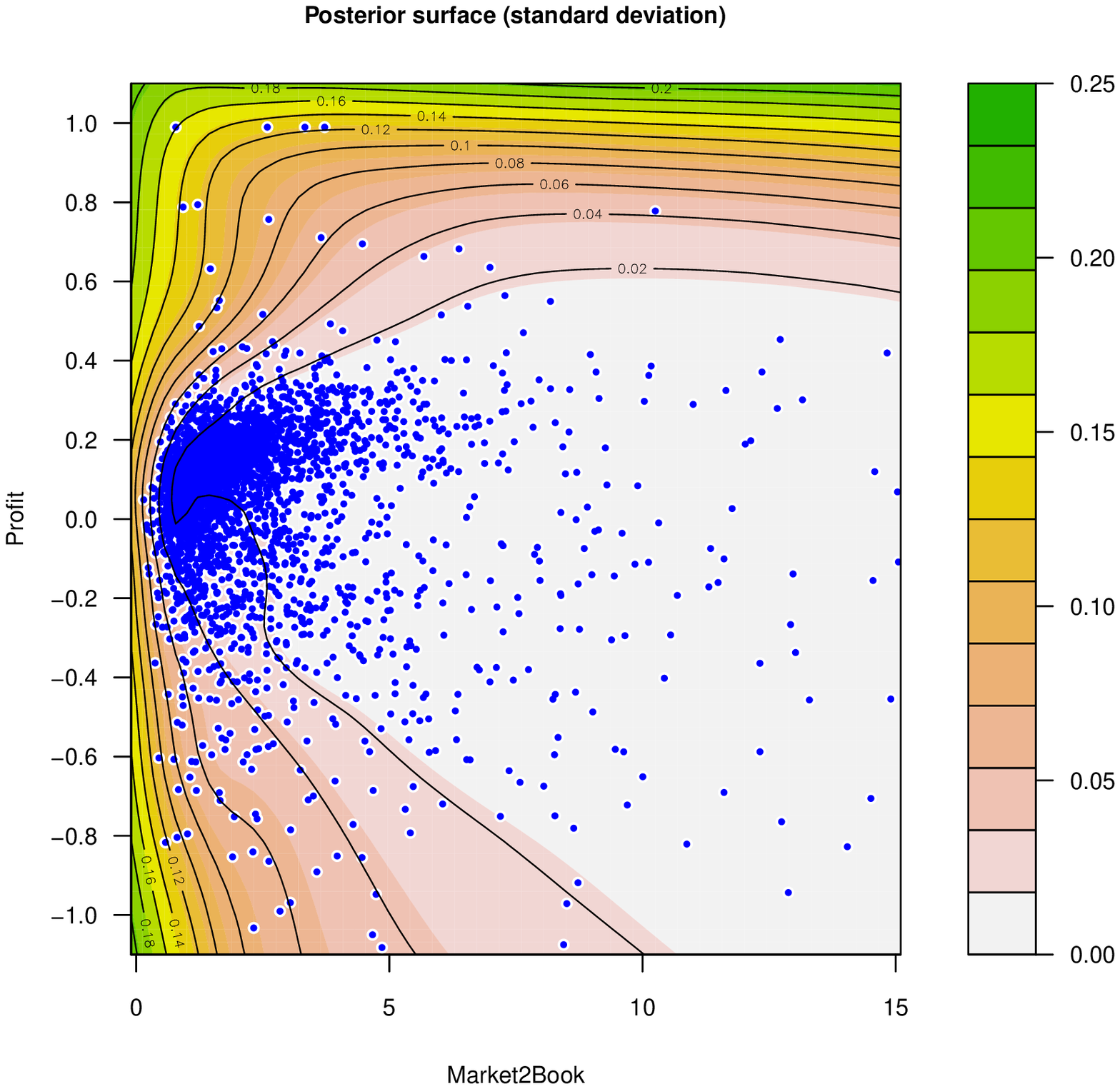}
    &\includegraphics*[trim = 30 0 110 50,height=0.25\textheight]{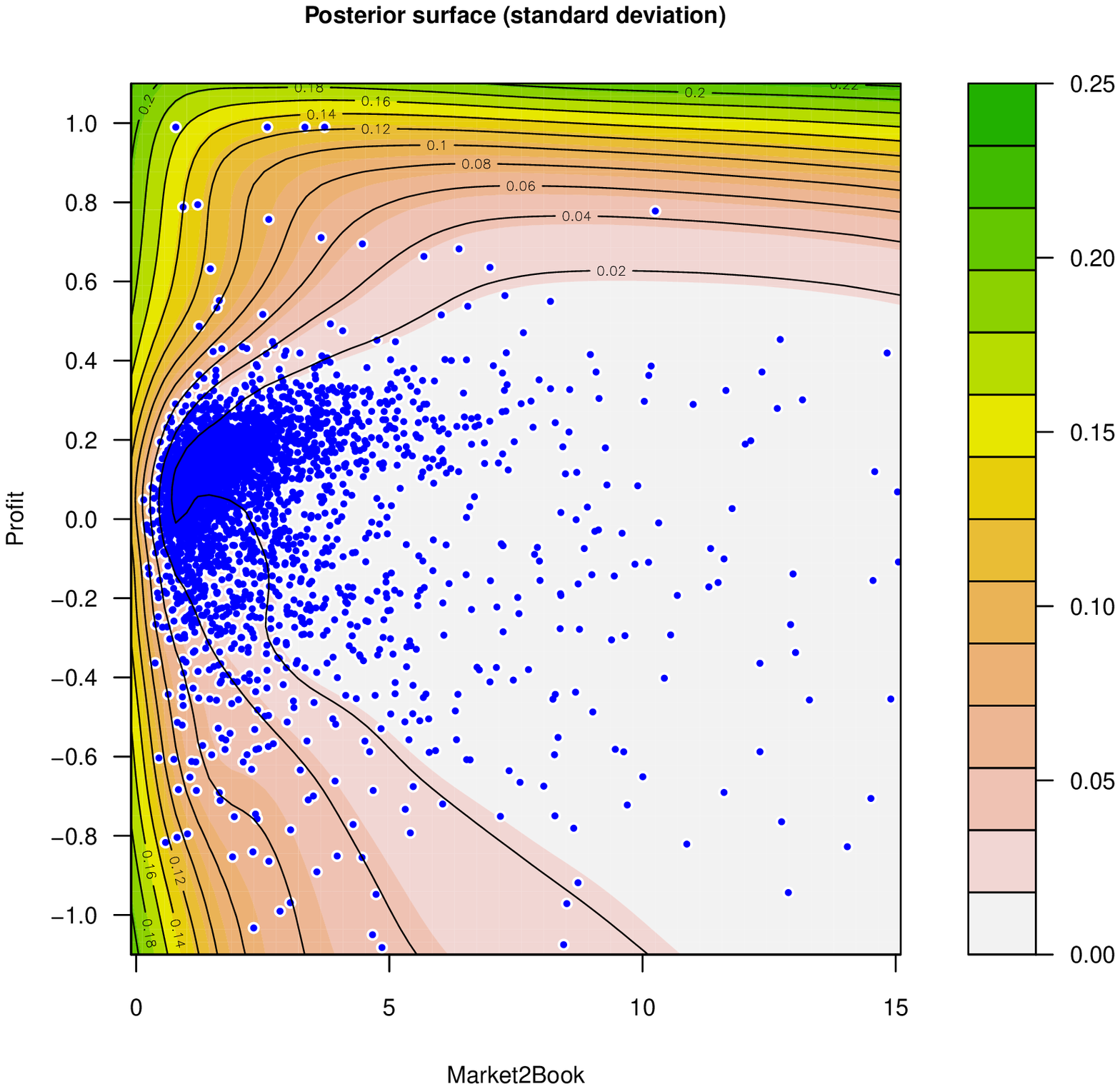}
    &\includegraphics*[trim = 30 0 0 50, height=0.25\textheight]{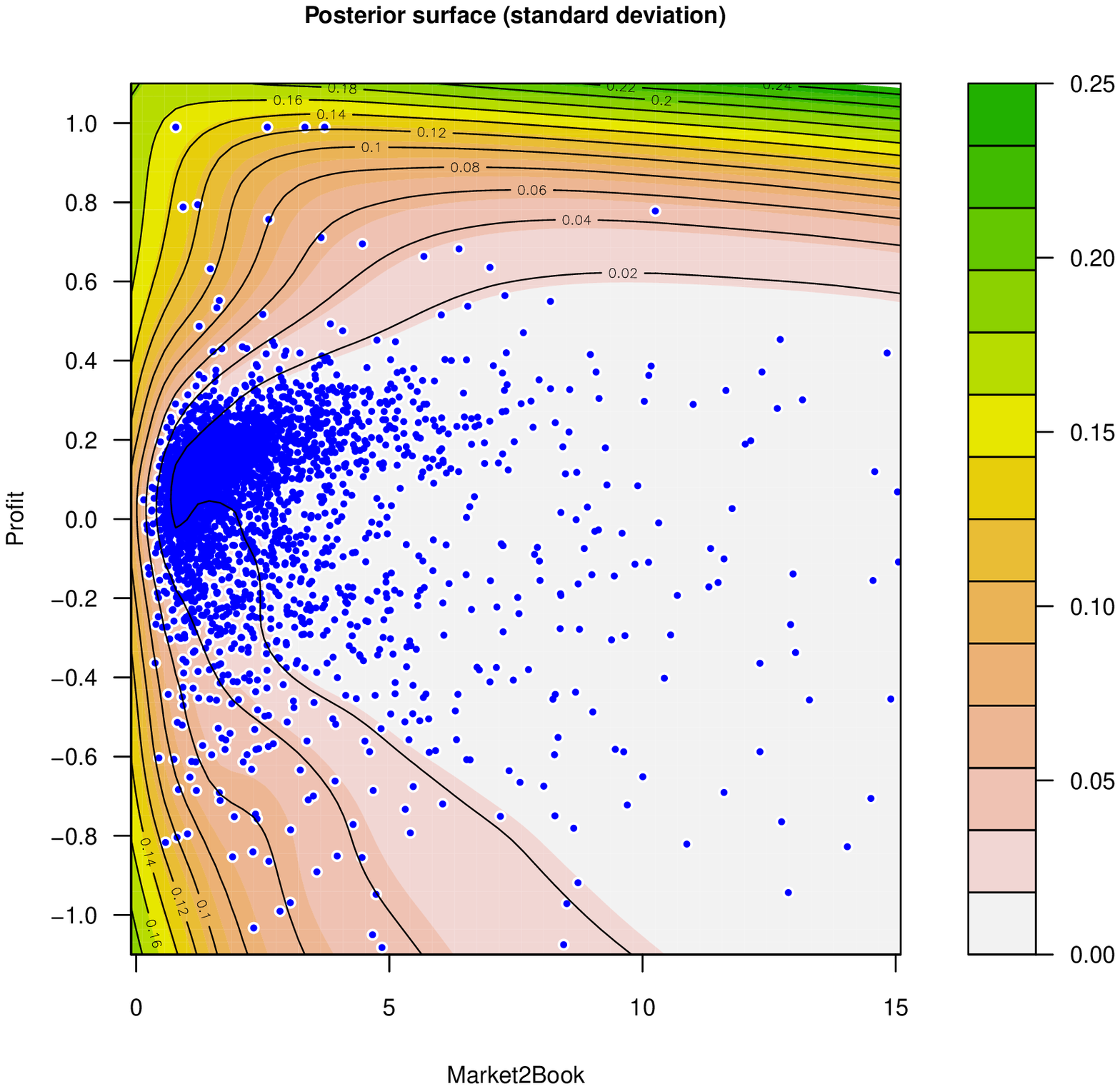}
    \\

  \end{tabular}

\caption{Posterior sensitivity with respect to changes in the prior variance
of the shrinkage factors. The first column shows the posterior of
the knot density (top row), the posterior mean surface (middle row)
and the posterior standard deviation surface (bottom row) using the
default prior in the paper. The second and third columns demonstrates
the effect on the posterior inferences when the prior variance is
half of the default value (second column) and twice the default value
(third column).}

\label{fig:post-shrinkage-comp}
\end{figure}

Figure \ref{fig:post-shrinkage-comp} displays the posterior sensitivity
of the knot locations, the posterior mean and standard deviation surfaces
to changes in the prior variance on the shrinkage factors. The posterior
and predictive results are clearly very robust to changes in this
particular aspect of the prior.

\subsection{The prior on the knot locations}

\begin{figure}
\centering %

  \begin{tabular}{rccc}
    & \small{\emph{K}-means location} & \small{Marginal Percentiles} & \small{Random} \\

    \begin{sideways}\hspace{1.8cm} \small{Default variance}\end{sideways}
    &\includegraphics*[trim = 0 0 110 50, height=0.25\textheight]{rajan_knots_1-1}
    &\includegraphics*[trim = 30 0 110 50, height=0.25\textheight]{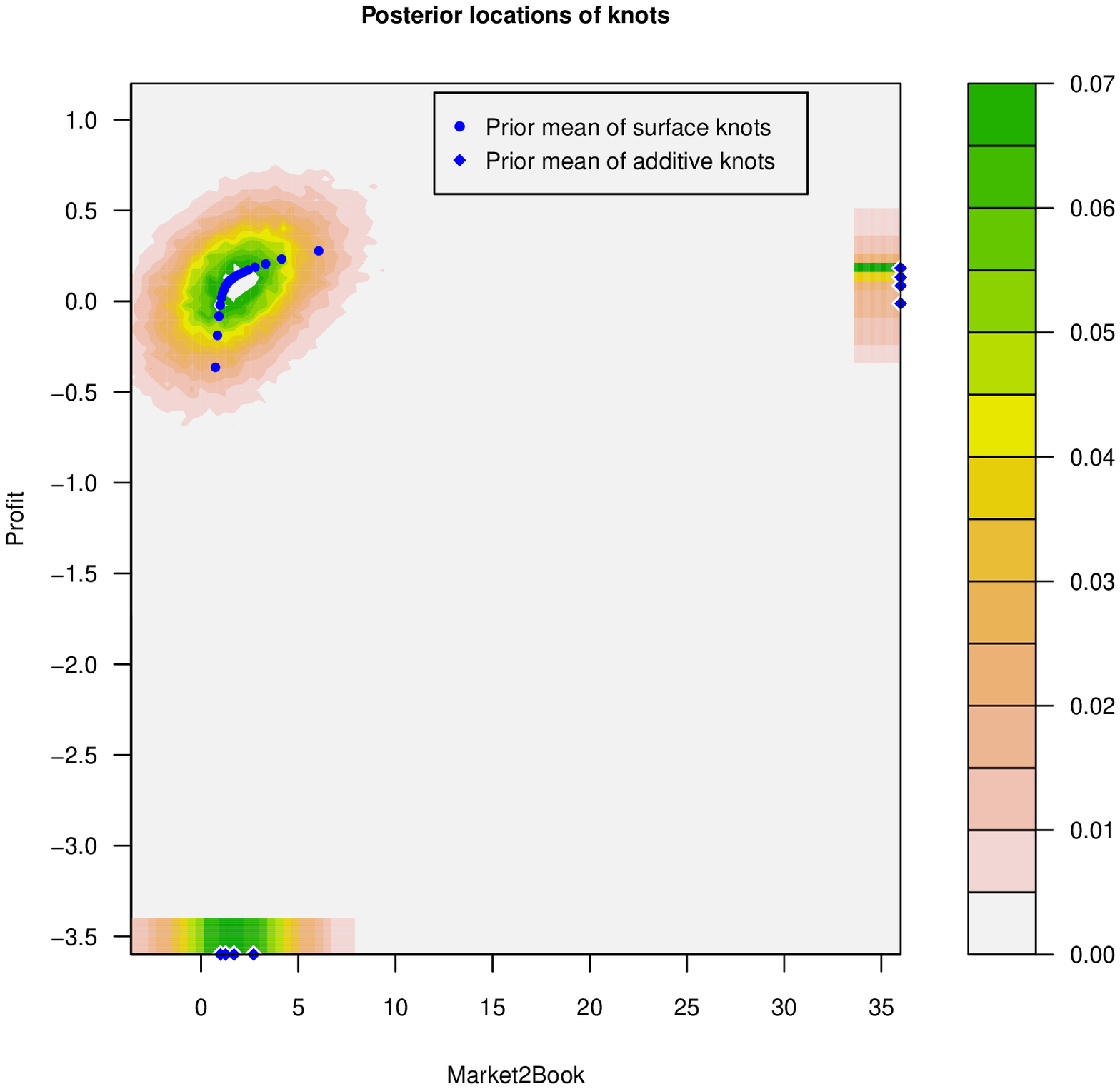}
    &\includegraphics*[trim = 30 0 0 50, height=0.25\textheight]{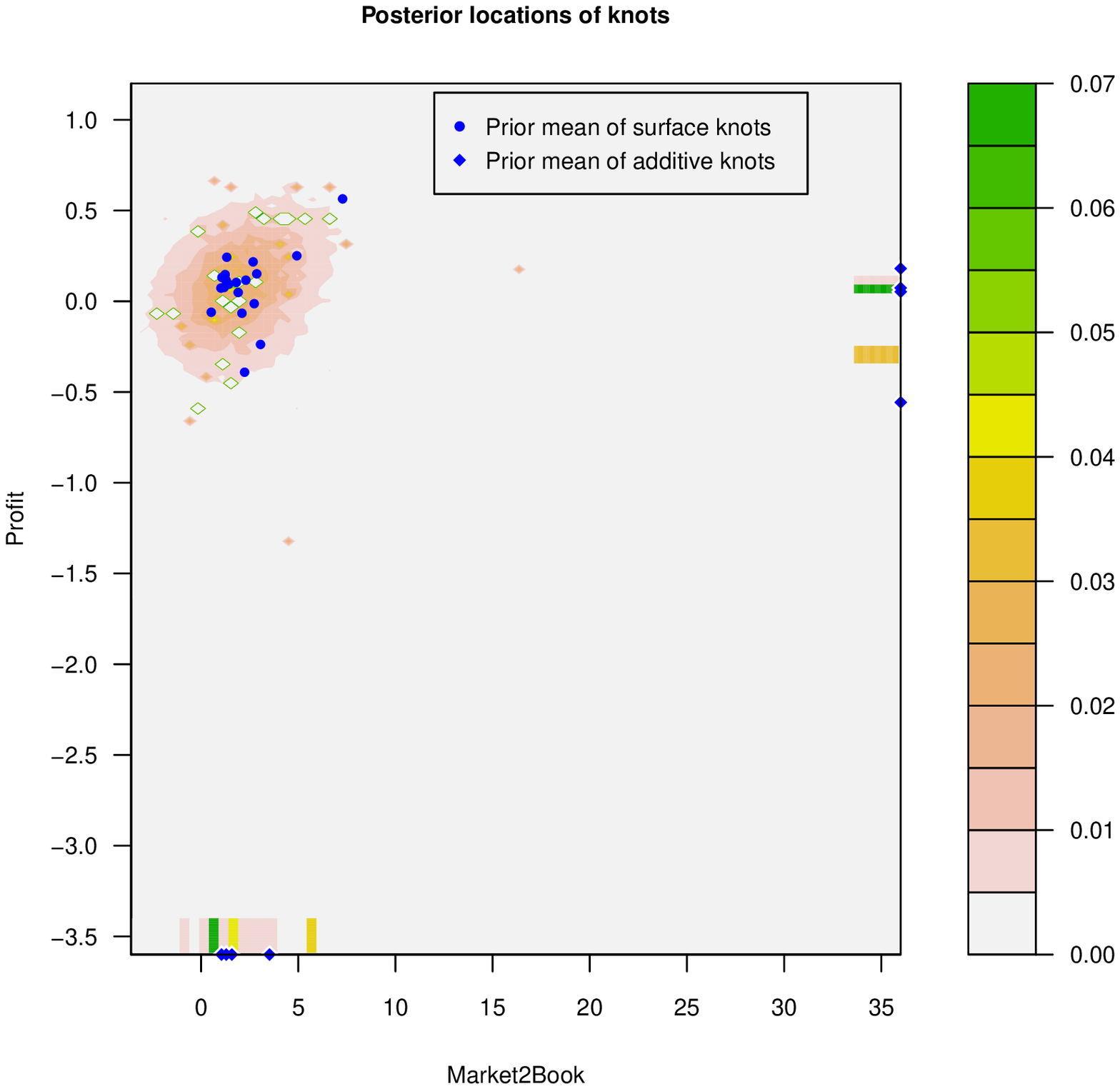}
    \\

    \begin{sideways}\hspace{1.2cm} \small{Half the default variance}\end{sideways}
    &\includegraphics*[trim = 0 0 110 50, height=0.25\textheight]{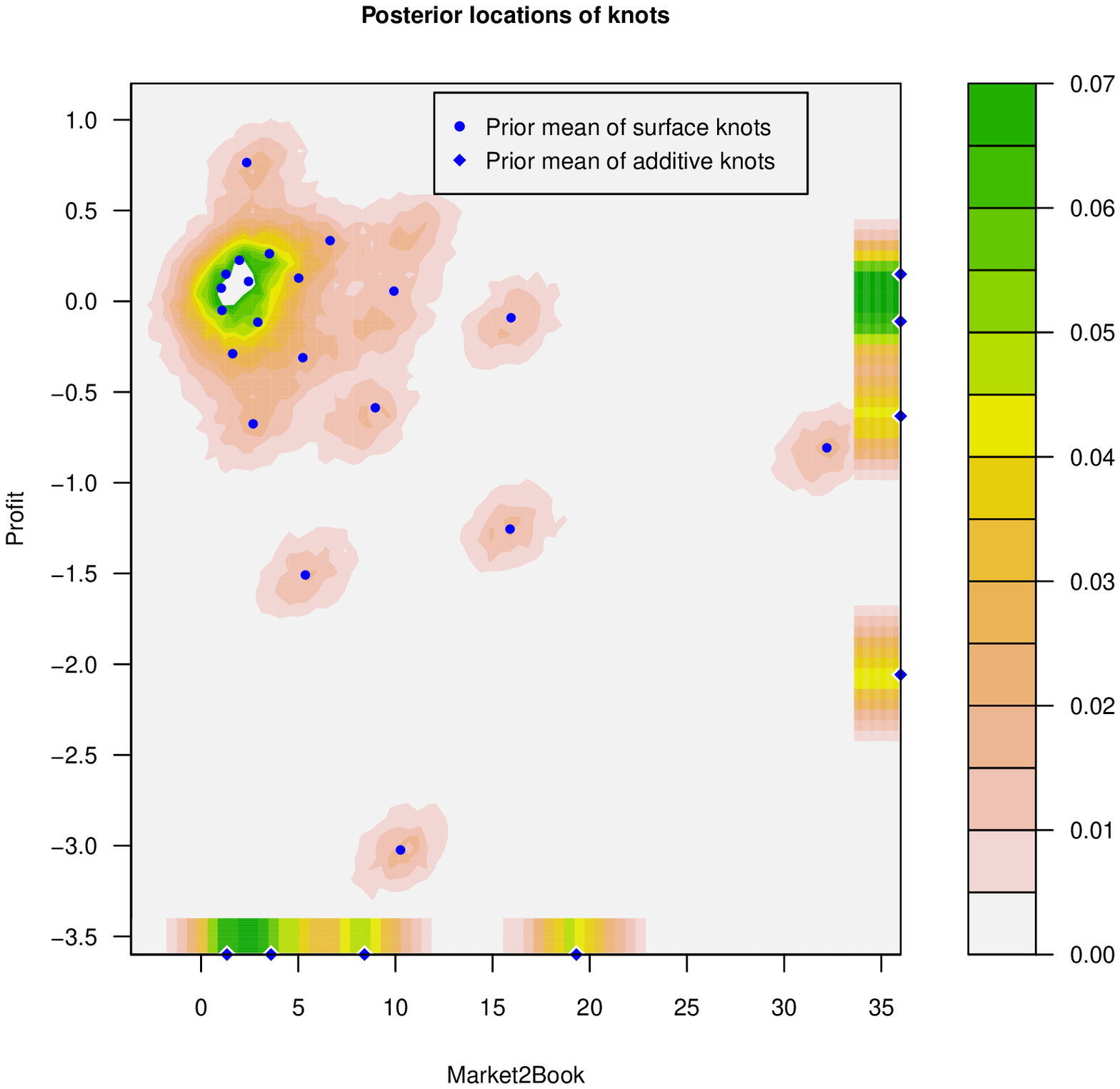}
    &\includegraphics*[trim = 30 0 110 50, height=0.25\textheight]{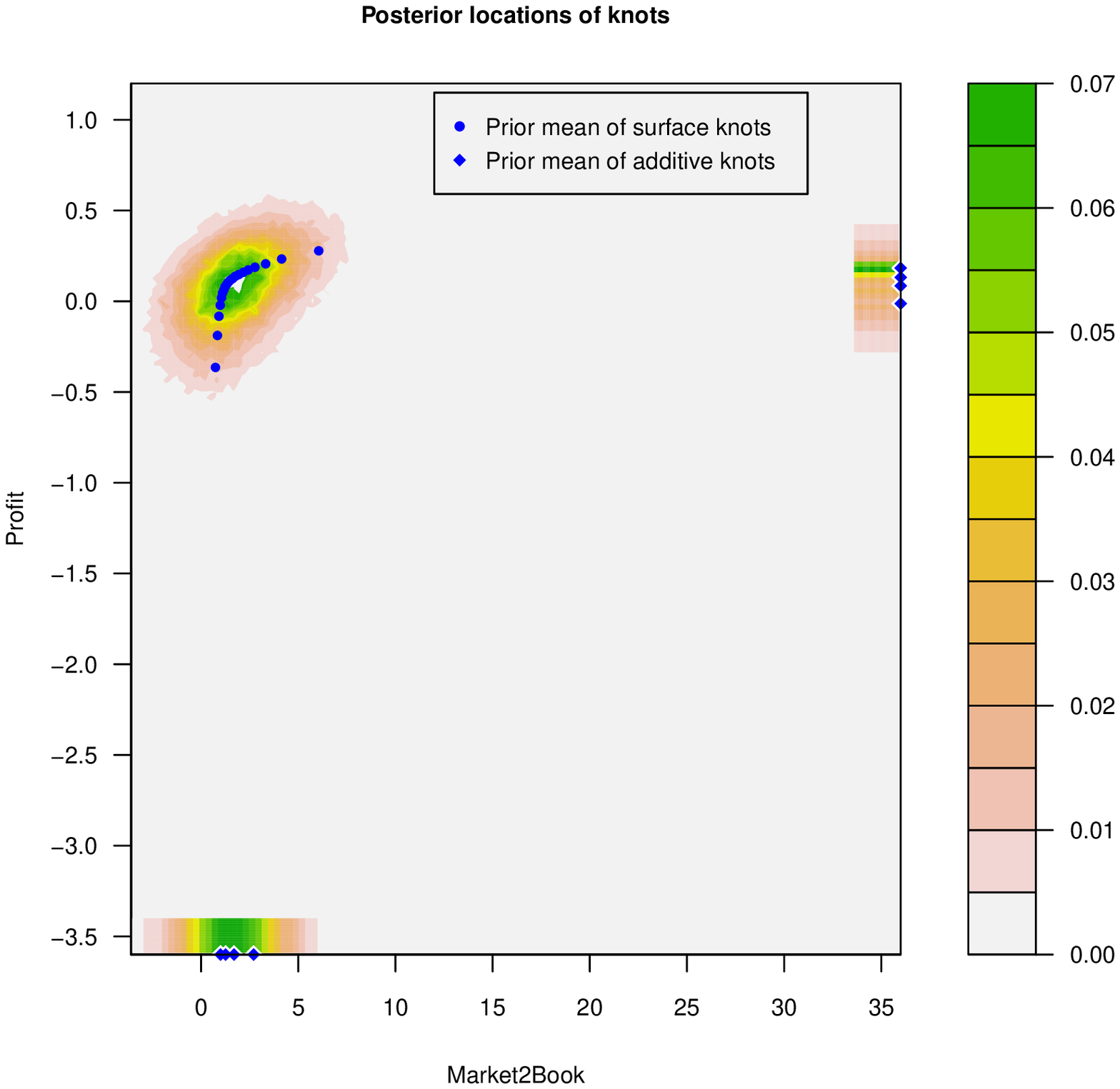}
    &\includegraphics*[trim = 30 0 0 50, height=0.25\textheight]{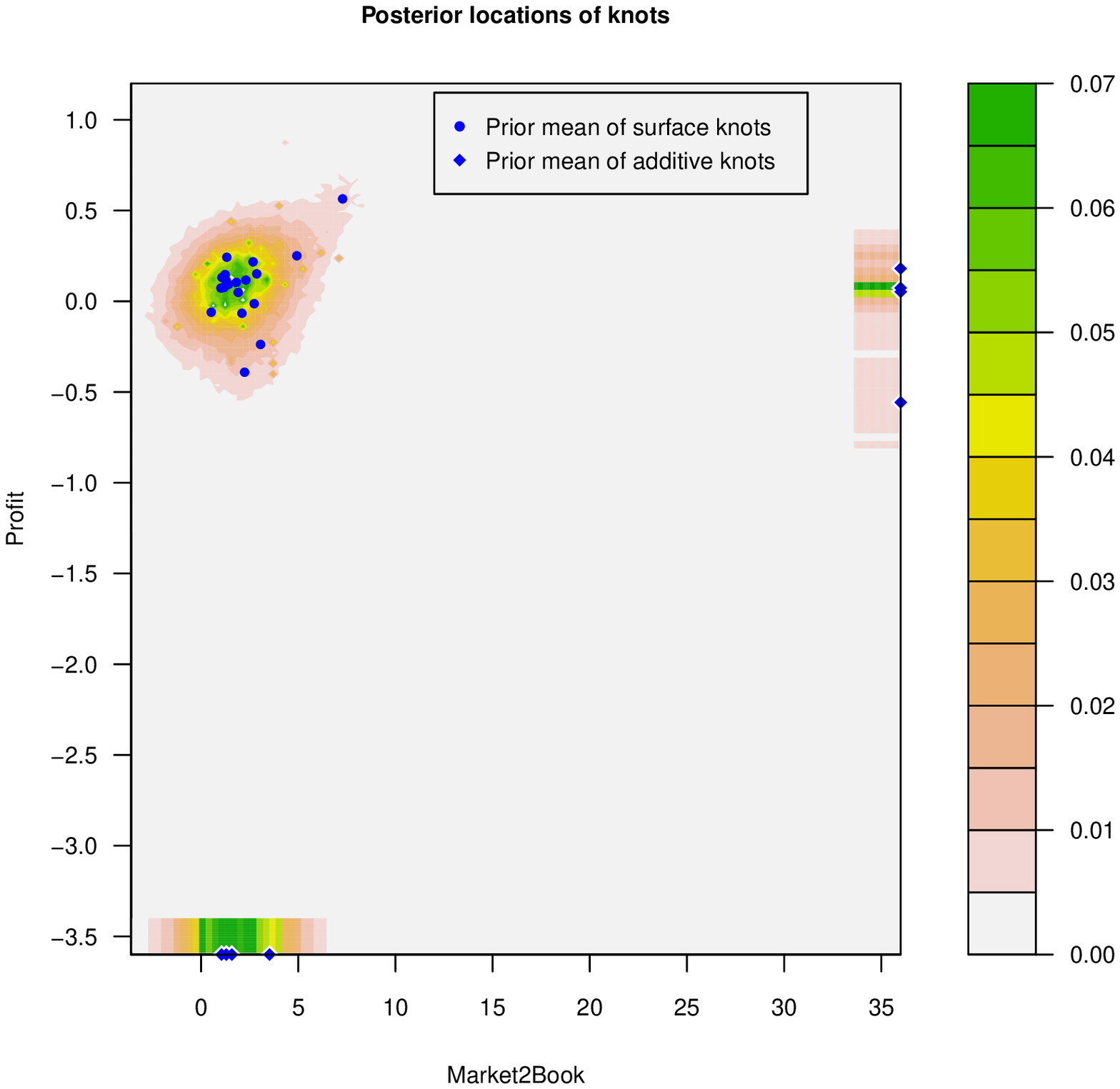}
    \\

    \begin{sideways}\hspace{1.2cm} \small{Twice the default variance}\end{sideways}
    &\includegraphics*[trim = 0 0 110 50, height=0.25\textheight]{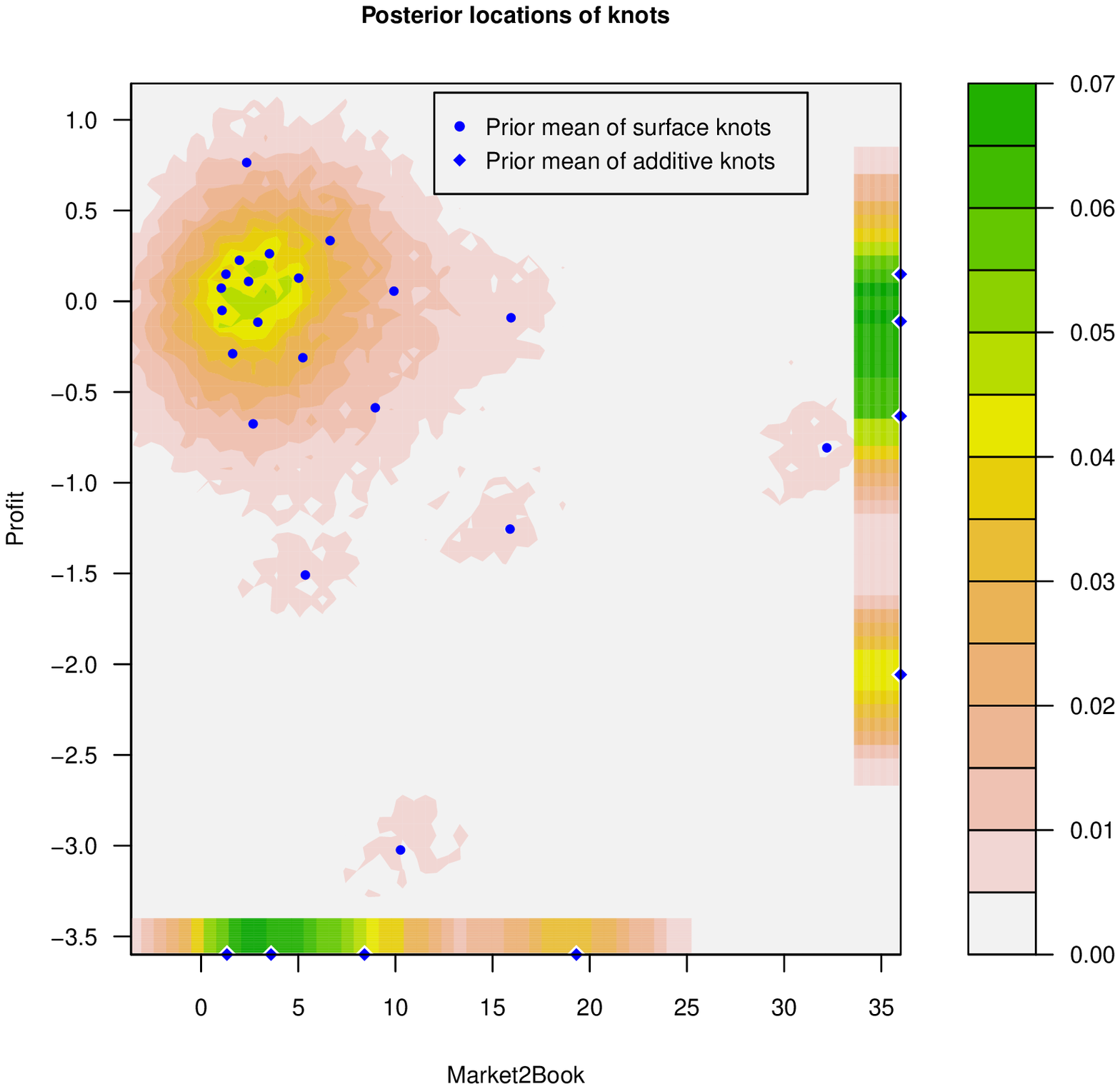}
    &\includegraphics*[trim = 30 0 110 50, height=0.25\textheight]{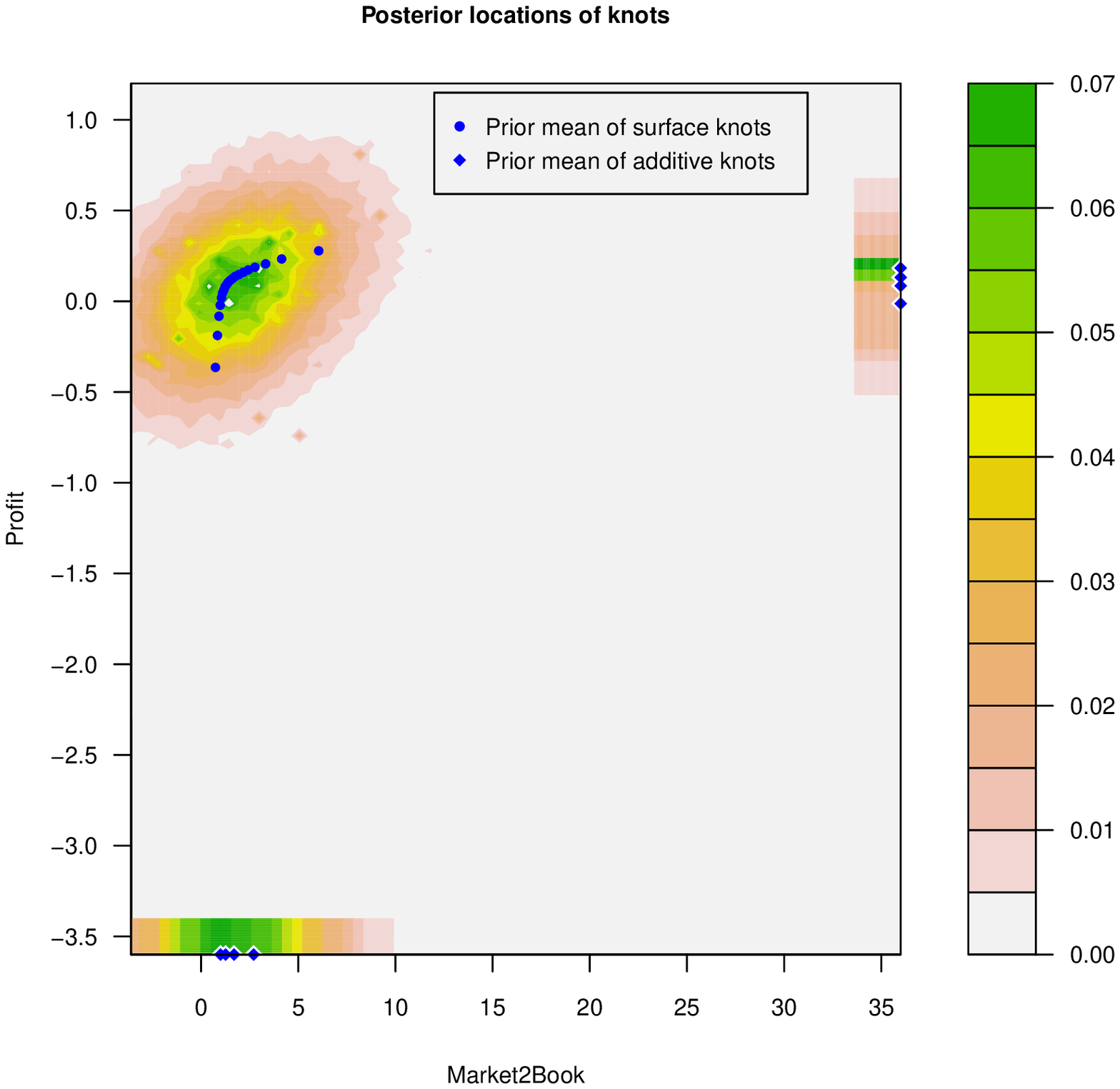}
    &\includegraphics*[trim = 30 0 0 50, height=0.25\textheight]{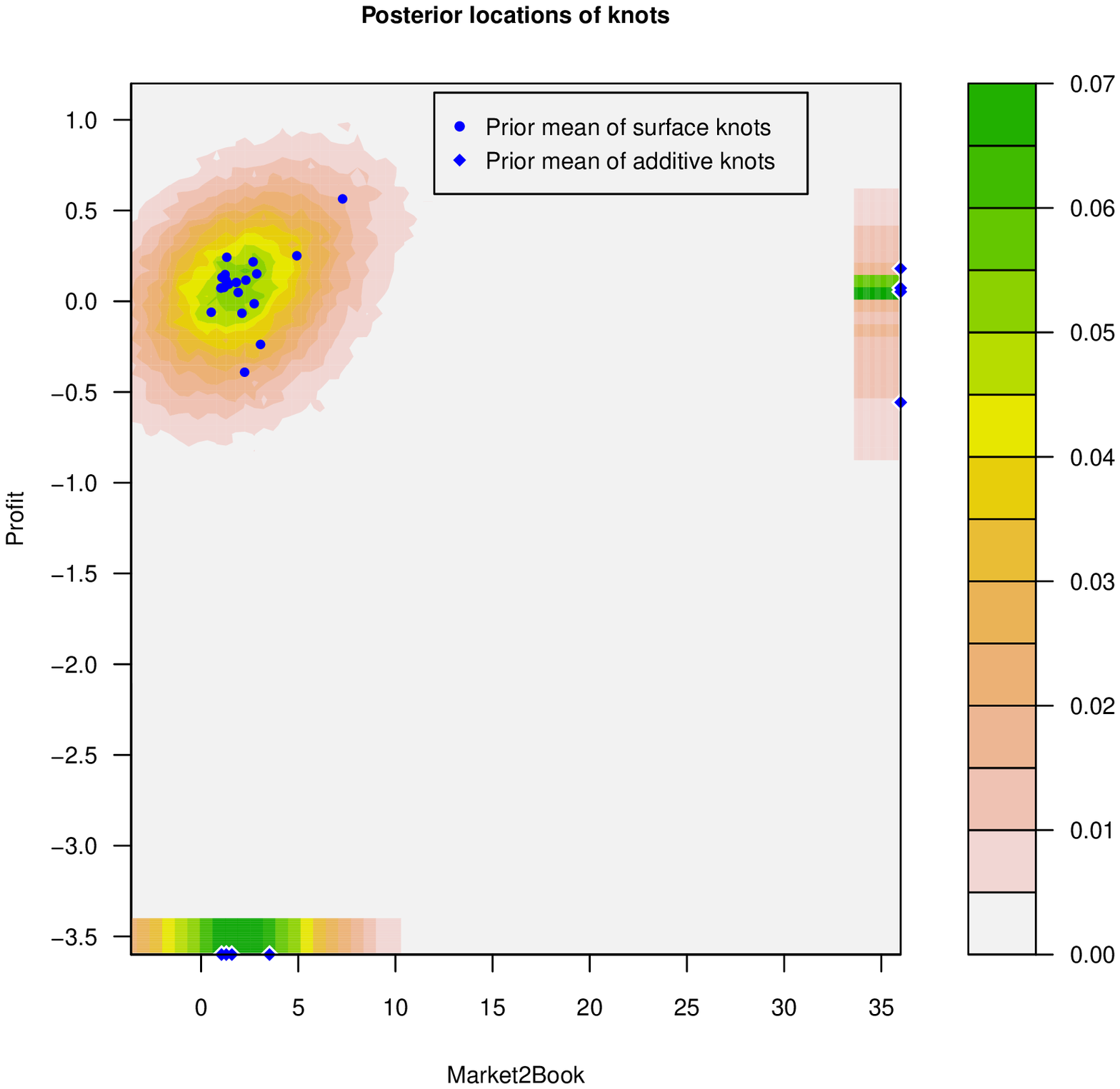}
    \\

  \end{tabular}

\caption{Sensitivity of the posterior knot density with respect to changes
in the mean (columns) and variance (rows) in the prior distribution
of the knot locations. The prior mean of the locations in the second
columns is chosen from the empirical marginal distribution of each
covariate, and the prior mean in the third column are random draws
without replacement among the data points.}

\label{fig:post-knots-comp}
\end{figure}

Figure \ref{fig:post-knots-comp} displays the effect on the posterior
knot density from changes in both the mean (columns) and the variance
(rows) of prior on the knot locations. While there are some differences
in the posterior knot densities when the prior variance changes (changes
across rows), there is much larger difference in the posterior of
the knots when the prior mean of the knot locations change. This is
partly explained by fact that the differences between the three ways
of placing the prior means are rather dramatic, but it is clear that
the prior mean of the knot locations are affecting where the knots
are located a posteriori. Considering the complexity in the inference
on the knot locations and the fact that many of the knots probably
correspond to regression coefficients that are close to zero, this
is perhaps not too surprising.

\begin{figure}
\centering

  \begin{tabular}{rccc}
    & \small{\emph{K}-means location} & \small{Marginal Percentiles} & \small{Random} \\

    \begin{sideways}\hspace{1.8cm} \small{Default variance}\end{sideways}
    &\includegraphics*[trim = 0 0 110 50, height=0.25\textheight]{rajan_mean_1-1}
    &\includegraphics*[trim = 30 0 110 50, height=0.25\textheight]{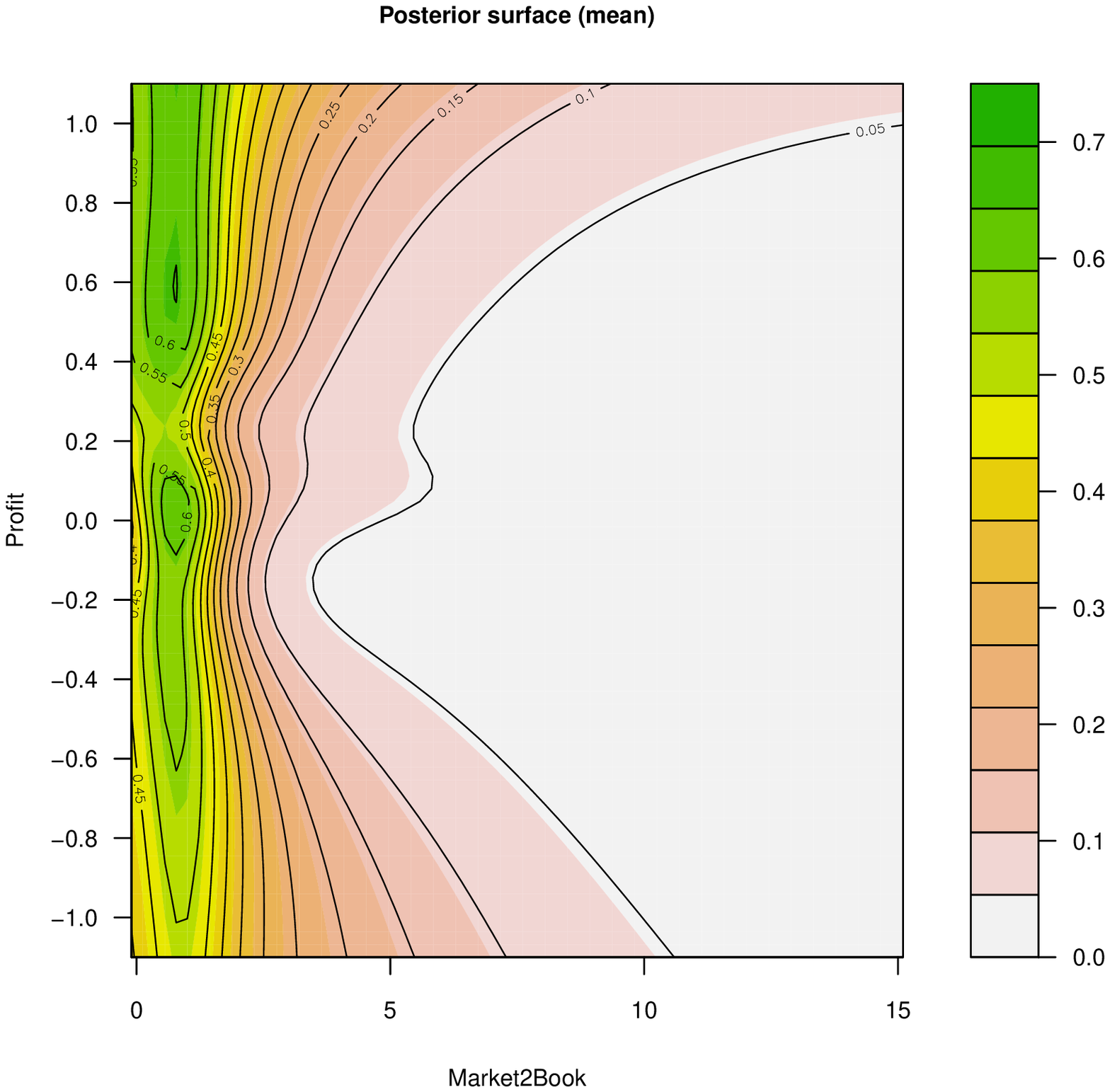}
    &\includegraphics*[trim = 30 0 0 50, height=0.25\textheight]{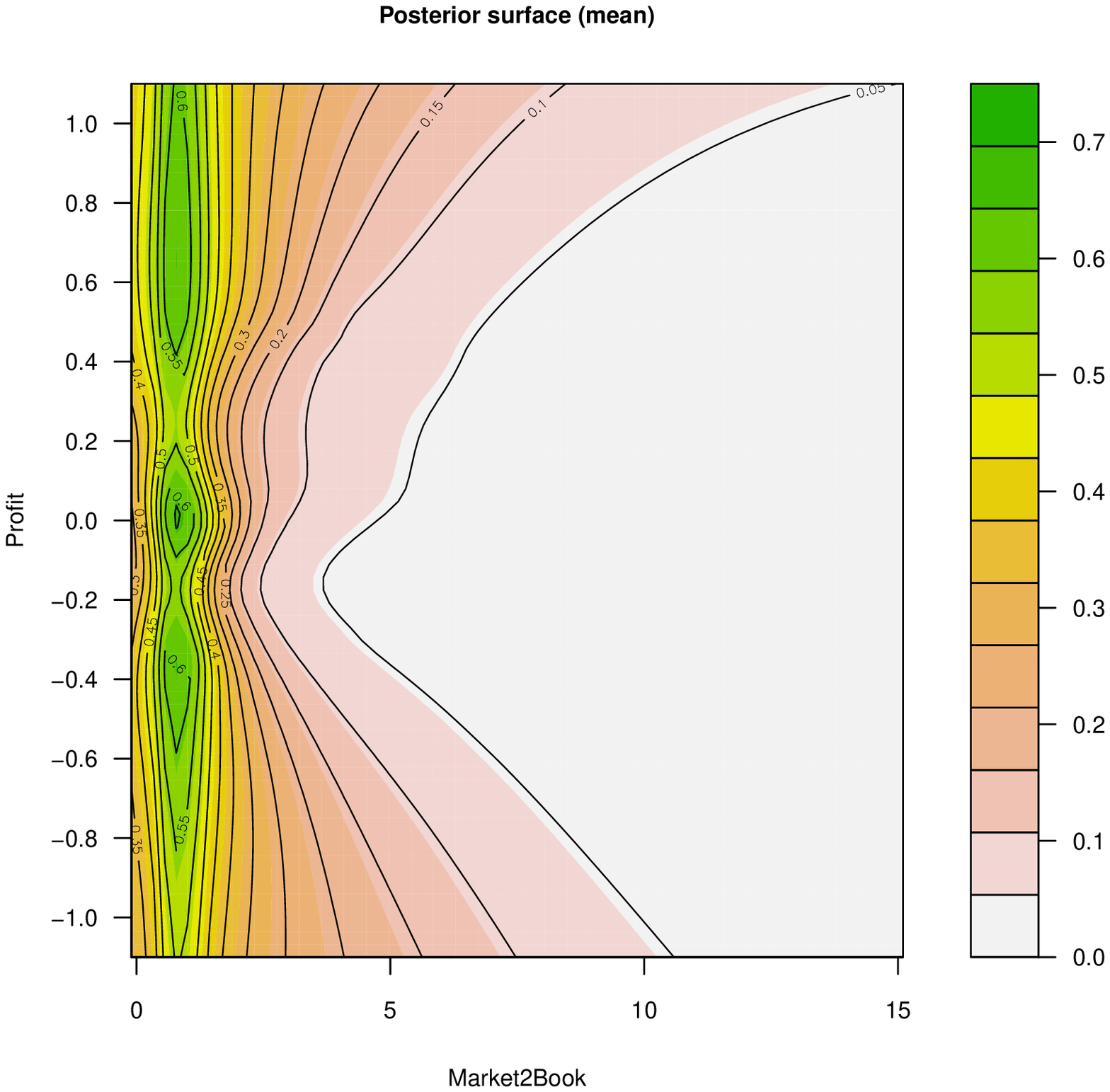}
    \\

    \begin{sideways}\hspace{1.2cm} \small{Half the default variance}\end{sideways}
    &\includegraphics*[trim = 0 0 110 50, height=0.25\textheight]{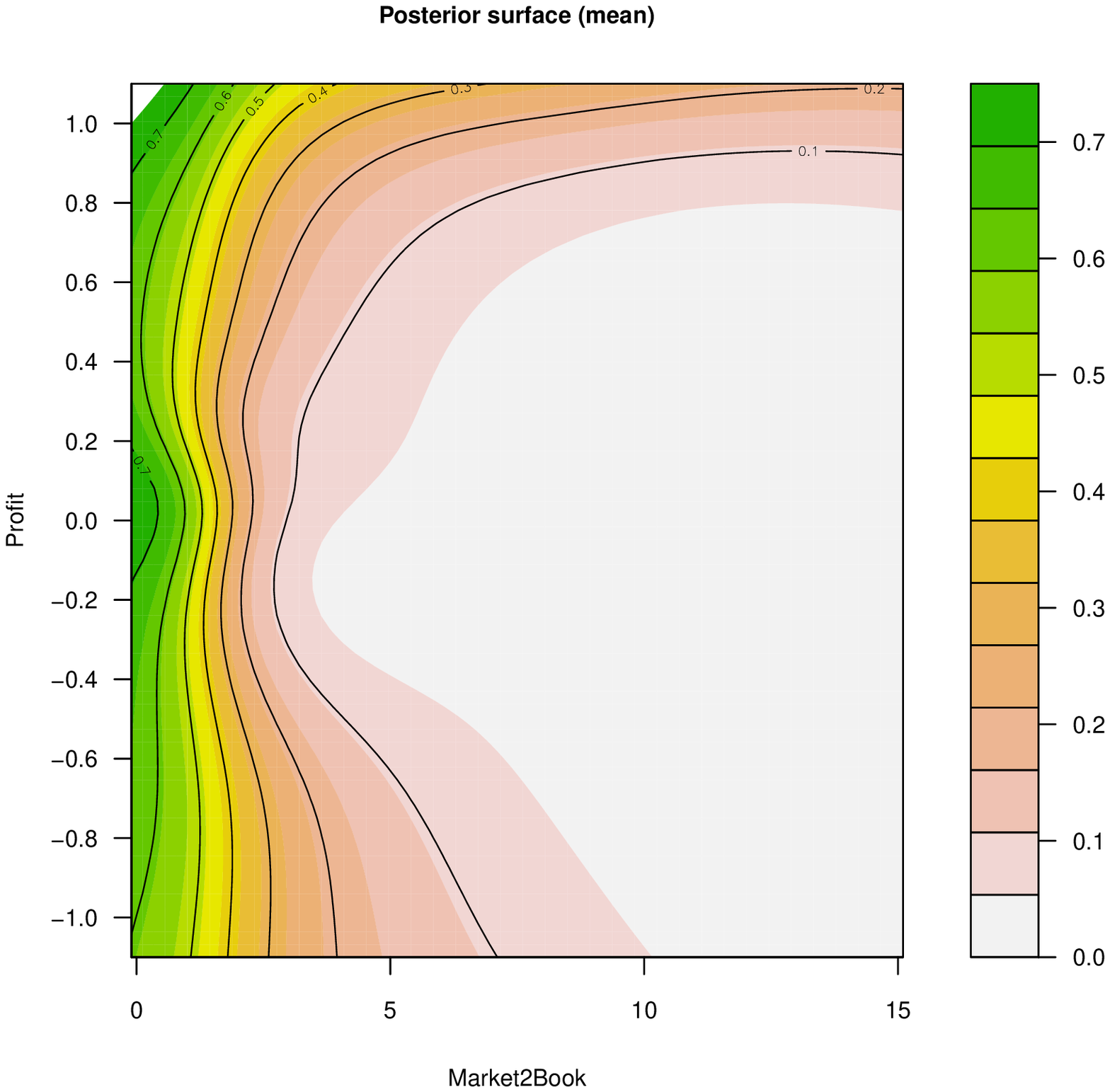}
    &\includegraphics*[trim = 30 0 110 50, height=0.25\textheight]{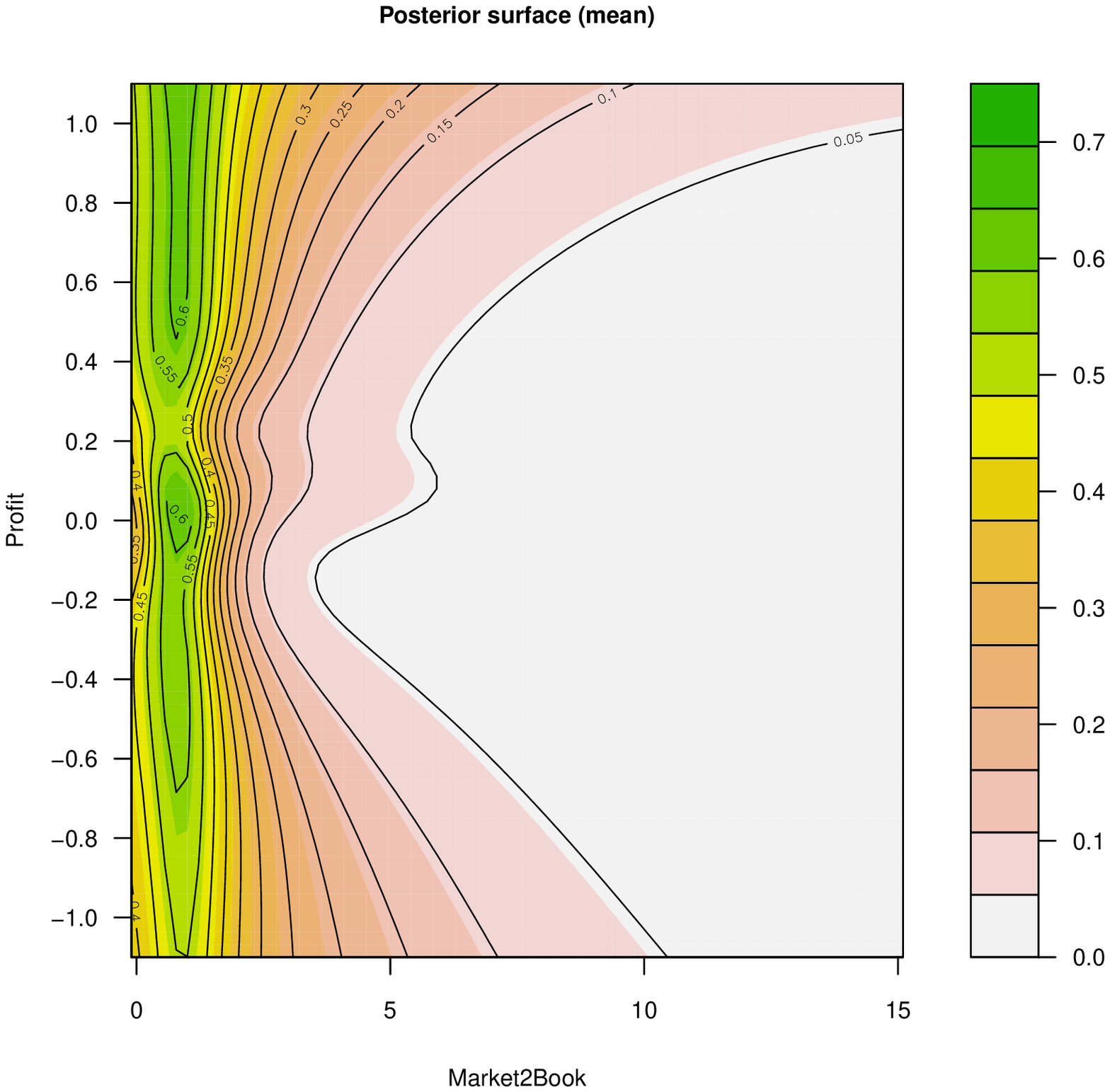}
    &\includegraphics*[trim = 30 0 0 50, height=0.25\textheight]{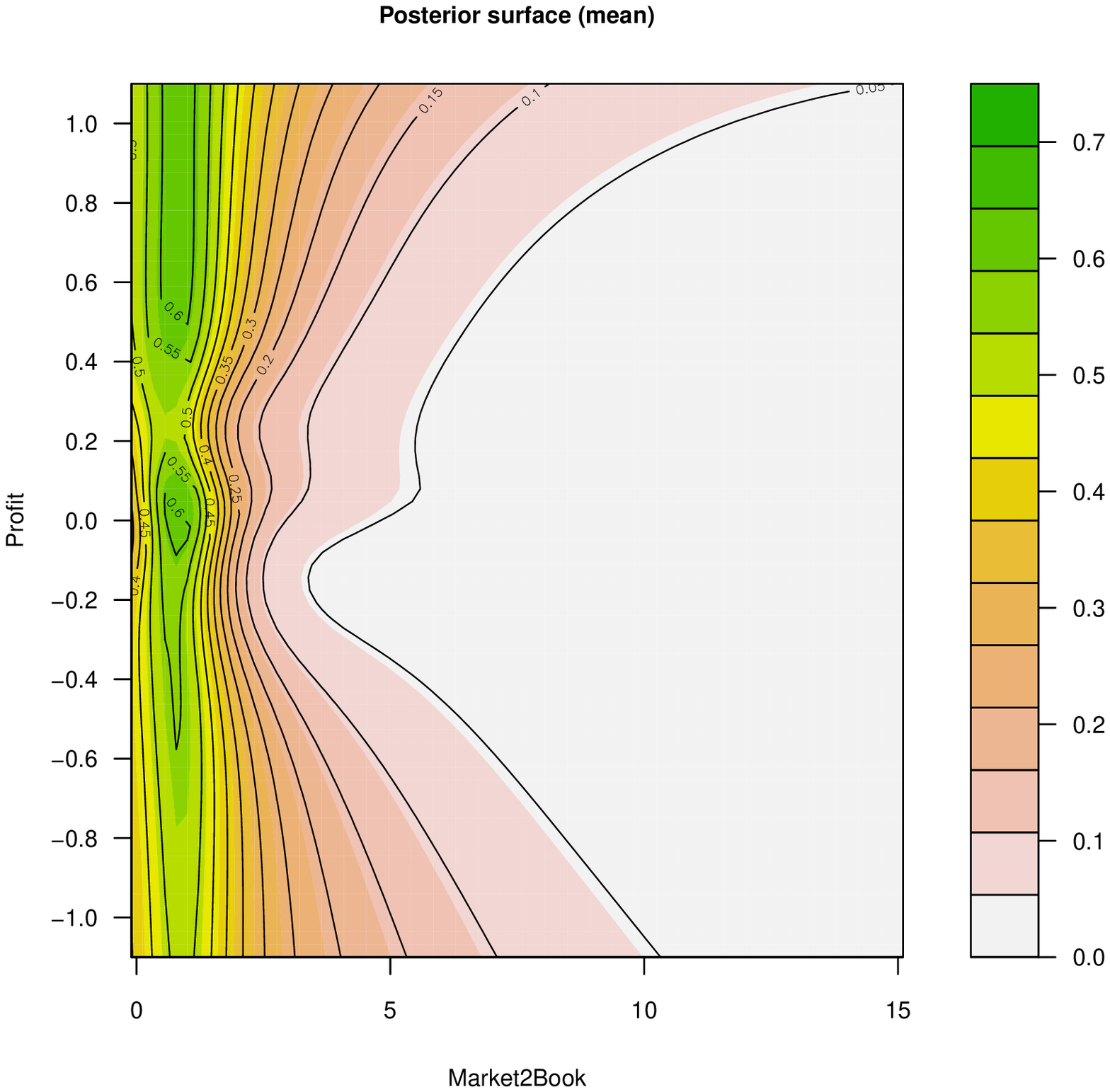}
    \\

    \begin{sideways}\hspace{1.2cm} \small{Twice the default variance}\end{sideways}
    &\includegraphics*[trim = 0 0 110 50, height=0.25\textheight]{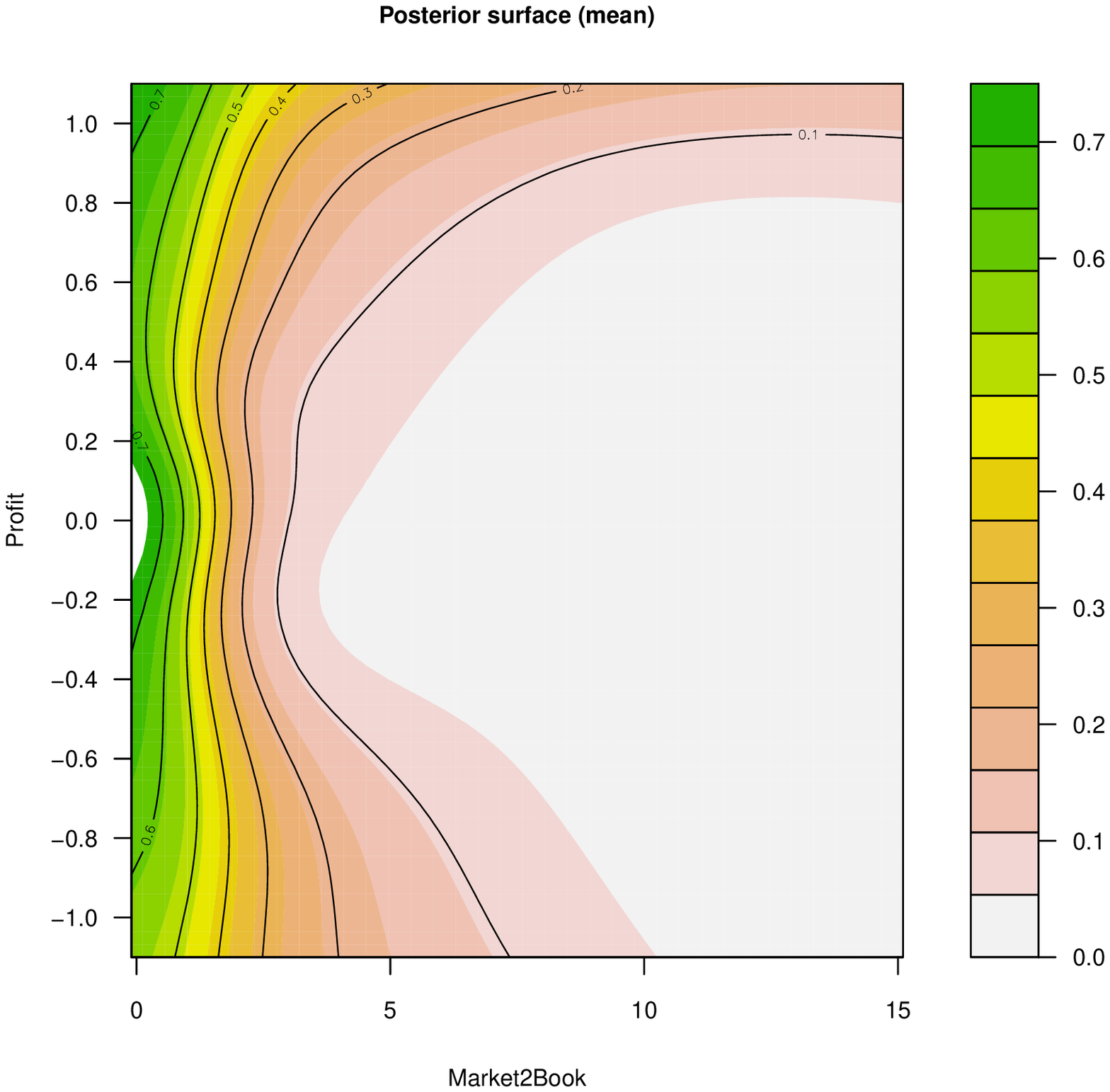}
    &\includegraphics*[trim = 30 0 110 50, height=0.25\textheight]{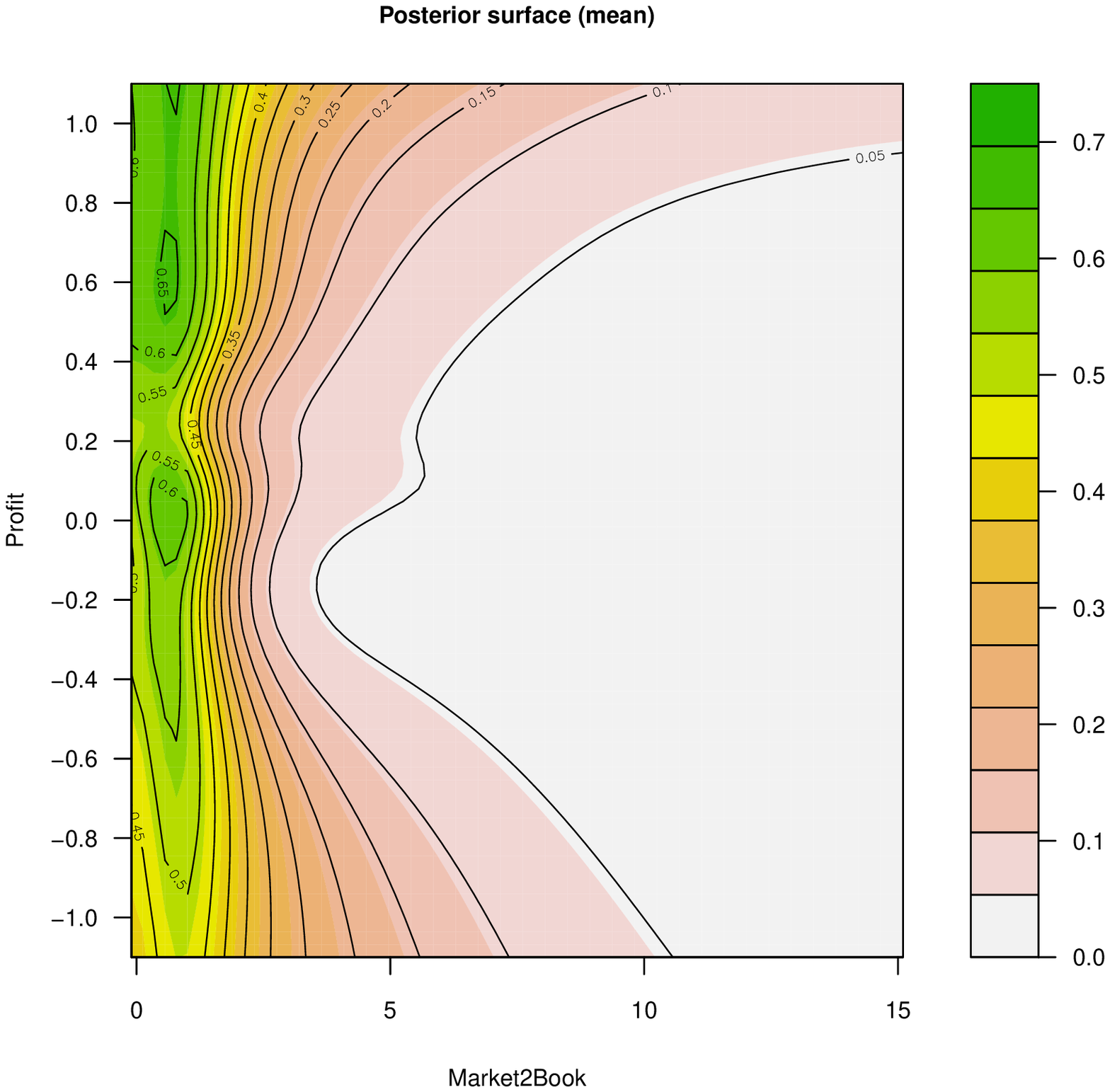}
    &\includegraphics*[trim = 30 0 0 50, height=0.25\textheight]{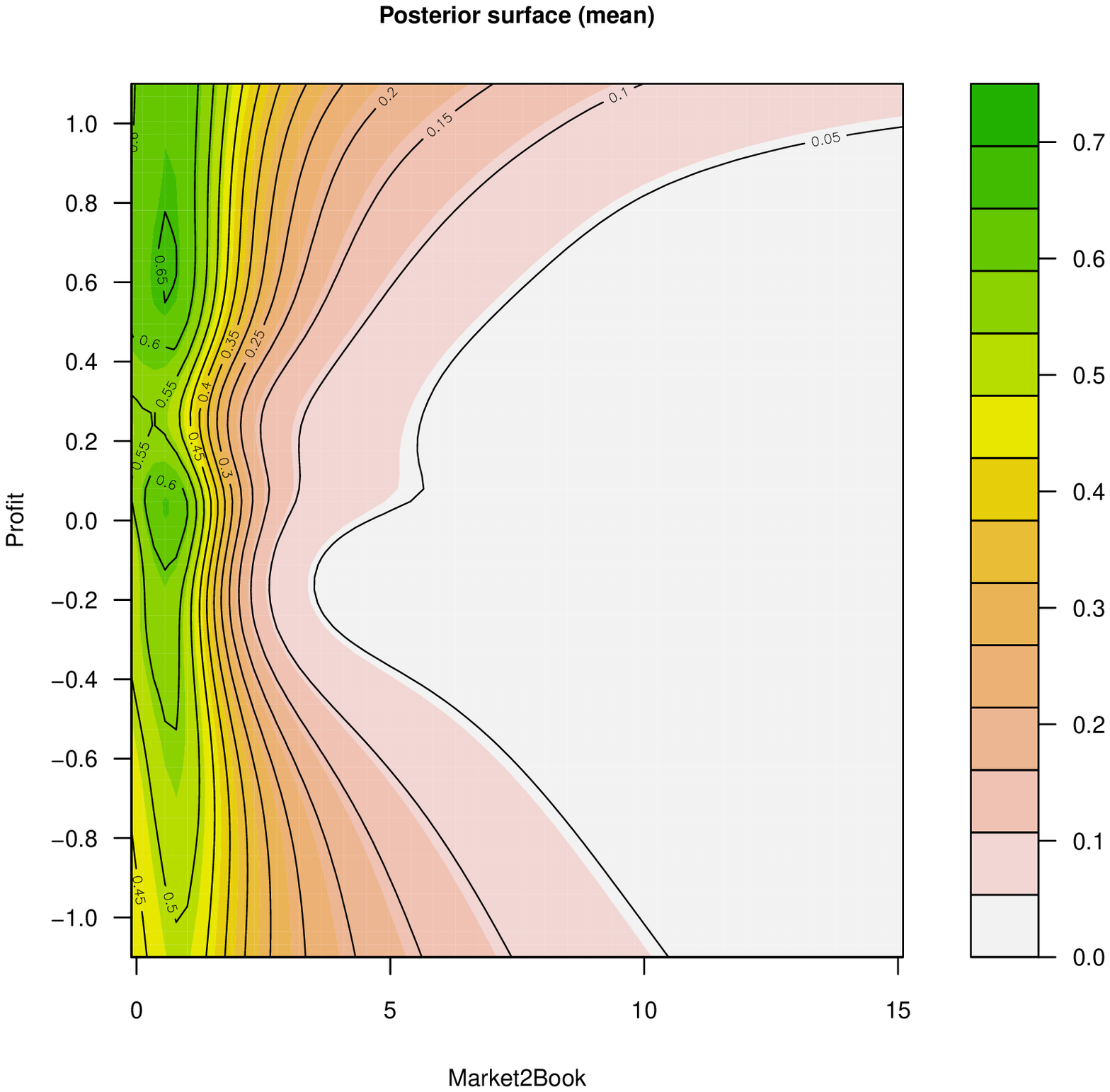}
    \\

  \end{tabular}

\caption{Sensitivity of the posterior mean surface with respect to changes
in the mean (columns) and variance (rows) in the prior distribution
of the knot locations. The prior mean of the locations in the second
columns is chosen from the empirical marginal distribution of each
covariate, and the prior mean in the third column are random draws
without replacement among the data points.}

\label{fig:post-mean-comp}
\end{figure}

\begin{figure}
\centering

  \begin{tabular}{rccc}
    & \small{\emph{K}-means location} & \small{Marginal Percentiles} & \small{Random} \\

    \begin{sideways}\hspace{1.8cm} \small{Default variance}\end{sideways}
    &\includegraphics*[trim = 0 0 110 50, height=0.25\textheight]{rajan_sd_1-1}
    &\includegraphics*[trim = 30 0 110 50, height=0.25\textheight]{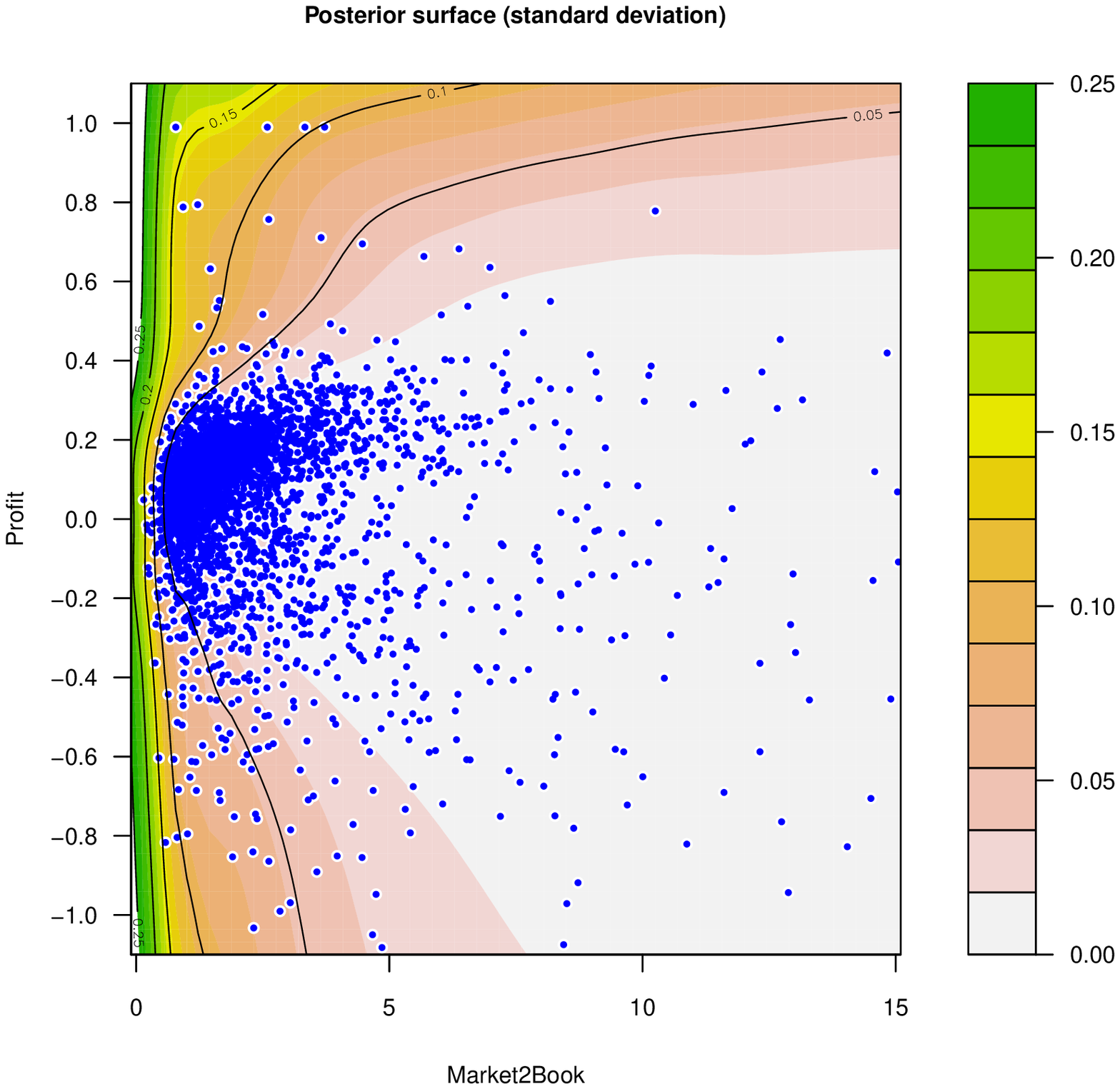}
    &\includegraphics*[trim = 30 0 0 50, height=0.25\textheight]{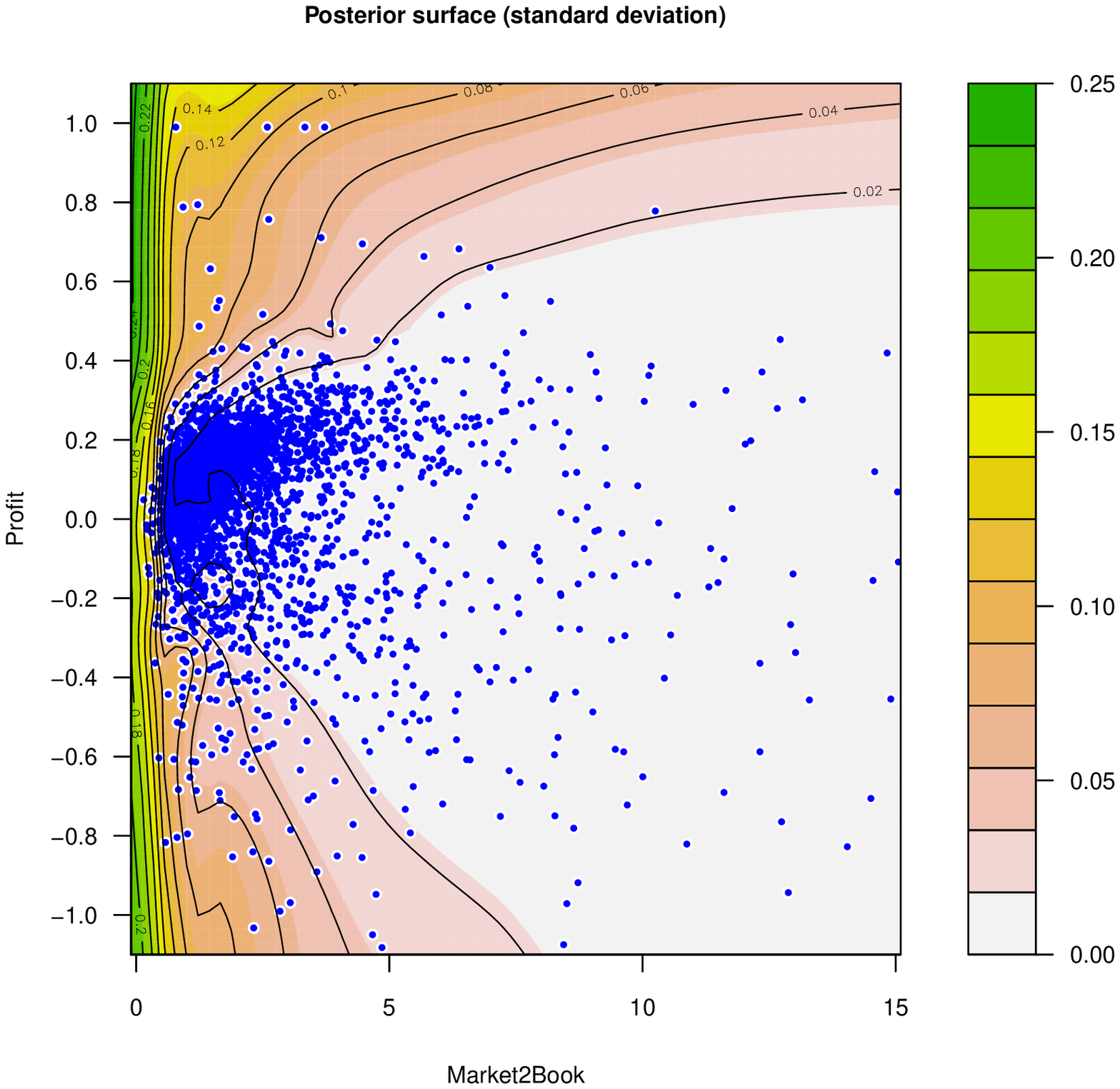}
    \\

    \begin{sideways}\hspace{1.2cm} \small{Half the default variance}\end{sideways}
    &\includegraphics*[trim = 0 0 110 50, height=0.25\textheight]{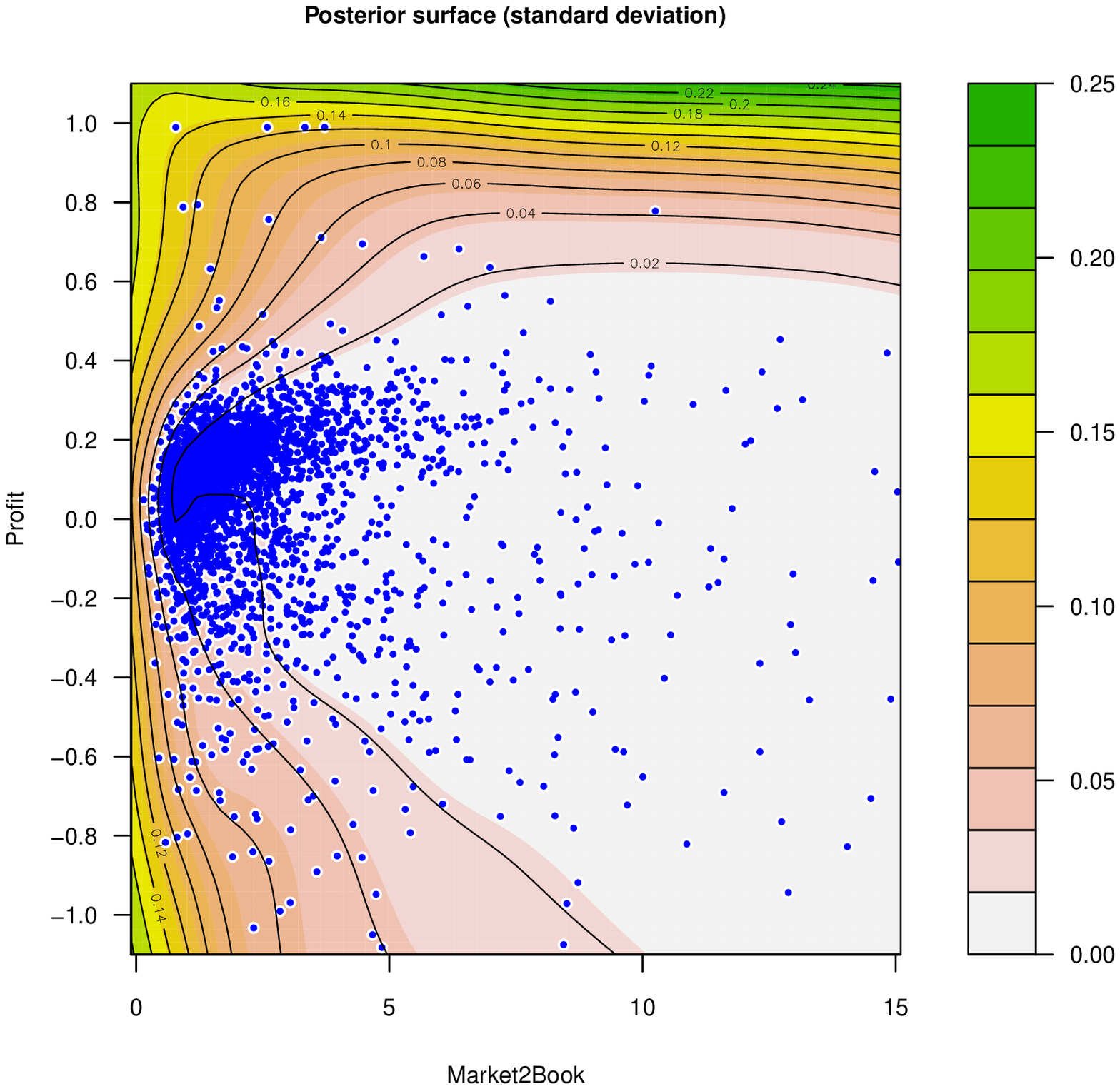}
    &\includegraphics*[trim = 30 0 110 50, height=0.25\textheight]{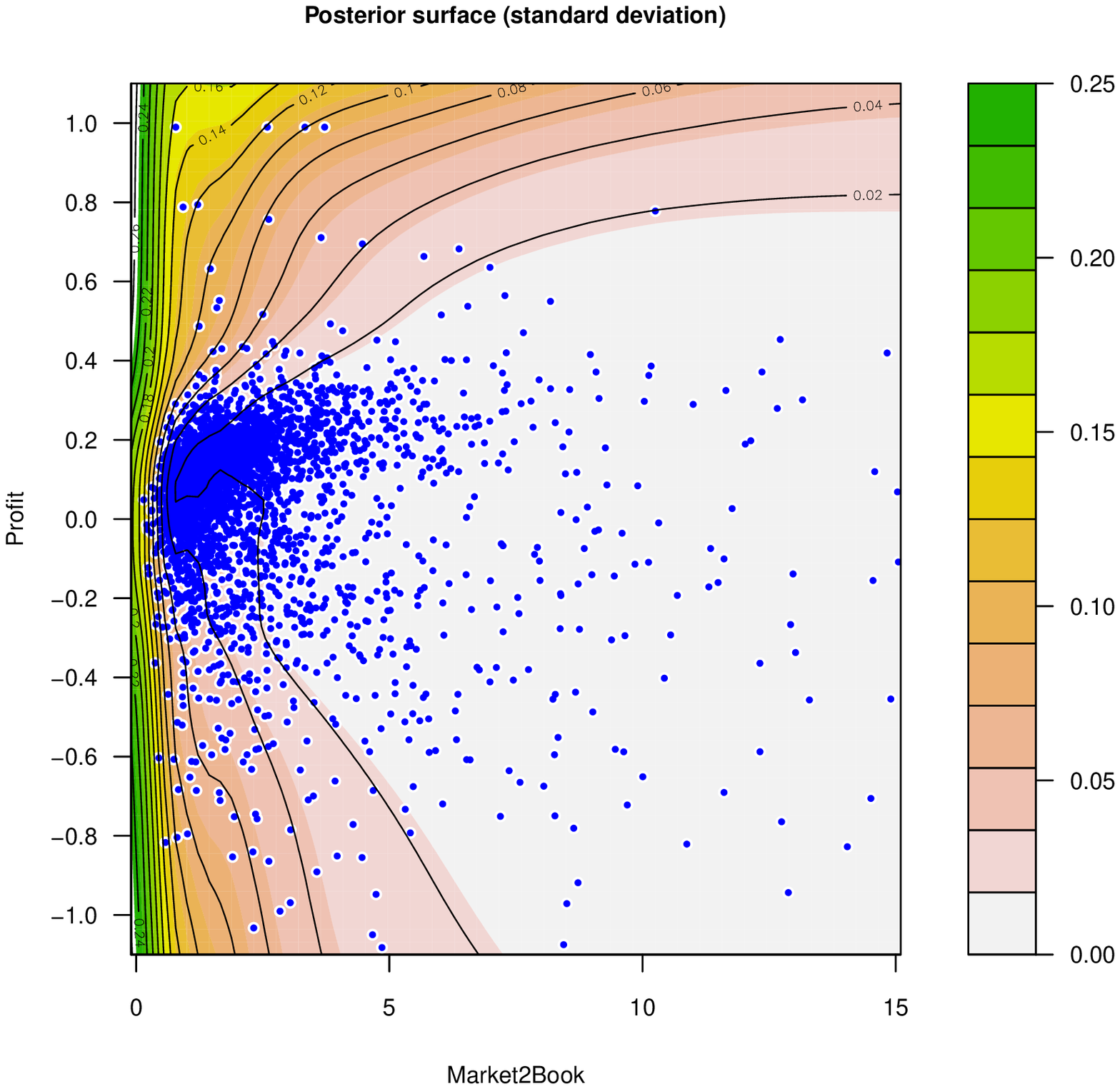}
    &\includegraphics*[trim = 30 0 0 50, height=0.25\textheight]{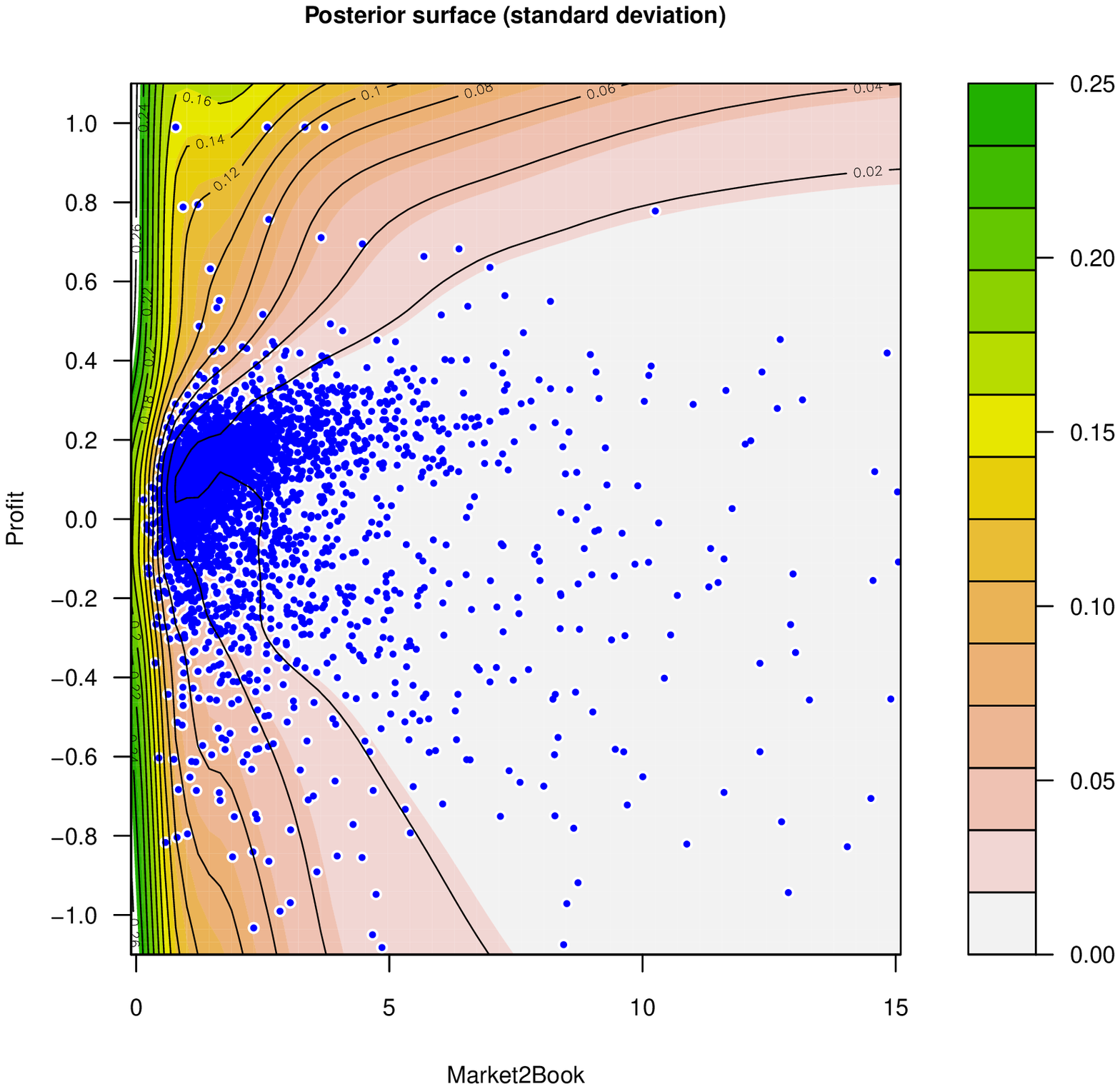}
    \\

    \begin{sideways}\hspace{1.2cm} \small{Twice the default variance}\end{sideways}
    &\includegraphics*[trim = 0 0 110 50, height=0.25\textheight]{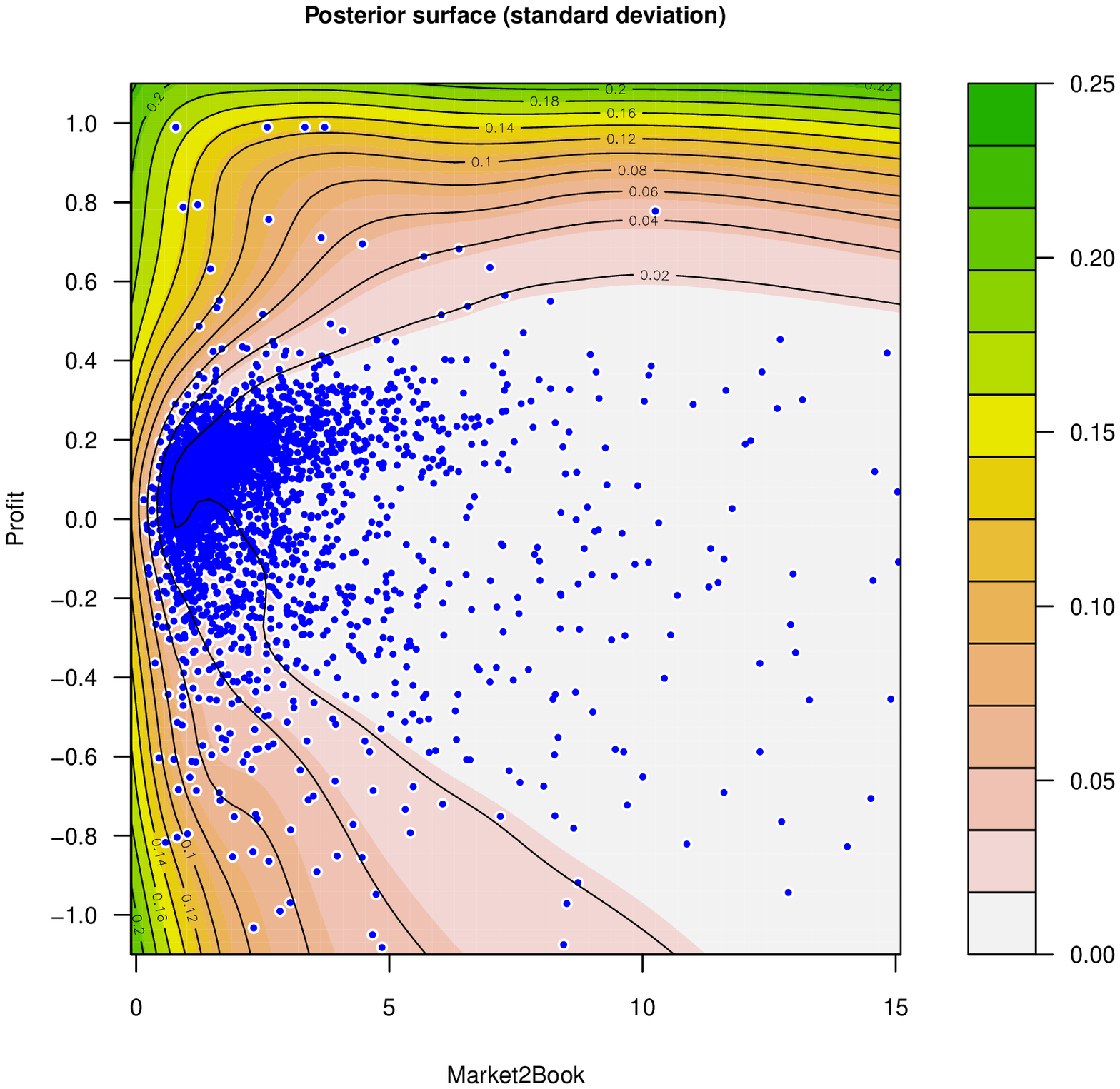}
    &\includegraphics*[trim = 30 0 110 50, height=0.25\textheight]{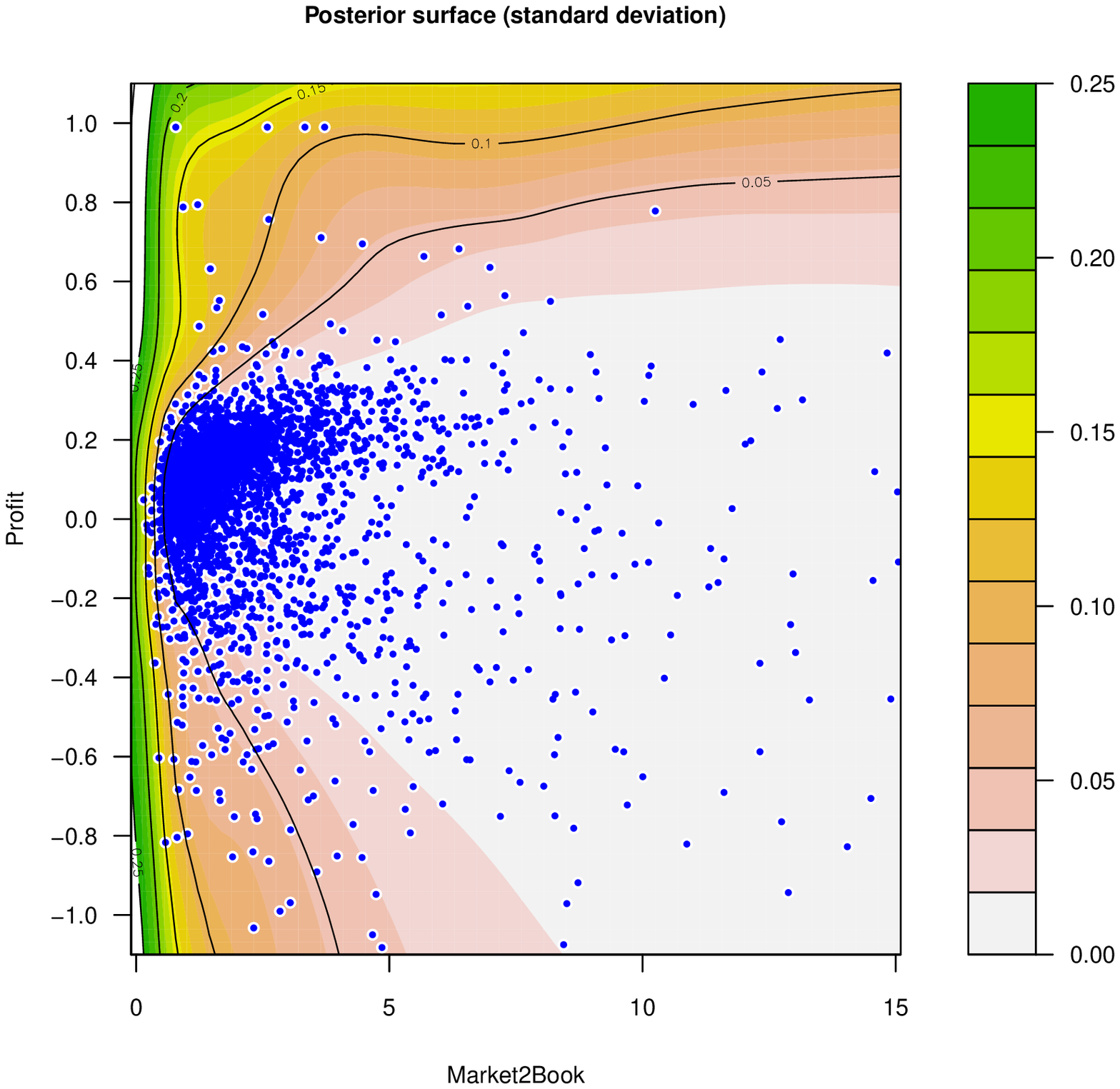}
    &\includegraphics*[trim = 30 0 0 50, height=0.25\textheight]{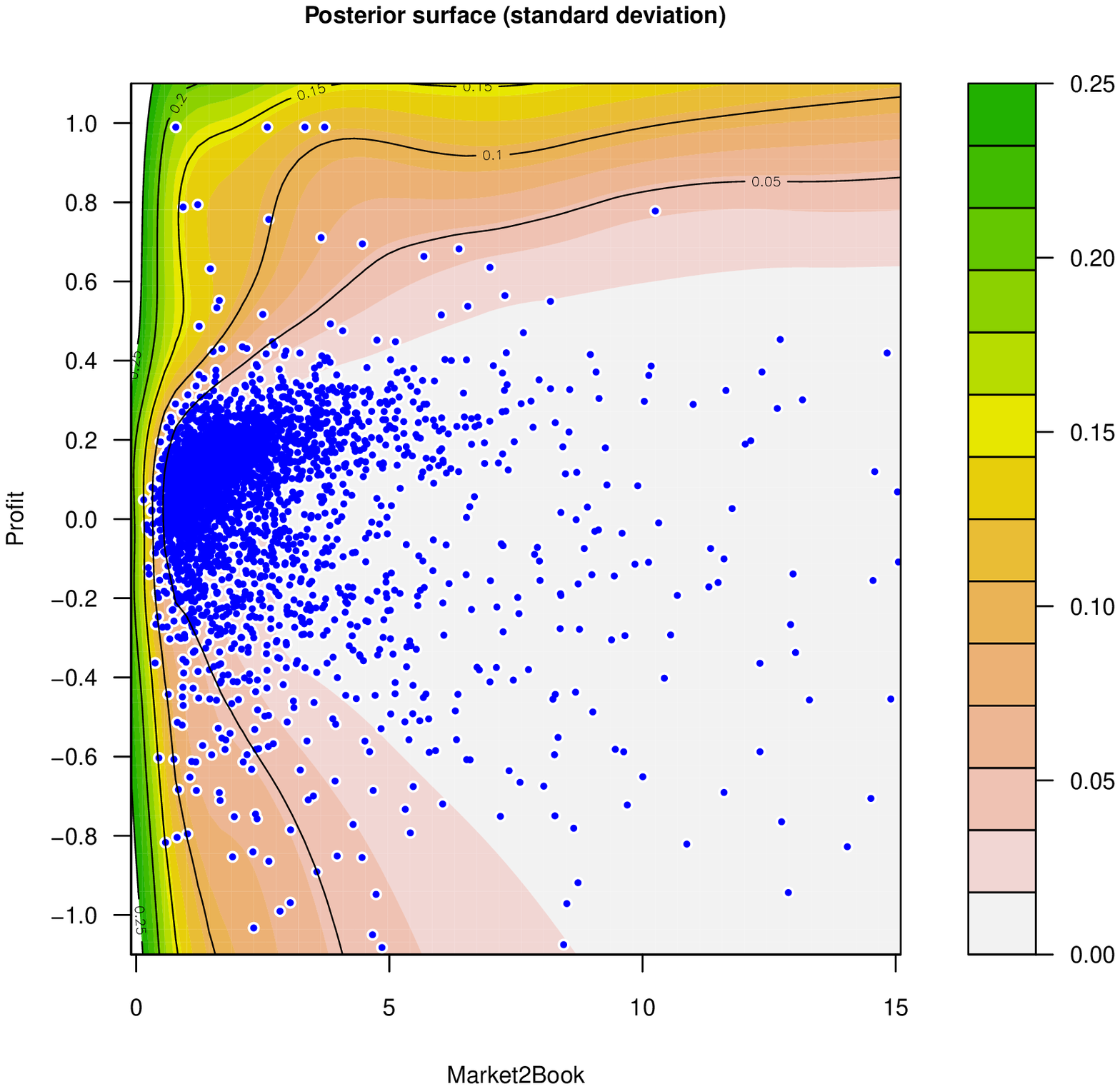}
    \\

  \end{tabular}

\caption{Sensitivity of the posterior standard deviation surface with respect
to changes in the mean (columns) and variance (rows) in the prior
distribution of the knot locations. The prior mean of the locations
in the second columns is chosen from the empirical marginal distribution
of each covariate, and the prior mean in the third column are random
draws without replacement among the data points.}

\label{fig:post-sd-comp}
\end{figure}

The posterior inference of the knot locations is typically not of
interest. What matters is the inferences on the conditional predictive
distribution $p(\mathbf{y}|\mathbf{x})$. Figure \ref{fig:post-mean-comp}
and \ref{fig:post-sd-comp} investigate the sensitivity of the posterior
mean and standard deviation surfaces to changes in the prior mean
and variance of the knot locations. Here the robustness to variations
in the prior is much larger. Both the predictive mean and the predictive
standard deviation remain largely unchanged, considering the magnitude
of the changes in the prior. The main differences in the prior mean
surface occur in points of covariate space where the uncertainty in
the predictive mean is large.

\section{Further simulation results\label{sec: Conclusions}}

\subsection{Additional results from the simulation study in Section 4 of the
paper}

\begin{figure}
\centering \includegraphics[width=0.8\textwidth]{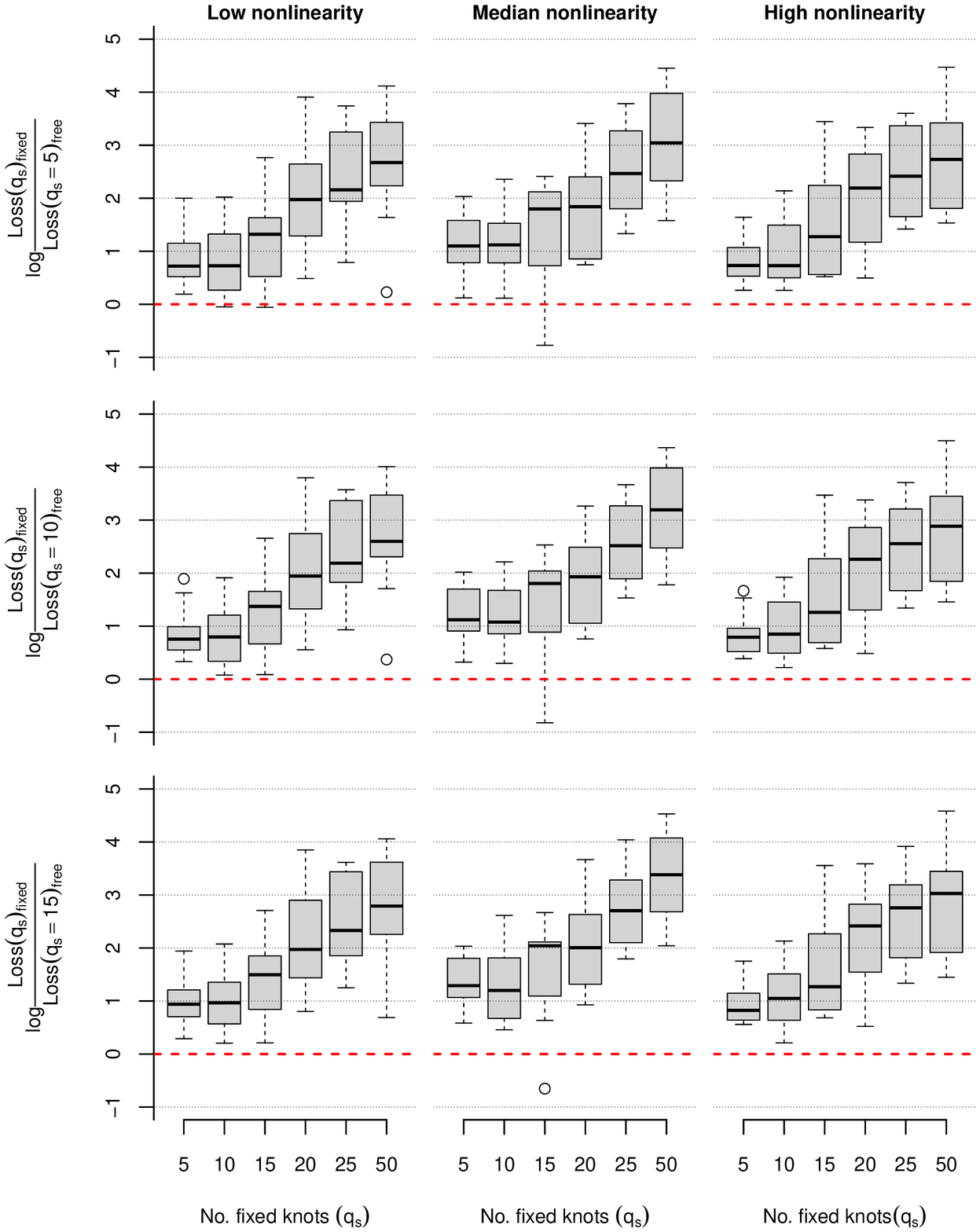}
\caption{Boxplot of the log loss ratio comparing the performance of the fixed
knots model with the free knots model for the DGP with $p=2$ and
$n=1000$. The three columns of the figure correspond to different
degrees of nonlinearity of the realized datasets, as measured by estimated
$\mathrm{DNL}$ in (5) in the paper. }

\label{fig:LOSS_1000n}
\end{figure}

\begin{figure}
\centering \includegraphics[width=0.7\textwidth]{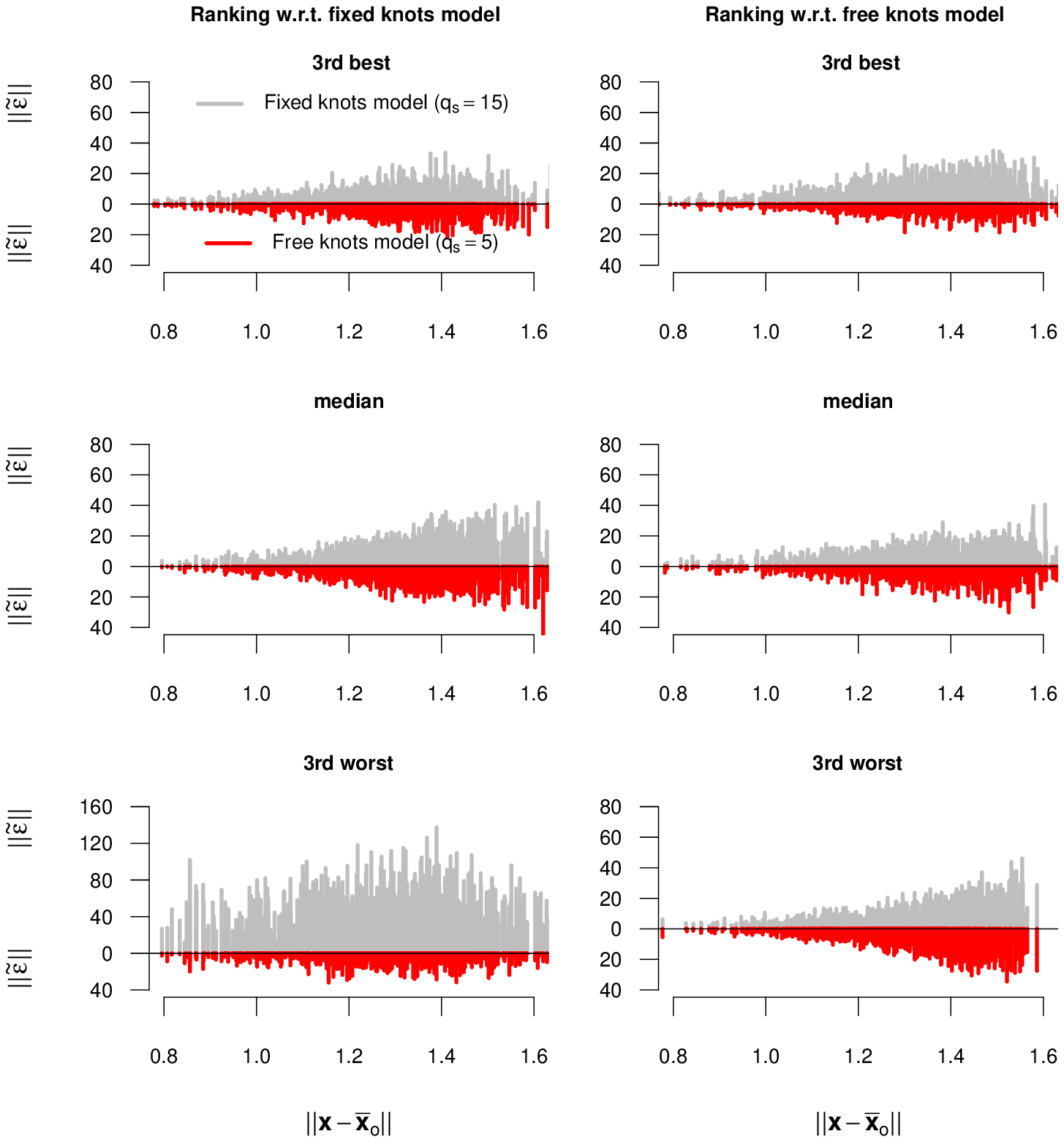}
\caption{Plotting the norm of the predictive multivariate residuals as a function
of the distance between the covariate vector and its sample mean.
The results are for the DGP with $p=2$ and $n=1000$. The lines in
each subplot are the normed residuals from the model with $15$ fixed
surface knots (vertical bars above the zero line), and the model with
$5$ free knots (vertical bars below the zero line). The column to
the left shows the results for three datasets chosen when performance
is ranked according to the fixed knots model, and the right column
displays the results for three datasets chosen when performance is
ranked according to the free knots model.}

\label{fig:Resid_1000_5_15}
\end{figure}

\begin{figure}
\centering \includegraphics[width=0.7\textwidth]{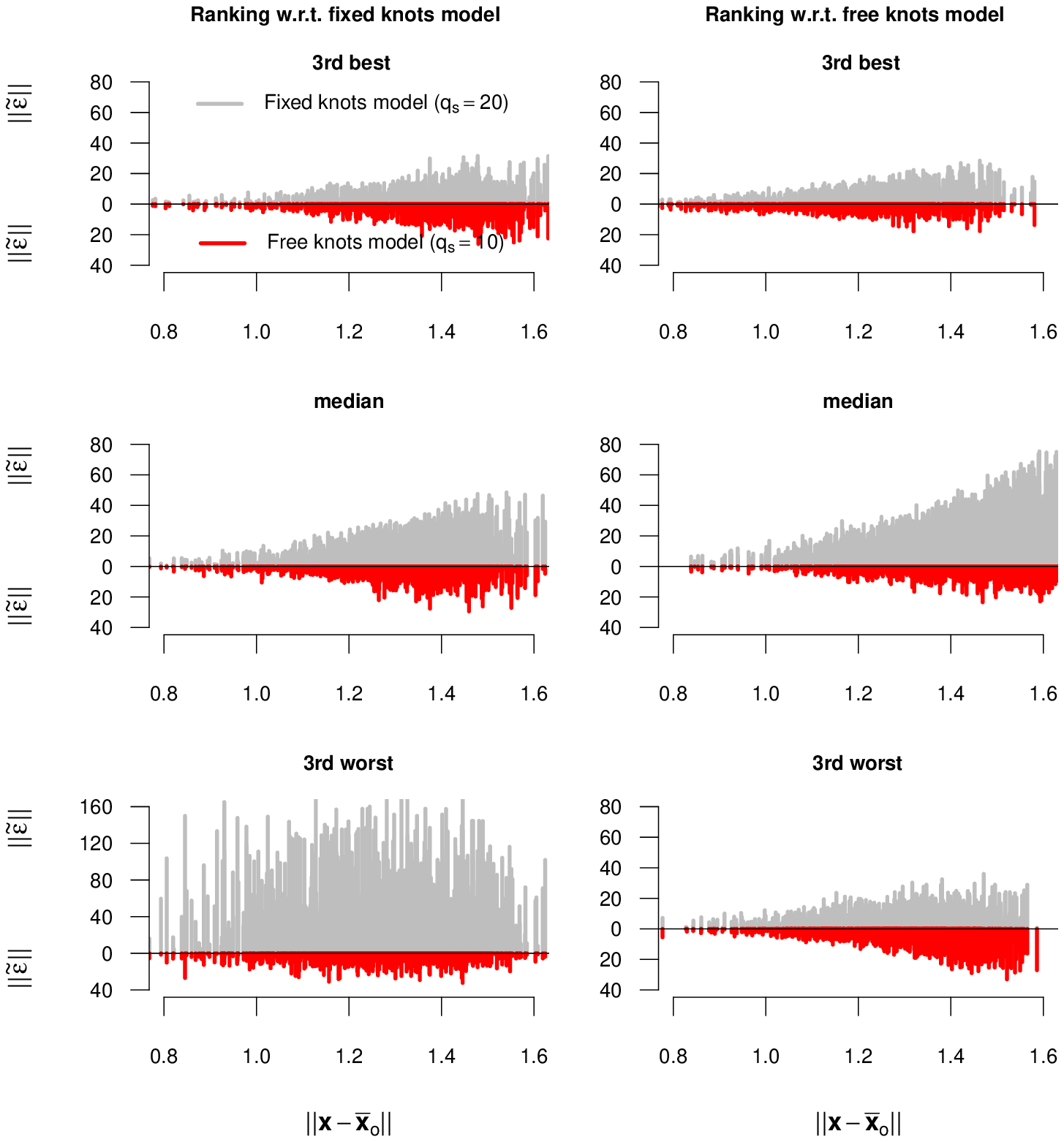}
\caption{Plotting the norm of the predictive multivariate residuals as a function
of the distance between the covariate vector and its sample mean.
The results are for the DGP with $p=2$ and $n=1000$. The lines in
each subplot are the normed residuals from the model with $20$ fixed
surface knots (vertical bars above the zero line), and the model with
$10$ free knots (vertical bars below the zero line). The column to
the left shows the results for three datasets chosen when performance
is ranked according to the fixed knots model, and the right column
displays the results for three datasets chosen when performance is
ranked according to the free knots model.}

\label{fig:Resid_1000_10_20}
\end{figure}

This section documents the simulation results for the simulation setup
with $p=2$ and $n=1000$. The simulation setup is identical to the
one in Section 4 of the paper with the exception that number of data
points is increased from $n=200$ to $n=1000$. Figure \ref{fig:LOSS_1000n}
compares the estimation loss from using fixed and free knots, respectively.
Figure \ref{fig:Resid_1000_5_15} compares the out-of-sample predictive
residuals from a models with $15$ fixed surface knots to a model
with $5$ free knots, and Figure \ref{fig:Resid_1000_10_20} does
the same type of comparison for a model with $20$ fixed surface knots
to a model with $10$ free knots.

\subsection{Simulation results from a situation where none of the two models
are correct}

In this section, we briefly describe a simple simulation example where
the true model is not nested in any of the two estimated models. We
generate Gaussian data around the following mean surface
\begin{equation}
\begin{split}y=42.659[ & 0.1+(x_{1}-0.5)(0.05+(x_{1}-0.5)^{4}\\
 & -10(x_{1}-0.5)^{2}(x_{2}-0.5)^{2}+5(x_{2}-0.5)^{2})]
\end{split}
\label{eq:radial}
\end{equation}
The function in Equation (\ref{eq:radial}) is called a radial function.
We generate $100$ datasets using $N(0,0.1)$ disturbances around
the mean. The number of observations are $1000$ in each dataset.
We use linear and surface components. The number of knots used in
the free knot models is $10$, $20$, and $40$, and the number of
knots used in the fixed knots model is $10$, $20$, $40$, $60$,
$80$ and $100$.

\begin{figure}
\centering \includegraphics[width=0.45\textwidth]{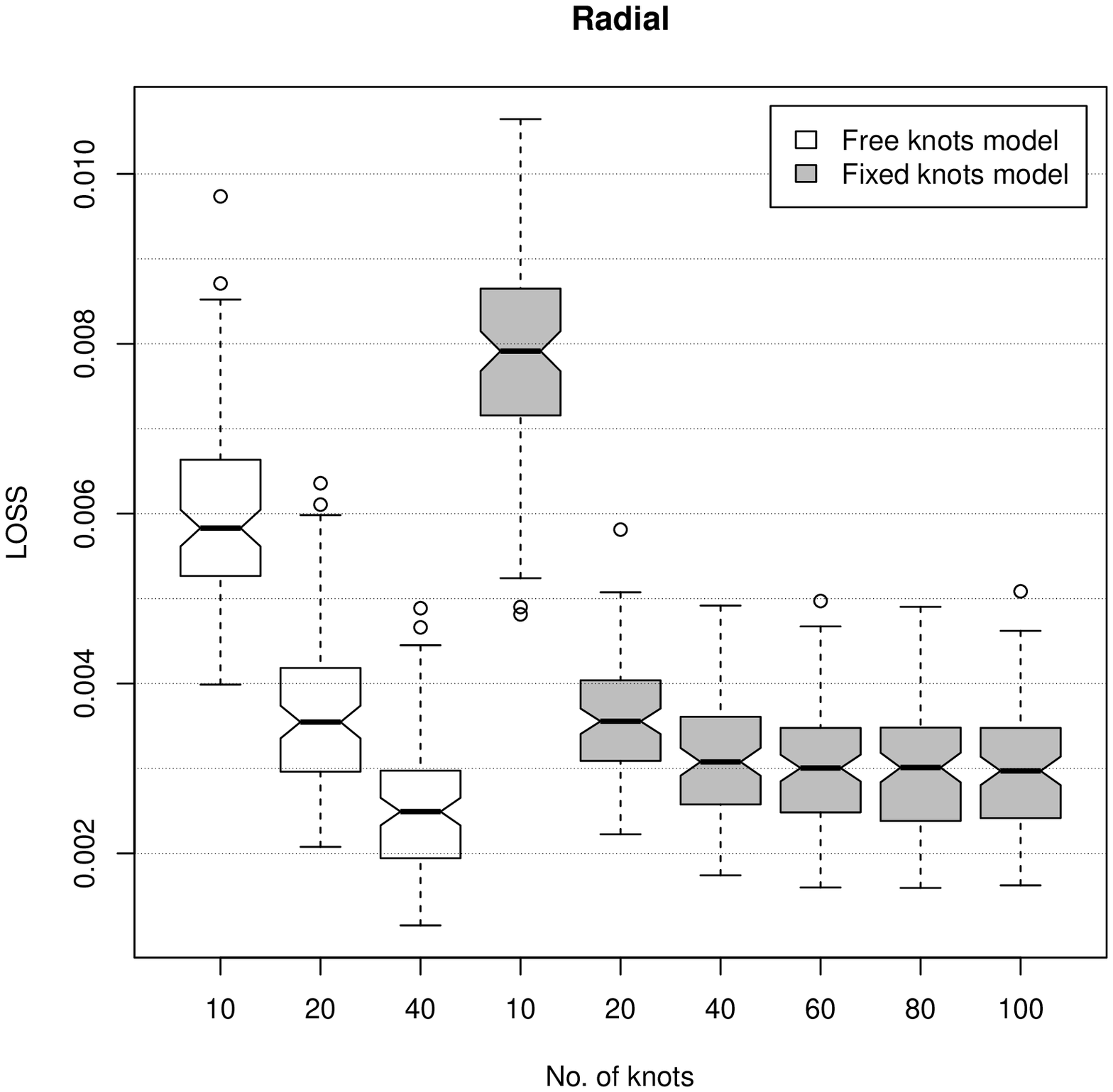}
\caption{Boxplots of the loss in the simulations for the radial mean function.}
\label{fig:hwang_loss}
\end{figure}

Figure \ref{fig:hwang_loss} displays boxplots of the losses for the
different number of knots in each model. The free knots model outperforms
the fixed knots model, but the improvement from using free knots are
not that large here since the covariate space is only two-dimensional,
which is small enough for the fixed knots to have a decent coverage.

